\documentclass[a4paper,11pt]{article}
\pdfoutput=1 

\usepackage{jcappub,color} 

\usepackage[T1]{fontenc} 
\newcommand{\black}{\color{black}}
\newcommand{\red}{\black}
\newcommand{\green}{\red}

\title{\black Search for anomalous \black features
in gamma-ray blazar spectra corrected for the absorption on the
extragalactic background light}

\author[a,b]{Alexander Korochkin,}
\author[a]{Grigory Rubtsov}
\author[a]{and Sergey Troitsky}
\affiliation[a]{Institute for Nuclear
Research of the Russian Academy of Sciences,\\
60th October Anniversary
Prospect 7a, Moscow 117312, Russia}
\affiliation[b]{%
APC, Universit\'e Paris Diderot,
B\^atiment Condorcet,
Case 7020,\\
75205 Paris Cedex 13,
France}

\emailAdd{st@ms2.inr.ac.ru}

\abstract{%
We consider the ensemble of very-high-energy gamma-ray sources observed at
distances and energies where a significant absorption of gamma rays is
expected due to pair production on the extragalactic background light
(EBL). Previous studies indicated that spectra of these sources,
upon correction for the absorption, exhibit unusual spectral hardenings
which happen precisely at the energies where the correction becomes
significant. Here, we address this subject with the most recent clean
sample of distant gamma-ray blazars, making use of published results of
imaging atmospheric Cerenkov telescopes and of the Fermi-LAT Pass~8
data, supplemented by the newest absorption models and individual
measurements of sources' redshifts. We perform a search for spectral
breaks at energies corresponding to unit optical depth with respect to the
absorption on EBL. These energies are different for distant and nearby
objects, and consequently, such features \red may not \black be related to
intrinsic properti\red{}e\black{}s of the sources. While in some spectra
such breaks are not seen, hardenings at distance-dependent energies are
present in many of them, though the overall statistical significance of
the effect is lower than reported in previous studies. The dependence of
the break strength on the redshift found earlier is not confirmed in the
new analysis. }

\begin{document}
\begin{flushright}
INR--TH--2018--025
\end{flushright}
\maketitle
\flushbottom
\section{Introduction}
\label{sec:intro}
It is well known since 1960s \cite{Nikishov} that energetic gamma rays
produce electron-positron pairs on the extragalactic background light
(EBL). Because of this process, gamma-ray flux from distant sources is
attenuated severely and the propagation distance is limited.
Very-high-energy (VHE) gamma rays (energy $E \gtrsim 100$~GeV) scatter
most efficiently on the infrared and visible background photons, and
information on the intergalactic absorption may be \red deduced \black from
VHE observations of emitters located at cosmologically large distances.

There exists some long-standing controversy in the determination of the
EBL intensity, with direct measurements indicating consistently
higher EBL than indirect constraints \red do\black. Direct observations
are very difficult to carry out because of strong foreground contamination
by the Zodiacal light and by the Galactic foreground. Nevertheless,
numerous attempts gave coherent, though quite uncertain, results
\cite{Sano:2015bsa, Sano:2015jih, Matsumoto:2015fma, Tsumura:2013iza,
CIBER}. At the same time, there are two groups of methods that set lower
and upper limits on the EBL. The first group is based on the deep field
observations of the Universe. This approach uses the method of galactic
counts and results in lower limits on the EBL intensity, given just by the
sum of contributions from guaranteed light sources
\cite{Madau:1999yh, Driver:2016krv, Keenan:2010}. On the other hand, upper
constraints come from gamma-ray observations of blazars
\cite{Ahnen:2016gog, Biteau, Abramowski:2013, Fermi-opacity}. As the VHE
part of the blazars spectra is highly absorbed by the EBL,
the requirement that the intrinsic spectrum is not too hard, or is a
 smooth extrapolation from lower energies, provides constraints on the
attenuation and thus on the EBL. These gamma-ray upper limits assume
standard absorption on the EBL and therefore cannot be used in a work
testing possible deviations from this standard picture.
Finally, numerical simulations include more theoretical input from the
star formation rate history but point to a similar range of intensities as
the galaxy-count methods. The situation remains unsettled: two independent
observations, published in 2017 and using different sophisticated
techniques to get rid of the foreground systematics~\cite{CIBER, notCIBER},
point consistently to the EBL intensities considerably higher than the
most recent theoretical models. For the purposes of the present study,
we need a conservative (low-absorption) model and use that of Korochkin
and Rubtsov (2018) \cite{Korochkin2018} as the most recent available
benchmark for the low-EBL theoretical models. Variations of the
absorption model will be discussed in the context of systematic
uncertainties of our results.

Distant blazars \red are \black observed in energetic gamma rays from the
early days of the VHE astronomy. It has been pointed out long ago that
some of them are seen from distances for which the optical depth with
respect to the pair production on EBL is significant. This apparent
tension with expectations was dubbed ``the infrared/TeV crisis''
\cite{IR-TeV-crisis}. However, subsequent improvements in EBL models
demonstrated that the tension is not that strong. \red While
\black apparent hardening of the intrinsic spectrum (corrected for the
pair-production attenuation, that is ``deabsorbed''), seen in several
individual sources, might be related to physical conditions in particular
objects, a successful model for these hardenings is missing
\cite{Aha-eating}.

The situation has been changed when the number of sources observed at
large optical depths became sufficient for studies of their ensembles.
The study of a sample of 7 sources observed by
imaging atmospheric Cerenkov telescopes (IACTs) at optical depths $\tau
\gtrsim 2$ by Horns and Meyer \cite{HornsMeyer} \red suggested \black that
the energy at which \red these \black hardenings happen is correlated with
the distance to the source in such a way that the spectra become harder
only when the correction for the pair-production absorption is
significant. This correlation can hardly be caused by any physical reason
and suggests that the optical depth is estimated incorrectly. The
statistical significance of this ``pair-production anomaly'' was estimated
at the $\sim 4 \sigma$ level
\label{sigma-footnote}\footnote{Hereafter, we quote statistical
significances in terms of standard deviations, $\sigma$, as it is
customary in the high-energy physics. One should always understand the
following meaning of these estimates: the probability that the observed,
or stronger, effect appears as a random fluctuation is the same as it
would happen for a normally distributed random quantity deviating from the
mean by this number of standard deviations. The underlying probability
distribution is never Gaussian in the analyses we discuss.}%
\footnote{See also the discussion of tests~\cite{Biteau} of this
result in Sec.~\ref{sec:disc:comparison:Biteau}.}.

The \red largest \black sample of sources \red detected \black at large
optical depths ($\tau \gtrsim 1$, 15 objects observed by IACTs and 5
objects observed by Fermi Large Area Telescope, LAT) has been considered
\cite{gamma} in 2014. There, we first confirmed the existence of significant
hardenings in deabsorbed spectra of a number of objects at the energies
where the correction for pair production becomes important. We then
considered the full sample of objects, including those with and without
statistically significant hardenings, and fit their deabsorbed spectra
with the broken power-law function, assuming the break at the energy
$E_{0}$, at which the pair-production optical depth $\tau(E_{0})=1$
(these energies are different for sources at different redshifts). We
found that the break strength, that is the difference between power-law
indices below and above $E_{0}$, grows with the source redshift $z$.
Assuming the growth linear in $\log z$, the statistical significance of
this distance dependence, again suggesting an unphysical origin of the
hardenings and hence an incorrect account of the absorption, was found
to be $\sim 12 \sigma$.

All these results are of great importance not only for the gamma-ray
astronomy, but also for other fields of physics and astrophysics. As it
will be discussed below, overestimation of the absorption can hardly be
understood without invoking new physical or astrophysical phenomena, and
\red its \black  most promising explanation requires the
existence of new light particles beyond those described by the Standard
Model of particle physics.

Since 2014, several important improvements changed the field. Firstly,
many new observations of distant VHE blazars have been performed by IACTs.
Secondly, the Fermi-LAT team not only accumulated additional years of
statistics, very important for faint distant VHE sources, but also issued
the new ``Pass~8'' reconstruction \cite{Pass8} improving the data
processing. Another related novelty is the 3FHL catalog \cite{3FHL} of
hard-spectrum sources making pre-selection of the sample easier and more
uniform than before. Thirdly, from the EBL side, new and improved
conservative theoretical models have become available~\cite{Korochkin2018,
Stecker, Franceschini2017}, but at the same time new sophisticated
observational studies strengthened the tension in the determination of the
background radiation intensity \cite{CIBER, notCIBER}. Finally, for a
number of objects, new information on their redshifts became available.
The main purpose of this work is to revisit the claims of
Refs.~\cite{HornsMeyer, gamma} with \red an extended \black high-quality
sample of sources, taking into account all the listed improvements and
making use of a more conservative analysis method. We also present some
tests demonstrating independence of our results from potential biases and
their stability with respect to systematic uncertainties. The result of
the present study
\red favours  the presence of \black
spectral features precisely at the energies for which the correction
becomes important\red, though with a significance of only $\sim 2\sigma$.
\black However, we do not confirm the dista\red{}n\black{}ce dependence of
the break strengths.

The rest of the paper is organized as follows. In Sec.~\ref{sec:data}, we
discuss gamma-ray data used in our study.
Section~\ref{sec:data:criteria} summarizes general criteria for the
sample construction. Section~\ref{sec:data:IACT} detalizes how they are
implemented for the objects observed by IACTs, while
Sec.~\ref{sec:data:Fermi} discusses the construction of the Fermi-LAT
subsample. We list the resulting sample of \red 31 \black sources and
discuss its general properties in Sec.~\ref{sec:data:sample}.

Section~\ref{sec:anal}  describes the data analysis procedure and presents
the main results of the paper. We describe how the correction to the
pair-production opacity is implemented and perform a combined analysis of
the full sample of \red 31 \black sources. We fit all the spectra with
absorbed broken power laws, assuming that the break energy $E_{b}$ is
a free parameter, and demonstrate that the fits suggest $E_{b}
\approx E_{0}(z)$. Then, we repeat the fits with fixed $E_{b}=E_{0}(z)$
and \red demonstrate that indeed \black
spectral features at \red the \black distance-dependent energies \red are
slightly favoured\black. This result
is discussed in detail in Sec.~\ref{sec:disc}. There, we first turn, in
Sec.~\ref{sec:disc:syst}, to potential biases and systematic uncertainties
which might affect our result. In Sec.~\ref{sec:disc:syst:absorb}, we
discuss variations in the absorption model: besides the most recent model
we use throughout the paper, Korochkin and Rubtsov
(2018)~\cite{Korochkin2018}, we repeat our analysis with three other
EBL models, namely: Gilmore et al.\ (2012) fixed \cite{Gilmore},
which was used in our 2014 paper; Franceschini and Rodighiero
(2017)~\cite{Franceschini2017}; and a toy upper-limit model artificially
normalized to the most recent direct observational data~\cite{CIBER,
notCIBER} on the background-light intensity. Next, in
Sec.~\ref{sec:disc:syst:Malmquist}, we discuss a possible bias related to
the fact that only brightest sources are seen at large distances, the
Malmquist bias. We extend our sample to include numerous sources not
detected in the highest-energy spectral bins, but satisfying at the same
time all other selection criteria. We use flux upper limits for the
undetected sources and demonstrate that they are consistent with the
results of Sec.~\ref{sec:anal}. In Sec.~\ref{sec:disc:syst:classes}, we
address another potential bias related to the fact that, generally,
blazars detected at higher redshifts are mostly flat-spectrum radio
quasars (FSRQs) while those at lower redshifts are mostly BL~Lac type
objects (BLLs). FSRQs might have intrinsic features in the spectrum which
mimic the distance dependence just because all observed FSRQs are distant.
We demonstrate that the problematic spectral features do not depend on the
synchrotron peak frequency often used to distinguish between different
classes of blazars. In Sec.~\ref{sec:disc:syst:tail}, we discuss the
effect of uncertainties in the energy determination, which may have
important consequences for the estimation of the spectral shape at the
highest energies for steeply falling spectra. We perform various tests and
conclude that this effect is unlikely to affect our results. Finally, in
Sec.~\ref{sec:disc:syst:z} we attempt to estimate a potential impact of
possible wrong determination of distances to the sources. Since 2014, new
redshift data removed \red 4 \black of 20 sources from the original sample
of Ref.~\cite{gamma}. We randomly remove 1, 2 or 3 sources from our new
sample of \red 31 \black objects and demonstrate that, in most cases, this
does not change significantly our results, therefore suggesting that (a
reasonable number of) erroneous redshifts can hardly cause the effect we
observe. Therefore, we conclude that none of the biases or uncertainties
we are aware of \red may affect \black the principal result of the paper,
and continue in Sec.~\ref{sec:disc:comparison} with a brief account of
novelties and differences of this result in comparison with previous
studies.

While the outline of our study and a summary of results are given in this
Introduction, we present a brief account of our conclusions and discuss
possible approaches to the interpretation of our results in
Sec.~\ref{sec:concl}. Appendix~\ref{sec:appendix} presents observed and
best-fit deabsorbed spectra of \red 31 \black blazars entering our main
sample.

\section{The data}
\label{sec:data}

\subsection{Selection criteria}
\label{sec:data:criteria}
\begin{figure}
	\centerline{\includegraphics[width=0.67\columnwidth]{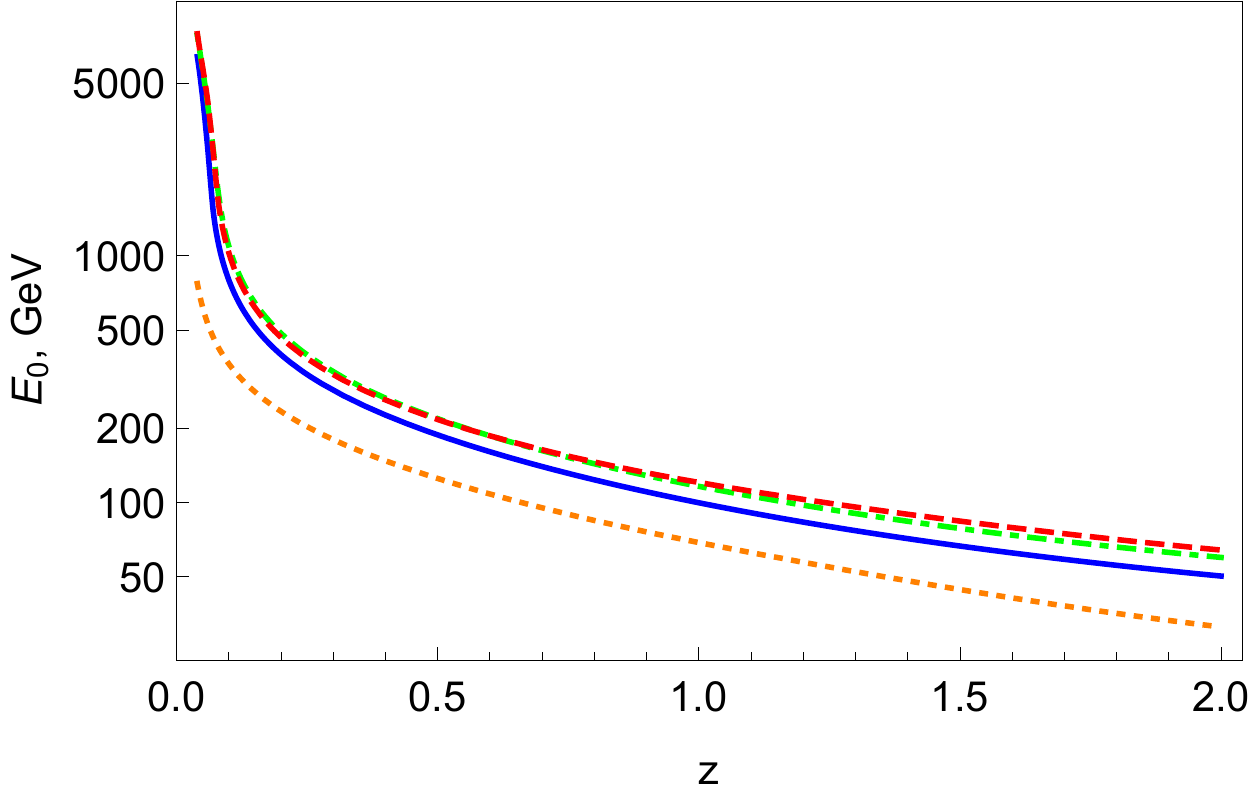}}
	\caption{\label{fig:E0}
			The energy $E_{0}$ corresponding to the pair-production  optical depth
			$\tau=1$ as a function of the redshift $z$ for different absorption
			models: Korochkin and Rubtsov (2018)~\cite{Korochkin2018} (full line);
			Franceschini et al.\ (2017)~\cite{Franceschini2017} (dashed line);
			Gilmore et al.\ (2012) fixed~\cite{Gilmore} (dash-dotted line);
			and the toy high-absorption model discussed in
			Sec.~\ref{sec:disc:syst:absorb} (dotted line). }
\end{figure}

The purpose of the present study implies the use of high-quality gamma-ray
spectra of sources located at large, confidently known distances. The
obvious candidates are blazars with firmly measured redshifts. For various
redshifts $z$, the absorption due to $e^{+}e^{-}$ pair production becomes
important at different energies of photons: for more distant sources, pair
production affects the spectra at lower energies\footnote{\label{fo}This
is true for energies below $\sim 100$~TeV which we discuss here.}. The
benchmark energy, at which the absorption becomes important, corresponds
to the optical depth $\tau=1$ and is denoted as $E_{0}$ hereafter,
\[
\tau (E_{0}, z) =1,
\]
which determines $E_{0}(z)$ for a given absorption model, see
Fig.~\ref{fig:E0}.

To study the behaviour of a spectrum in the region of strong absorption,
one needs observations at $E \gtrsim E_{0}$\red; therefore, \black
the use of different instruments \red is required \black for objects
located at different redshifts \red: \black for large $z$, $E_{0}$ is
typically dozens of GeV, and the relevant instrument is Fermi LAT, while
for less distant objects, $E_{0}$ is of order of several hundred GeV, and
the data of IACTs should be used. Though the data are published in quite
different ways, it is essential to treat them on equal footing. Here we
discuss our general requirements governing selection of the objects, which
will be implemented in the next two subsections for the Fermi-LAT and IACT
data.

\underline{Redshift criteria.} Incorrectly determined or uncertain
redshifts may hurt any of blazar studies. This happens quite often,
especially for BLLs whose spectra do not possess strong emission lines. We
require a firm spectroscopic redshift for the objects included into our
sample. More specifically, we start from a preselected sample of gamma-ray
blazars and check, for every object, the relevant redshift information and
references in the NASA/IPAC Extragalactic Database
(NED)\footnote{Available at \url{http://ned.ipac.caltech.edu}.}.
The redshifts are accepted if they satisfy (R1) and one of (R2A), (R2B),
(R2C) criteria:

(R1) the redshift is spectroscopic (not photometric) and is derived
from \textit{emission} lines\red\footnote{\red Since the absorption-line
redshift gives a lower limit to the actual one, inclusion of objects with
emission lines in the spectra would be conservative in terms of the
search of anomalies; however, we removed 4 such objects from the sample
following the Reviewer's comment.\black}\black%

AND

EITHER (R2A) the redshift is determined in a dedicated study (if several
dedicated studies give different results, the latest one is used),

OR (R2B) the redshift quoted in NED is determined in 2dF Galaxy Redshift
Survey \cite{2dF}  or 6dF Galaxy Survey \cite{6dF},

OR (R2C) the redshift quoted in NED is determined in the Sloan Digital Sky
Survey (SDSS), the result is unique and does not change from one release
to another.

These criteria are based on the previous experience in the use of redshift
data and are of course ad hoc ones. Their application removes many
uncertain redshifts.

\underline{Spectral criteria.}
In our work, we use binned gamma-ray spectra, because only this
information is publicly available for the objects observed by IACTs. We
impose the following criteria for the spectra:

(S1) information for at least 5 energy bins is available;

(S2) the lowest energy $E_{\rm last}$ of the last spectral bin
satisfies
$E_{\rm last}>E_{0}$ AND the lowest energy
of the third from below spectral bin is below $E_{0}$.

These criteria are aimed at the reconstruction of spectra at $E$ both
below and above $E_0$ with reasonable accuracy.

Several comments are in order. Firstly, (S2) depends on the assumed
absorption model, and therefore the samples we use in
Sec.~\ref{sec:disc:syst:absorb}, when studying the sensitivity of our
results to the choice of this model, are slightly different \red one from
\black another. Secondly, while we require a nonzero measurement in the
last bin for our main study, we allow for an upper limit when studying
potential effects of the Malmquist bias in
Sec.~\ref{sec:disc:syst:Malmquist}.

\subsection{IACT subsample}
\label{sec:data:IACT}
To construct the sample of blazars observed by IACTs, we start with
the database consolidating results published by different observatories,
the TeVCat online source catalog~\cite{TeVCat}\footnote{Available at
\url{http://tevcat.uchicago.edu}.}.
The preselected sample includes objects classified as blazars there
(classes ``HBL'', ```IBL'', ``LBL'', ``BL Lac (class uncertain)'',
``FSRQ'' and ``blazar''), \red 72 \black objects in total as of
201\red9\black. Then, for each object in the sample, we checked the
redshift selection rules (R1) and ((R2A) or (R2B) or (R2C)), using
references from NED, and availability of gamma-ray spectra satisfying
(S1), using references from TeVCat.

Some objects have been observed several times, often by multiple
instruments. To avoid double counting of information, which may
artificially increase or dilute observed effects, we use only one
observation for each source. It is chosen on the basis of better
statistics (larger effective exposure in cm$^{2}\cdot$s). Because of
strong variability of many blazars, we never combine any two spectra
from observations performed at different epochs to \red extend \black the
energy coverage. We also never combine Fermi-LAT spectra with those
obtained from IACT observations because of potential systematic
differences between the instruments' energy scales. These requirements
represent a major refinement with respect to some of preceding studies.

The next step requires to fix the absorption model.
We use the most recent EBL model \cite{Korochkin2018} as our baseline
low-absorption model; when variations of the model are considered
(Sec.~\ref{sec:disc:syst:absorb}), this step is repeated. At this step, we
calculate the value of $E_0$ for every source and check the condition
(S2). We are left with \red 26 \black objects which, together with the
sources observed by Fermi LAT, are listed in Table~\ref{tab:list}.

\subsection{Fermi-LAT subsample}
\label{sec:data:Fermi}

For blazars observed by Fermi LAT, our starting point is the 3rd
Fermi-LAT Catalog of High-Energy Sources,
3FHL~\cite{3FHL}\footnote{Available at
\url{http://fermi.gsfc.nasa.gov/ssc/data/access/lat/3FHL/} .},
presenting a list of sources detected confidently above 10~GeV. From the
catalog, we select 1212 objects classified as blazars (classes ``BLL'',
``bll'', ``FSRQ'', ``fsrq'', ``bcu'').
Since the redshifts presented in the catalog may
be doubtful or erroneous for distant sources, we check them manually to
satisfy (R1) and ((R2A) or (R2B) or (R2C)) criteria, using references from
NED, and apply \red a preselection \black constraint $z \ge 0.2$,
\red which is a consequence of (S2) and the Fermi-LAT energy coverage:
\black for less distant sources, $E_{0} \gtrsim 500$~GeV, \red that
\black is
beyond the limit of reasonable sensitivity of Fermi LAT. This procedure
results in a list of 309 blazars. For each of them, we use publicly
available Fermi-LAT \cite{LAT} data\footnote{Available at
\url{https://heasarc.gsfc.nasa.gov/FTP/fermi/data/lat/weekly/photon/} .}
to construct binned spectra satisfying (S1) automatically {\red  with the
help of standard routines from Fermi Science Tools 1.0.1. To this end, we
use Fermi Pass~8 \cite{Pass8} Release 3 data (version 303). We use
``SOURCE'' class events with version 2 instrumental response functions
``P8R3$\underline{~}$SOURCE$\underline{~}$V2'', recorded between August 4,
2008, and February 26, 2018 (Mission Elapsed Time interval 239557417 --
541338374). The background models used are
``gll$\underline{~}$iem$\underline{~}$v06.fits'' for the Galactic
component and
``iso$\underline{~}$P8R3$\underline{~}$SOURCE$\underline{~}$V2.txt'' for
the isotropic one. \black The background model also includes other sources
from 3FGL catalog within $5^\circ$ angular distance from the given source.
\black For most of the sources (305 of 309), the coordinates of the source
are taken from the 3rd Fermi-LAT Source Catalog, 3FGL \cite{3FGL}.
However, not all 3FHL sources have 3FGL counterparts, and for the
remaining 4 sources, we used 3FHL coordinates.

\red
Firstly, we have performed an independent fit with the standard
{\tt gtlike} routine with the power-law spectrum model ``PowerLaw2'' in
each of 8 energy bins with the width of 0.3 dex, for $E\ge 2$~GeV. The
lower energy limit of 2~GeV was chosen to cut the inverse-Compton peak
energy range in order to improve broken power-law fits used in our
subsequent analysis.

Despite the fact that the use of {\tt gtlike} is the preferred method of
constructing sources' spectra, it demonstrates an unstable
behavior when the estimated number of counts in the energy bin is
lower than one. In particular, the fit may not converge or converge
to zero flux with unphysical value of the flux upper limit. Moreover,
even when the number of photons is less than three, the estimated
confidence interval for the flux follows the simple symmetric Gaussian
distribution but not the Poisson one, as it should. Clearly these features
may affect further analysis and to avoid the peculiar behavior of the fit
optimization procedure we have stepped one level back to use the Fermi
Science Tools routines which provide an input for {\tt gtlike}. Namely,
the standard {\tt gtsrcprob} routine assigned the weight to each photon
which is the probability to originate from a given source within the given
source model. The sum of the corresponding weights provides an estimate
for the number of photons from the given source and from each of the
background sources.

At the next step, we fix the absorption model and apply the criterion (S2)
to the obtained spectra. For the Fermi-LAT subsample, the criterion is used
as follows: the source is accepted if it has at least one
photon in the bin above $E_{0}$ and the probability that this photon is
associated with the source is more than 99\% (according to the {\tt gtsrcprob}
output). Only 5 of 309 blazars satisfy (S2) and enter our final sample.

Finally, for the bins which contain less than three photons we have
replaced the {\tt gtlike} flux with one obtained from summing weights from
{\tt gtsrcprob}. Since the background photons are considered separately, this
sum may be treated as the number of signal events with zero
background. We round this number to an integer (0 or 1 in most cases) and
estimate the Poisson-based 68\% CL interval for the flux in these bins
by division by the exposure calculated with the {\tt gtexpcube} routine.
We have checked that the above procedure results in the value which
agrees within $10\%$ with the output of {\tt gtlike} is cases when the
latter converges to a physical value. }

\subsection{The sample}
\label{sec:data:sample}

The list of \red 31 \black objects selected in the way described above,
Sec.~\ref{sec:data:IACT}, \ref{sec:data:Fermi}, is presented in
Table~\ref{tab:list}.
\begin{table}
\centering
\caption{\label{tab:list}
List of blazars selected for the main sample as described in the text.
Classes: B$=$BLL, F$=$FSRQ. Instruments: H$=$HESS, L$=$Fermi LAT,
M$=$MAGIC, V$=$VERITAS.
Note: the symbol in the last column means the
source does not satisfy the (S2) criterion for \red both \black the
Franceschini et al.\ (2017) and Gilmore et al.\ 2012 fixed  ($*$) or for
Korochkin and Rubtsov 2018 ($\dagger$)
EBL models, respectively.
} \smallskip
\begin{tabular}{ccccccc}
\hline
Name&Redshift&Class& Instrument & Ref.\ (spec.) & Ref.\ ($z$) & Note \\
\hline
B2~2114$+$33     & 1.596 & F &
L&&\cite{2012-ApJ-764-135S}&$*$\\
GB6~J0043$+$3426 & 0.966 & F & L & &\cite{2012-ApJ-748-49S}&\\
 PKS~1441$+$25   & 0.939 & F & M &
\cite{1512.04435}&\cite{2012-ApJ-748-49S}&\\
 4C~$+$55.17     & 0.896 & F &
L&&\cite{2005-AJ-130-367S}&$*$\\
PKS~0537$-$441   & 0.892 & F & L
&&\cite{1997-ApJ-207L-5P}&$*$\\
PG~1246$+$586    & 0.847 & B & L & &\cite{2009-ApJS-182-543A}&\\
\red Ton~0599    & \red 0.725 & \red F & \red M &
\red \cite{Ton0599-TeVPA} &\red  \cite{1003.3017}&\\
PKS~1424$+$240   & 0.605 & B & V &
\cite{0912.0730} &\cite{1701.04305}&\\
3C~279           & 0.536 & F & M &
\cite{0807.2822} &\cite{1996-ApJS-104-37}&\\
4C~$+$21.35      & 0.432 & F & M &
\cite{1101.4645} &\cite{1987-ApJ-323-108}&\\
PKS~1510$-$089   & 0.360 & F & M &
\cite{1806.05367}&\cite{1990-PASP-102-1235}&\\
\red OT~081 & \red 0.322 & \red B & \red H &
\red\cite{1708.01083} & \red\cite{1988-AnA-191-L16}& \\
OJ~287           & 0.306 & F &
V&\cite{1708.02160}&\cite{2010-AnA-516-60}&\\
1ES~0414$+$009   & 0.287 & B & H &
\cite{1201.2044}&\cite{1991-AJ-101-818}&\\
\red 1RXS~J023832.6$-$311658 &\red 0.233 &\red B& \red H &
\red\cite{1708.09612} & \red\cite{2012-ApJ-764-135S} & \\
1ES~1011$+$496  & 0.212 & B & M &
\cite{1602.05239}&\cite{2007-ApJ-667-21}&\\
1ES~1218$+$304   &0.182 & B & V &
\cite{1307.7051}&\cite{2012-ApJS-203-21}&\\
H~2356-309      & 0.165 & B & H &
\cite{1004.2089}&\cite{2014-ApJ-795-57}&\\
1ES~0229$+$200  & 0.140 & B & V &
\cite{1312.6592}&\cite{1993-ApJ-412-541}&\\
1ES~0806$+$524  & 0.138 & B & M &
\cite{1504.06115}&\cite{2012-ApJS-200-17}&\\
1ES~1215$+$303  & 0.131 & B & M &
\cite{1203.0490}&\cite{1701.04305}&\\
\red H~1426$+428$ & \red 0.129 & \red B & \red HEGRA &
\red \cite{HEGRA-H1426} & \red \cite{0911.0423} & \\
PKS~2155$-$304  & 0.116 & B & H &
\cite{1005.3702}&\cite{1993-ApJ-411L-63}&\\
1ES~1312$-$423  & 0.105 & B & H & \cite{1306.3186}
&\cite{2000-ApJ-999-L1}&\red $\dagger$\\
W~Com           & 0.102 & B & V &
\cite{0808.0889}&\cite{1701.04305}&$*$\\
RGB~J0152$+$017 & 0.080 & B & H &
\cite{0802.4021}&\cite{1998-ApJS-118-27}&\\
PKS~2005$-$489  & 0.071 & B & H &
\cite{0911.2709}&\cite{1987-ApJ-318-L39}&\\
PKS~0548$-$322  & 0.069 & B & H &
\cite{1006.5289}&\cite{astro-ph/0605448}&$*$\\
\red PGC~2402248&\red 0.065 & \red B & \red M &
\red \cite{EHBL}& \red\cite{ATel-11621}&\red $*$\\
1ES~1959$+$650  & 0.048 & B & HEGRA &
\cite{AnA-406-2003-9}&\cite{1206.0031}&\\
Mrk~501         & 0.034 & B & H &
\cite{1509.04458}&\cite{1993-AnAS-98-393}&\\
\red Mrk~421    &\red 0.031 & \red B & \red HEGRA &
\red \cite{0205499} & \red \cite{CfA} & \\
\hline
\end{tabular}
\end{table}
For IACT observations, references for binned spectra used in this work are
given there. We also list classes of the blazars (BLL or FSRQ) as
determined in the catalogs used to compile the samples, TeVCat and 3FHL.
In addition\footnote{Following a request of the anonymous
referee.}, we give also references to original spectroscopic measurements
of redshifts.

\begin{figure}
	\centerline{\includegraphics[width=0.67\columnwidth]{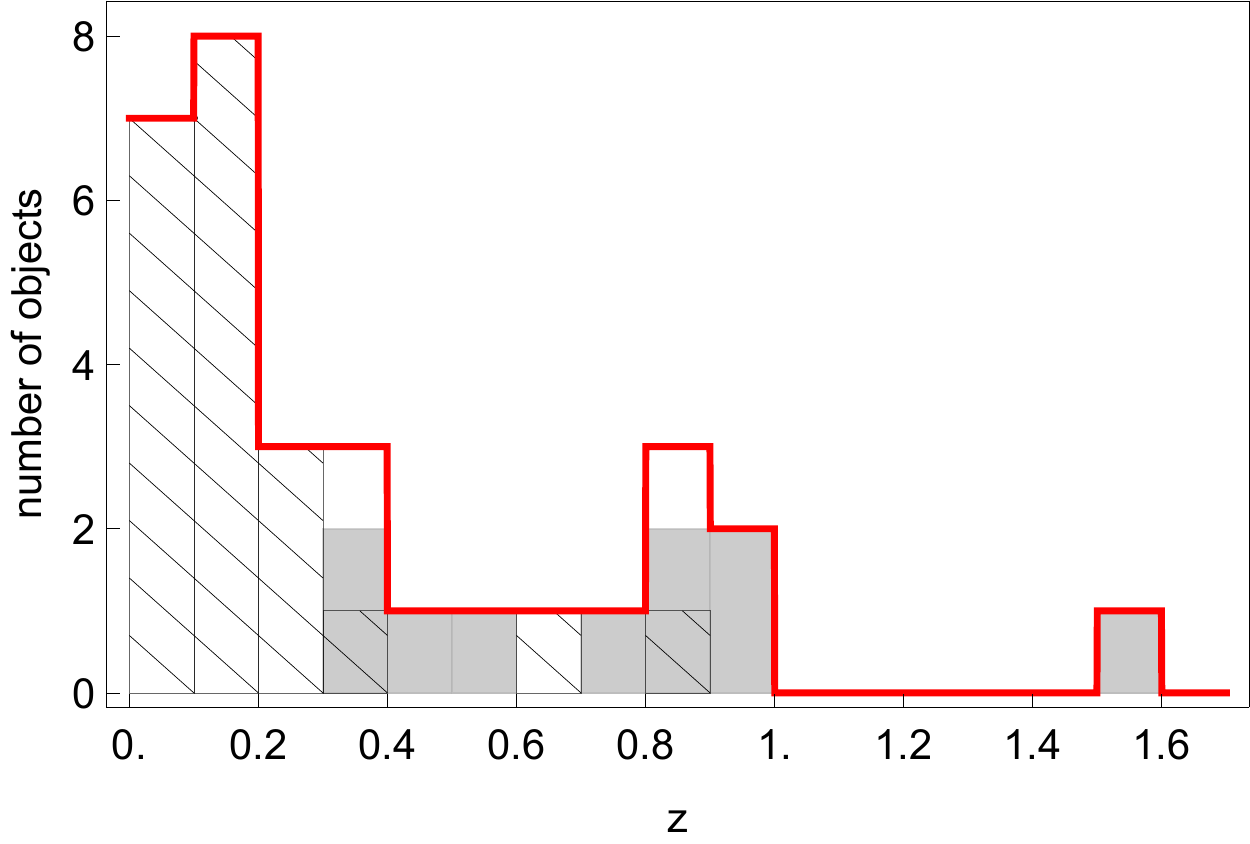}}
	\caption{\label{fig:z-distr}
			Distribution of \red 31 \black objects from the sample in redshift (full
			red histogram). Hatched and shaded histograms give separate distributions
			for BLLs and FSRQs, respectively.}
\end{figure}

Figure~\ref{fig:z-distr}
represents the distribution of blazars in our main sample in redshift. One
may note that, though BLLs dominate at short and moderate distances and
FSRQ\black{}s \black are in general farther away, both classes cover large
ranges of redshifts. We will return to this subject in
Sec.~\ref{sec:disc:syst:classes}.

\section{Data analysis and results}
\label{sec:anal}

The key point of our study is the use of blazar spectra corrected for the
pair production in a minimal-absorption model. Since, for IACT
observations, only binned spectra are available, we adopt our analysis
procedure for this case even for Fermi LAT, in order to process all
observations in a uniform way. In contrast with many previous studies,
we do not use bin-by-bin deabsorption
because it can introduce systematic biases in the highest-energy bins for
the following reasons. The optical depth $\tau$ is a
function of the incoming photon energy $E$ and the source redshift $z$. It
is important to note that the absorption is not uniform within the bin:
photons of higher energy are absorbed stronger\footnote{See footnote on
page~\pageref{fo}.}. For the energies and distances at which the absorption
is significant, this effect may affect strongly the result. It can be
taken into account if the shape of the observed spectrum within the bin is
known, as we have done in Ref.~\cite{gamma}; however, in practice,
statistical uncertainties in the highest energy bins often make the
determination \red of the shape \black uncertain. The uncertainty of the
spectral shape within the bin translates  into the error of
deabsorbed flux. The latter effect is enhanced in bins, \red within \black
which the absorption effect grows dramatically. It is shown that for wide
energy bins of 0.3 dex used for the analysis of Fermi LAT sources, the
uncertainty related to deabsorption contributes significantly to the flux
error in energy bins above $E_0$. Therefore the absorption of the model
spectra instead of the deabsorption of the observed one is performed in
our analysis.

\begin{figure}
	\centerline{\includegraphics[width=0.75\columnwidth]{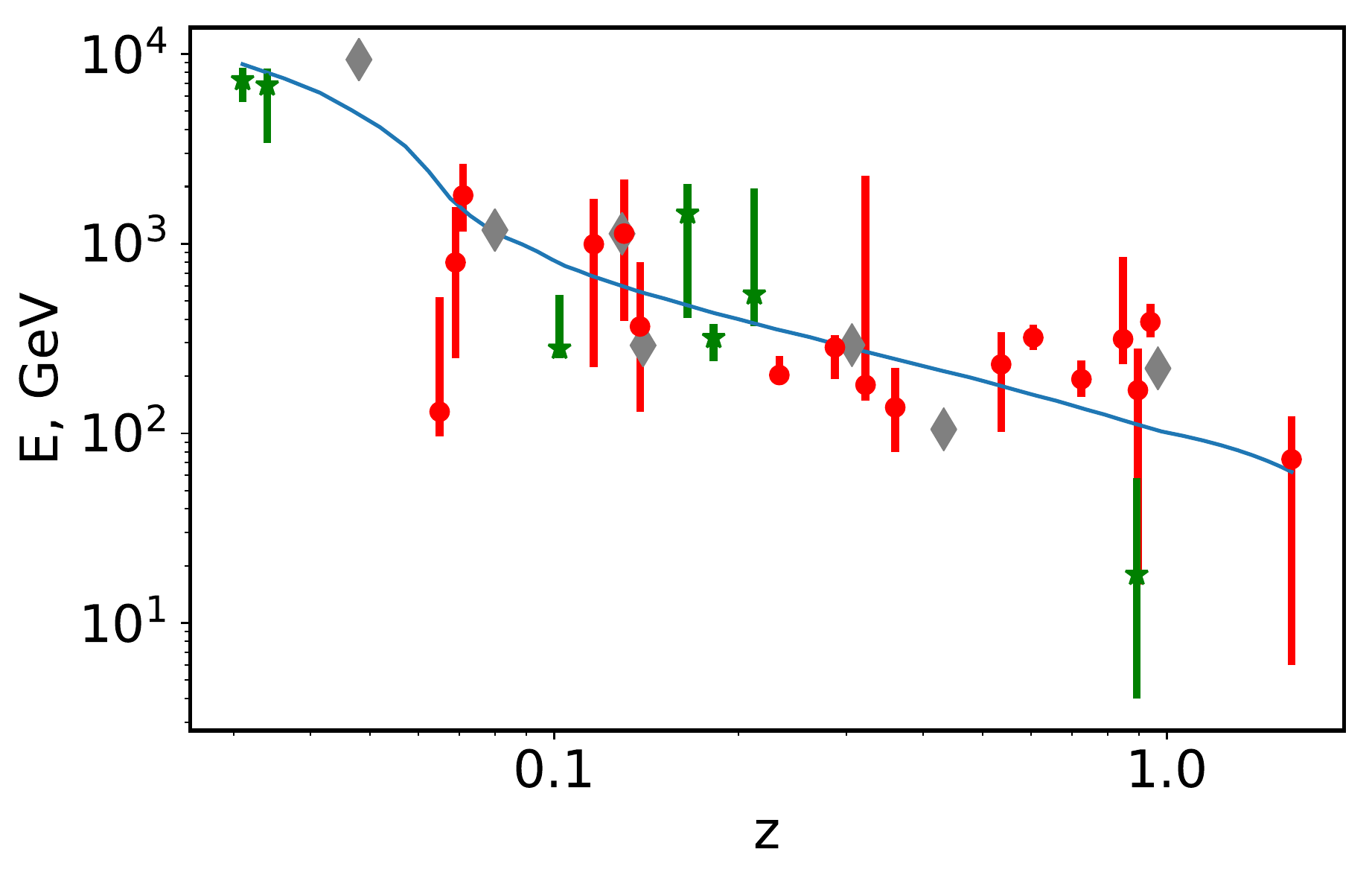}}
	\caption{\label{fig:break-positions}
Positions of the break energies (red points: hardenings, green
    asterisks: softening), treated as free parameters in independent
    fits, compared to $E_{0}(z)$ (red line). Objects for which \red
    the broken power-law fit is not better than the power-law fit by
    $\Delta\chi^{2}<1$, that is for which best-fit break is consistent
    with zero, are shown as gray diamonds (since the error bars are
    determined by $\Delta\chi^{2}=1$ \cite{NR}, they are infinite for
    these points.\black}
\end{figure}

Following the logic of previous studies, we start from the assumption
about the intrinsic spectrum of a blazar, for which we use a broken power
law. First, we make the break energy $E_{\rm b}$ a free parameter of the
fit, so that it is adjusted independently for every source. We account for
the absorption with the selected EBL model, integrate the obtained spectrum
over the energy bins and compare resulted bin-by-bin fluxes with the
observed data points. In this way, we fit the data with the four parameters
of the intrinsic spectrum, that is the overall normalization, the break
position $E_{\rm b}$, the spectral index at $E<E_{\rm b}$ and $\Delta \Gamma$,
the difference between spectral indices at $E>E_{\rm b}$ and $E<E_{\rm b}$.
Technically, all our fits are performed with the method described in detail
in Ref.~\cite{NR}.
Figure~\ref{fig:break-positions}
compares the fitted break positions $E_{\rm b}$ with the value of $E_{0}(z)$;
we see that they agree well to each other: averaged over the sample,
$\log_{10} \left(E_{\rm b}/E_{0}(z) \right)= \red -0.02 \pm 0.37$ \black
(for \red 22 \black sources with positive breaks, i.e. \ spectral
hardenings, we get \red $0.01 \pm 0.36$\black).

\begin{figure}
	\centerline{\includegraphics[width=0.67\columnwidth]{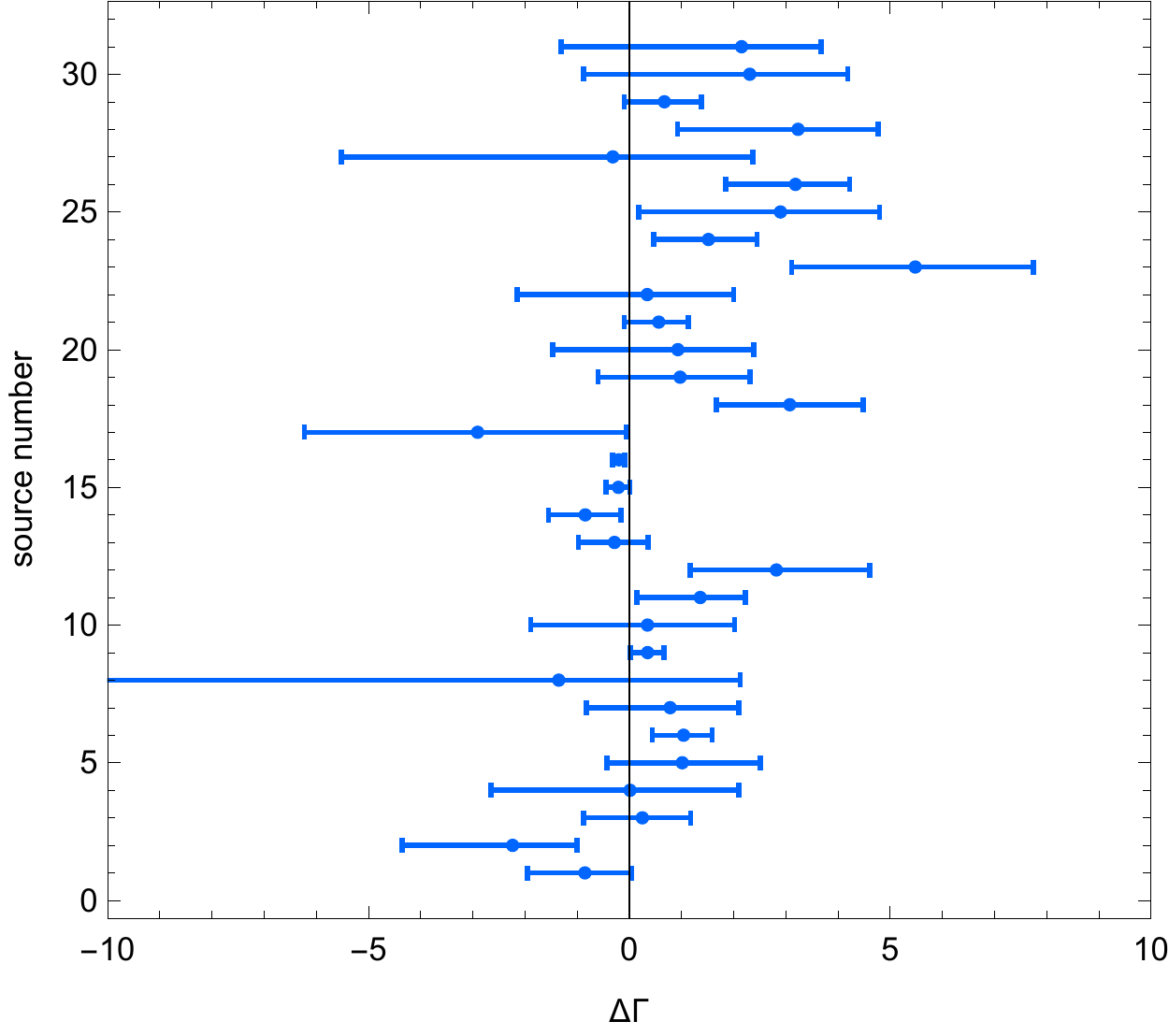}}
	\caption{\label{fig:main}
			\black Breaks in deabsorbed blazar spectra at $E_{\rm b}=E_{0}(z)$.
			See the text for details.}
\end{figure}

Then, we turn to similar fits with the break energy fixed at
$E_{\rm b}=E_{0}(z)$. Recall that this energy depends on the redshift of
the source $z$, cf.\ Fig.~\ref{fig:E0}, and therefore we assume breaks at
different energies for different sources.
Values of $\Delta \Gamma$ obtained in this way for individual sources
are given in Appendix~\ref{sec:appendix} and graphically presented in
Fig.~\ref{fig:main}.
One sees that while some spectra have $\Delta \Gamma$ consistent with
zero, indications for $\Delta \Gamma >0$ exist in many cases. For the
entire sample, the assumption of $\Delta\Gamma=0$ results in \green
$\chi^{2}=44.0$ for 31 \black degrees of freedom,
corresponding to the probability $P \simeq \green 0.06 \black$
for the observed values of $\Delta \Gamma$ at $E_{0}(z)$ to
occur as a result of a random fluctuation. Therefore, our results
disagree with the hypothesis of the absence of breaks at $E_{0}(z)$
at \green 1.9 \black standard deviations (recall the footnote at
p.~\pageref{sigma-footnote} for our conventions)\footnote{For these
probability estimates, we assume the chi-squared distribution. The
latter is valid for Gaussian \red errors \black and can be used here as
many independent factors contribute to uncertainties.}.

\begin{figure}
	\centerline{\includegraphics[width=0.67\columnwidth]{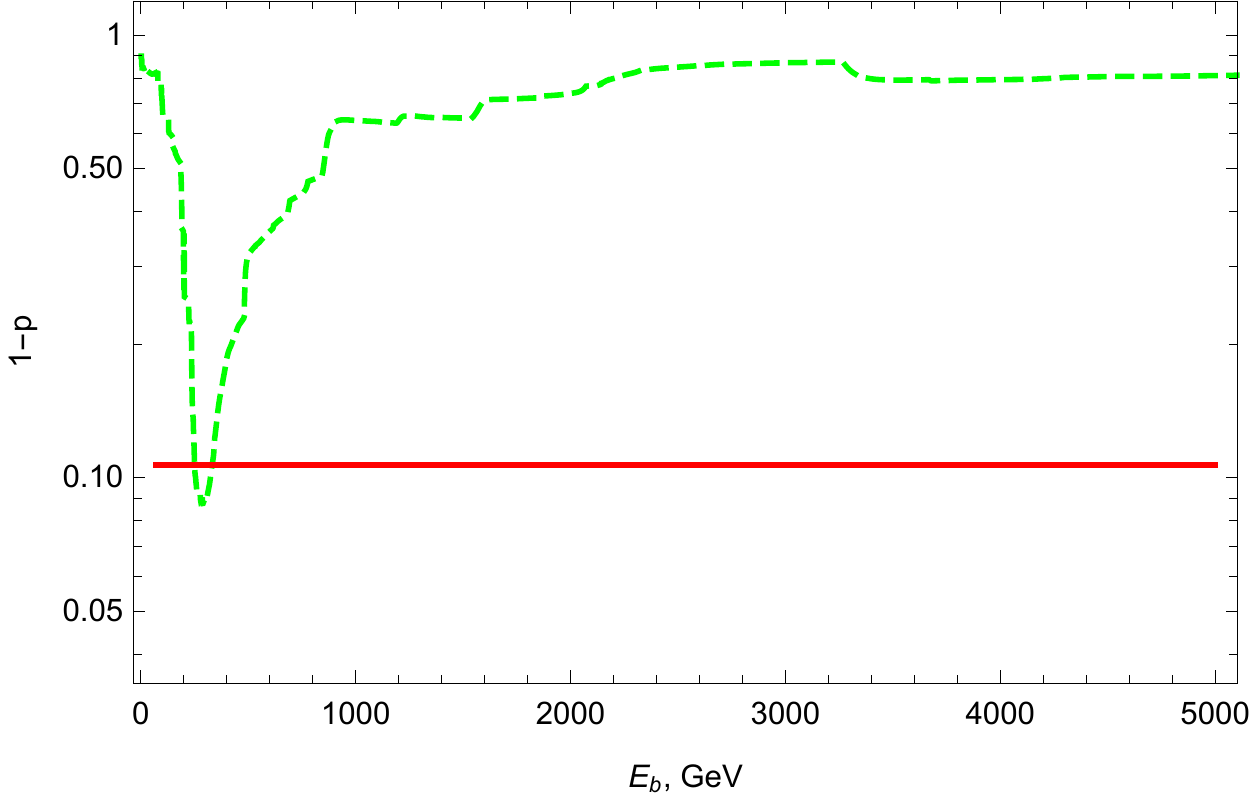}}
	\caption{\label{fig:pval}
			Comparison of the fit quality (good fits correspond to low $(1-p)$) for the
			assumptions of spectral breaks at the fixed energy $E_{\rm b}$ (dashed
			green line, as a function of $E_{\rm b}$) and at the distant-dependent
			energy $E_{0}(z)$ (full red line).}
\end{figure}

\black
We can compare the assumption of breaks at $E_{0}(z)$ with that of breaks
at some distance-independent energy, common to all sources. To this end, we
perform fits of all spectral data by absorbed broken power laws, fixing
the common break energy $E_{\rm b}$ for all objects in the sample, for
various values of $E_{\rm b}$. For every $E_{\rm b}$, we determine the
goodness of the fit, $p$, from the chi-squared distribution with the
corresponding number of degrees of freedom (for \red 225 \black data
points, \red 31 \black spectra fitted with 3 parameters each, the number of
degrees of freedom is \red 132\black). The plot of $(1-p)$ as a function of
$E_{\rm b}$ is shown in Fig.~\ref{fig:pval}
(the best possible fit would correspond to $(1-p)=0$, a very bad fit -- to
$(1-p)\approx 1$), together with the same quantity calculated for fits
with breaks at $E_{0}(z)$. The fit with breaks at distance-dependent
energies $E_{0}(z)$ is better than a fit with any fixed $E_{\rm b}$ \red
except for a narrow range $E_{\rm b} \sim (200 - 300)$~GeV.
\black

\section{Discussion}
\label{sec:disc}
In this section, we
perform a study of possible systematic effects affecting our results. We
will see that the effect we observed cannot be explained by any potential
bias or systematics we are aware of. Then we compare our result with
those of previous studies.

\subsection{Systematic uncertainties}
\label{sec:disc:syst}

\subsubsection{Absorption models}
\label{sec:disc:syst:absorb}
The results of the present study may indicate some problems with the
absorption model used. For this study, we used the most recent published
model by Korochkin and Rubtsov (2018), Ref.~\cite{Korochkin2018}. In this
section, we repeat our study with several other representative models to
see that the effect we found is present independently of the choice of the
model. Besides the baseline model of Ref.~\cite{Korochkin2018}, we
considered the following three models:

(i)~Gilmore et al.\ (2012) fixed~\cite{Gilmore}, used in our previous
study~\cite{gamma};

(ii)~Franceschini et al.\ (2017)~\cite{Franceschini2017};

(iii)~a toy upper-limit model normalized to the most recent
direct observations of EBL~\cite{CIBER, notCIBER}.

\begin{figure}
	\centerline{\includegraphics[width=0.67\columnwidth]{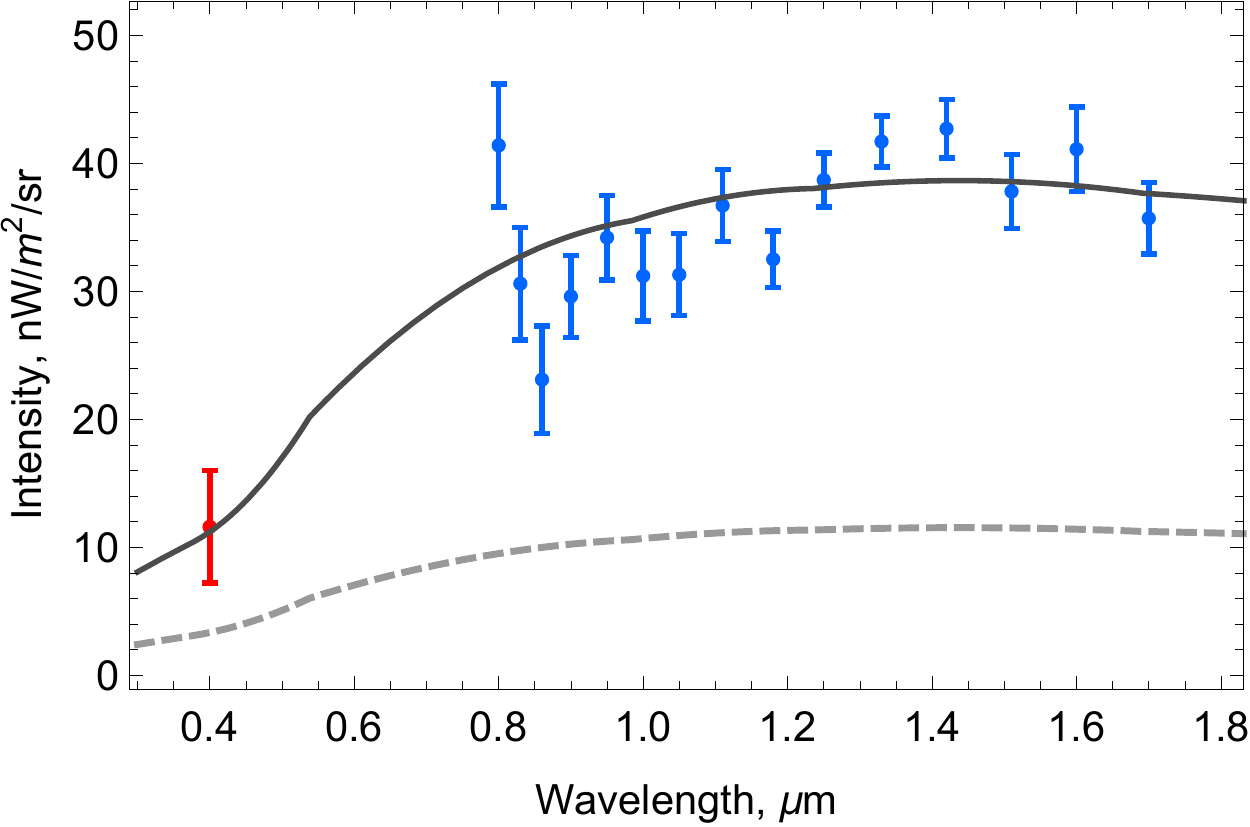}}
	\caption{\label{fig:EBLscaling}
			Measurements of the extragalactic background light from Ref.~\cite{CIBER}
			(blue) and Ref.~\cite{notCIBER} (red, point at 0.4~$\mu$m), together with
			the best-fit scaled intensity of Ref.~\cite{Franceschini2017} (full line).
			This scaling corresponds to the toy upper-limit model used to
			study systematic uncertainties in Sec.~\ref{sec:disc:syst:absorb} of this
			work. The original intensity of Ref.~\cite{Franceschini2017} is shown by a
			dashed line for comparison.}
\end{figure}

The latter model was obtained by scaling the model (ii) by a
multiplication factor fitted to the data of Refs.~\cite{CIBER, notCIBER}.
These two independent studies used different techniques to distinguish the
extragalactic background light from the foreground contribution:
Ref.~\cite{CIBER} used spectral templates, which are different for EBL
and for foregrounds, while Ref.~\cite{notCIBER} benefited from
observations of an absorbing cloud to separate foregrounds experimentally.
Figure~\ref{fig:EBLscaling}
presents 15 data points of Ref.~\cite{CIBER} and a single data point of
Ref.~\cite{notCIBER} together. All the points, corresponding to different
background-light wavelengths, are fitted nicely by the intensity of the
model~\cite{Franceschini2017} multiplied by a
wavelength-independent factor $3.35 \pm 0.07$. For the toy high-absorption
model we therefore take the model of Ref.~\cite{Franceschini2017}
uniformly upscaled by 3.35. \black The toy absorption model \black should
\black be considered as an upper limit on the opacity.

As it has been discussed above, the energies $E_{0}$, at which the
absorption becomes important, depend on the EBL model, and therefore
samples of blazars selected for the study are different for different
models. For the models~\cite{Franceschini2017, Gilmore}, some blazars from
Table~\ref{tab:list} do not enter the sample \red while one extra blazar
joined the set, see Notes \black in Table~\ref{tab:list}. For the toy
\black upper-limit \black model, the list of \red 29 \black sources in
the sample is available from the authors by request.

We present the resulting significance of the observation of spectral
features at $E_{0}(z)$ for various absorption models in
Table~\ref{tab:EBLmodels-sigma}.
\begin{table}
\centering
\caption{\label{tab:EBLmodels-sigma}
Results of the analysis performed with various models of pair-production
opacity discussed in the text. Number of objects in the sample and the
significance of breaks at $E_{0}(z)$ are given.
}
\smallskip
\begin{tabular}{ccc}
\hline
EBL model&Number of objects&Significance, $\sigma $\\
\hline
Korochkin and Rubtsov (2018) &\red  31 & \red 1.9 \black\\
\hline
Gilmore et al. (2012) fixed &\red 26 & \red 1.3 \black\\
Franceschini et al. (2017) & \red 26 & \red 1.5 \black\\
\black Upper-limit \black EBL normalized to~\cite{CIBER, notCIBER} &\green
29 & \black >10 \black\\
\hline
\end{tabular}
\end{table}
We see that for the low-absorption models, the
\black results are similar to those \black obtained for the model of
Ref.~\cite{Korochkin2018}.
Clearly, a 3.35 times increase in the opacity, suggested by
Refs.~\cite{CIBER, notCIBER}, had to result in strengthening the effect,
and we quantified this consistently within our approach\footnote{The
fact that the increase of the absorption up to the values
suggested by Ref.~\cite{CIBER} sharpens the problems of interpretation
of the blazar gamma-ray observations has been illustrated in
Ref.~\cite{ax-CIBER} for two particular sources.}.

\subsubsection{The Malmquist bias}
\label{sec:disc:syst:Malmquist}
\black Given the approach we use (selecting objects detected beyond
$\tau=1$), demonstrating that there is no Malmquist bias is a doable but
not-so-easy task. Indeed, \black inclusion of sources to our sample is
limited by observational capabilities of the instruments through the (S2)
criterion of Sec.~\ref{sec:data:criteria}. Therefore, sources
intrinsically more and more luminous are included at larger distances. In
principle, sources may have spectral features correlated with their
luminosity, and the flux selection might mimic the observed effect (a
particular model for this is not known). In this hypothetical scenario,
weaker objects not detected above $E_{0}$ would not have
spectral hardenings which we observe for \red some of \black the sources
included in our sample. This suggests a way to test this possible bias by
a study of objects not detected above $E_{0}$.

\begin{figure}
	\centerline{\includegraphics[width=0.67\columnwidth]{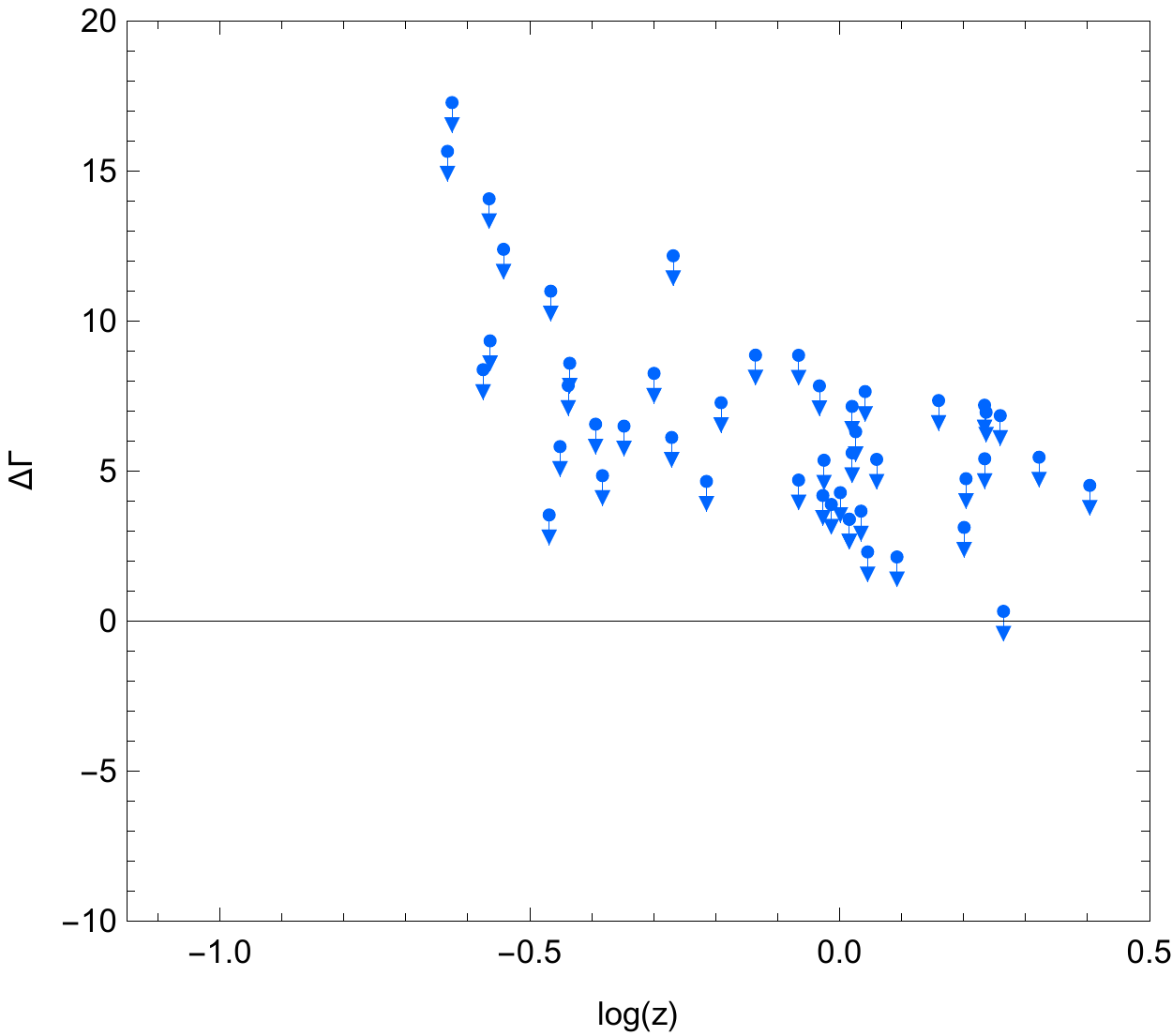}}
	\caption{\label{fig:UL}
			Estimated upper limits (95\%~CL, downward blue arrows) on the break
			strength $\Delta \Gamma$ for Fermi-LAT sources not detected at energies
			above $E_{0}$.}
\end{figure}

To this end, we consider the same pre-selected 3FHL sample described in
Sec.~\ref{sec:data:Fermi}  but relax the condition (S2) of
Sec.~\ref{sec:data:criteria}.
Note that we \red perform the test \black for Fermi-LAT sources only, not
including the sources observed only by IACT. \red The main reason for the
above limitation is that the flux upper limits are not available for the
most of the sources observed by IACTs. \black
We obtain
a sample of \green 47 \black blazars with redshifts between \green 0.233
\black and 2.534, which are detected significantly in \red the \black
spectral bin containing $E_{0}$, but not above. For each of the objects,
we determine upper limits on the strength $\Delta \Gamma$ of assumed
breaks at $E_{0}$ for these sources in the way described in
Sec.~\ref{sec:anal}.
\green Following the procedure described in Sec.~\ref{sec:data:Fermi} we
estimate the Poisson-based flux in the last bins and then calculate
95\% CL upper limits on $\Delta \Gamma$. The results of the calculation
of these upper limits are presented in Fig.~\ref{fig:UL}. \black
\red We see that the results obtained with the extended data set do not
contradict to those obtained with the main sample, and can conclude that
the Malmquist bias does not dominate the results. \black

\subsubsection{Source classes}
\label{sec:disc:syst:classes}
Gamma-ray sources observed at large distances, the blazars, belong to
various classes. In this section, we consider a possibility that distant
sources are different from nearby ones, and hence may have systematically
different spectral features. In general, broadband spectral energy
distributions (SEDs) of blazars have two wide bumps, often associated with
the synchrotron and inverse-Compton emissions. Relative positions of the
bumps are correlated, which may be explained if the same population of
relativistic electrons is responsible for both processes. Therefore, the
SEDs are, to the first approximation, characterized by the
position of the synchrotron peak so that all
blazars form ``the blazar sequence'' \cite{blazar-sequence}. Flat-spectrum
radio quasars (FSRQs) have the synchrotron peak frequency, $\nu_{\rm
peak}$, in the radio band, while blazars with $\nu_{\rm peak}$ from
infrared to X-ray bands represent various subclasses of BL Lac type
objects (BLLs). On average, FSRQs are more powerful and less numerous,
which results in a potential bias in flux-limited blazar samples: bright
FSRQs are seen at large distances while more numerous weaker BLLs dominate
the sample at relatively low redshifts. This bias might indeed affect our
observations because intrinsic absorption in (distant)
FSRQs might introduce spectral features not present in (nearby) BLLs.

\begin{figure}
\centerline{\includegraphics[width=0.67\columnwidth]{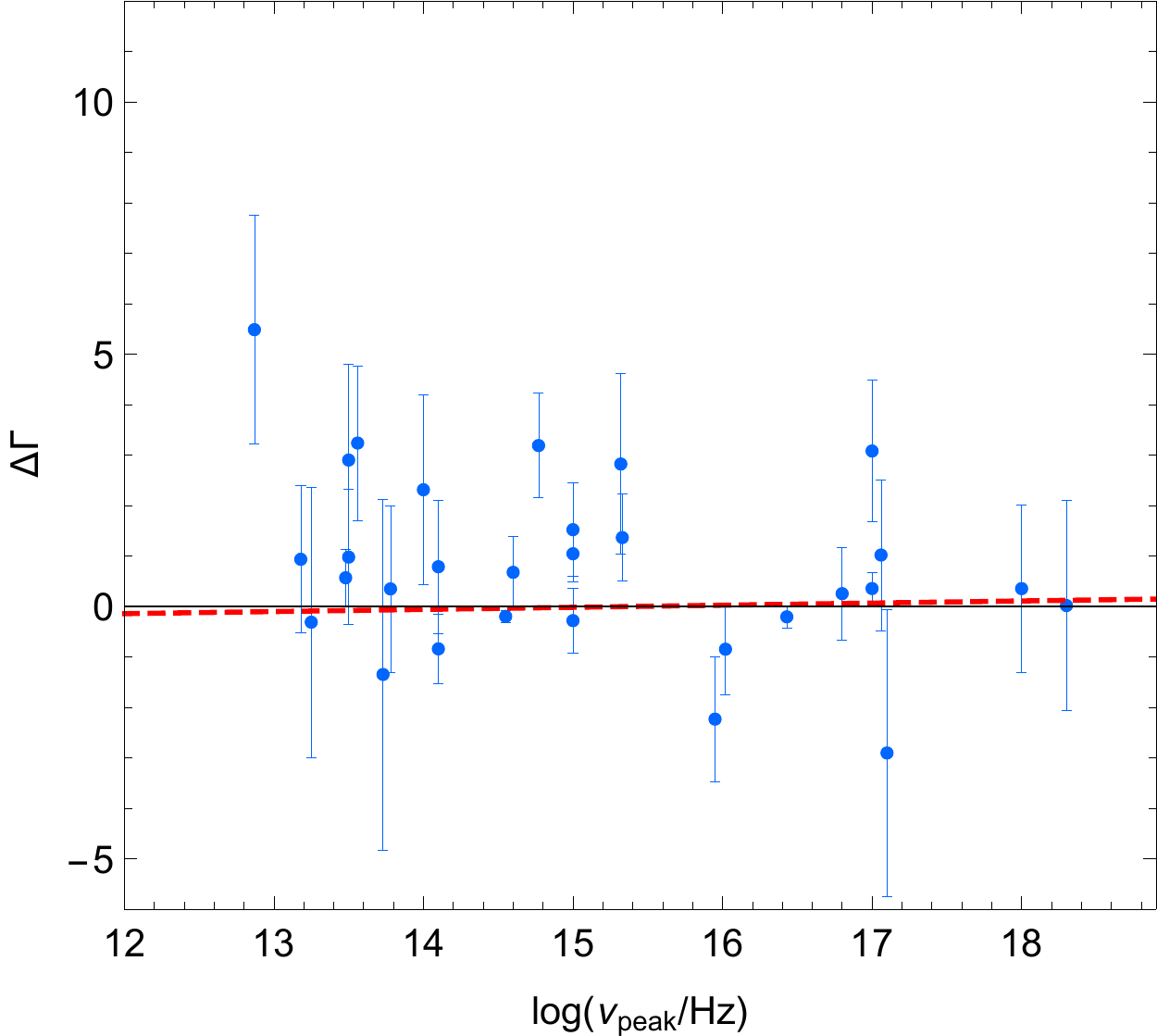}}
\caption{\label{fig:nu-peak}
       Breaks in the deabsorbed spectra of blazars versus the approximate
             synchrotron peak frequency.
               }
\end{figure}

The classification of FSRQ
or BLL given in Table~\ref{tab:list} is based on the information given in
the 3FHL catalog for Fermi-LAT sources and in TeVCat for sources observed
by IACTs. Since the classification is uncertain and various authors may
use different criteria for claiming a source is a FSRQ or a BLL, we
obtained approximate values of $\nu_{\rm peak}$ for objects in the sample,
based on SEDs available in NED and also on Refs.~\cite{1107.5105,
1501.03504, 1609.01095, 1903.07972}. We counted objects with
$\nu_{\rm peak}<10^{14}$~Hz as FSRQs and others as BLLs, and found that
this was in accordance with the classification given in
Table~\ref{tab:list} for all objects. Fig.~\ref{fig:nu-peak}
demonstrates that the same quantity $\Delta \Gamma$ does not depend on the
synchrotron peak frequency.
The analyses therefore support the
conclusion that potential BLL/FSRQ selection effects described in the
beginning of this section are not \red important\black.
Indirectly, this may suggest that the intrinsic absorption in FSRQs is a
subdominant effect with respect to the absorption on EBL.

\subsubsection{Tail of the falling spectrum}
\label{sec:disc:syst:tail}
At the energies under discussion, spectra are determined from
event-by-event information about observed individual photons. Determination
of the photon energies is subject to statistical and systematic
uncertainties. Statistical uncertainties are usually taken into
account in the spectrum reconstruction procedures; however, they might be
underestimated, while systematics may affect the overall energy scale of
an instrument. We are interested in the highest energies, where, because
of steeply falling spectra, the flux is often estimated from observations
of a few photons only. In the case of a systematic shift of their energies
upwards, or if the statistical uncertainty is larger than expected,
overestimation of the energies of these few photons may affect
significantly the shape of the spectrum.

We first address potential systematic errors. These are
instrument-dependent, and we take advantage of having objects observed by
four different instruments in our sample. Table~\ref{tab:instruments}
\begin{table}
\centering
\caption{
\label{tab:instruments}
Significance of the distance-dependent spectral hardenings for
subsamples including objects observed by different instruments.} \smallskip
\begin{tabular}{ccc}
\hline
Instruments & Number of objects & Significance\\
\hline
MAGIC $+$ HESS $+$ VERITAS &\red  23 & \red 1.9\black$\sigma$\\
MAGIC $+$ HESS $+$   LAT   &\red  23 & \red 2.4\black$\sigma$\\
MAGIC $+$ VERITAS $+$ LAT  &\red  19 & \red 1.7\black$\sigma$\\
HESS $+$ VERITAS $+$ LAT   &\red  19 & \red 1.7\black$\sigma$\\
\hline
\end{tabular}
\end{table}
presents significances of the observed distance-dependent  spectral
hardenings in subsamples of objects with data of one of the instruments
dropped. The observed significance is stable and agrees well with the
expectations from the sample size. Therefore, only a coherent upward
systematic error in the energy estimation by the three IACTs and Fermi
LAT may result in the effect we found. We note that this situation is
unlikely.

We turn now to the possibility of underestimated statistical errors, which
may result in occasional overestimation of energies of particular photons.
In the case of the published spectra, this effect is normally already taken
into account by the observing collaboration by making use of the
``unfolding'' procedure, see e.g.\ Ref.~\cite{MAGIC-unfolding}.
In any case, this potential systematics would be strongly suppressed for
the objects where the spectral hardening happens at energies for which a
sufficient number of photons is observed. To select those from our sample,
we replace the (S2) selection criterion from Sec.~\ref{sec:data:criteria}
by a much stronger one, requiring two (instead of one) spectral bins above
$E_{0}$. We stress that this does not directly select objects
observed at \red larger \black optical depths, nor intrinsically brighter
sources; this just reduces statistical errors in the determination
of $\Delta \Gamma$. Only \red 17 of 31 \black objects in our sample
satisfy this criterion. Performing our analysis for these \red 17 \black
objects, we find the significance of $\sim \red 2.7\black\sigma$ for this
purified sample. This disfavours the relation of our observation to
potentially underestimated errors in the energy determination; moreover,
the fact that a purified, though depleted, sample exhibits a \red somewhat
stronger \black effect \red supports its physical origin\black.

\subsubsection{Redshifts}
\label{sec:disc:syst:z}
\begin{figure}
	\centerline{\includegraphics[width=0.67\columnwidth]{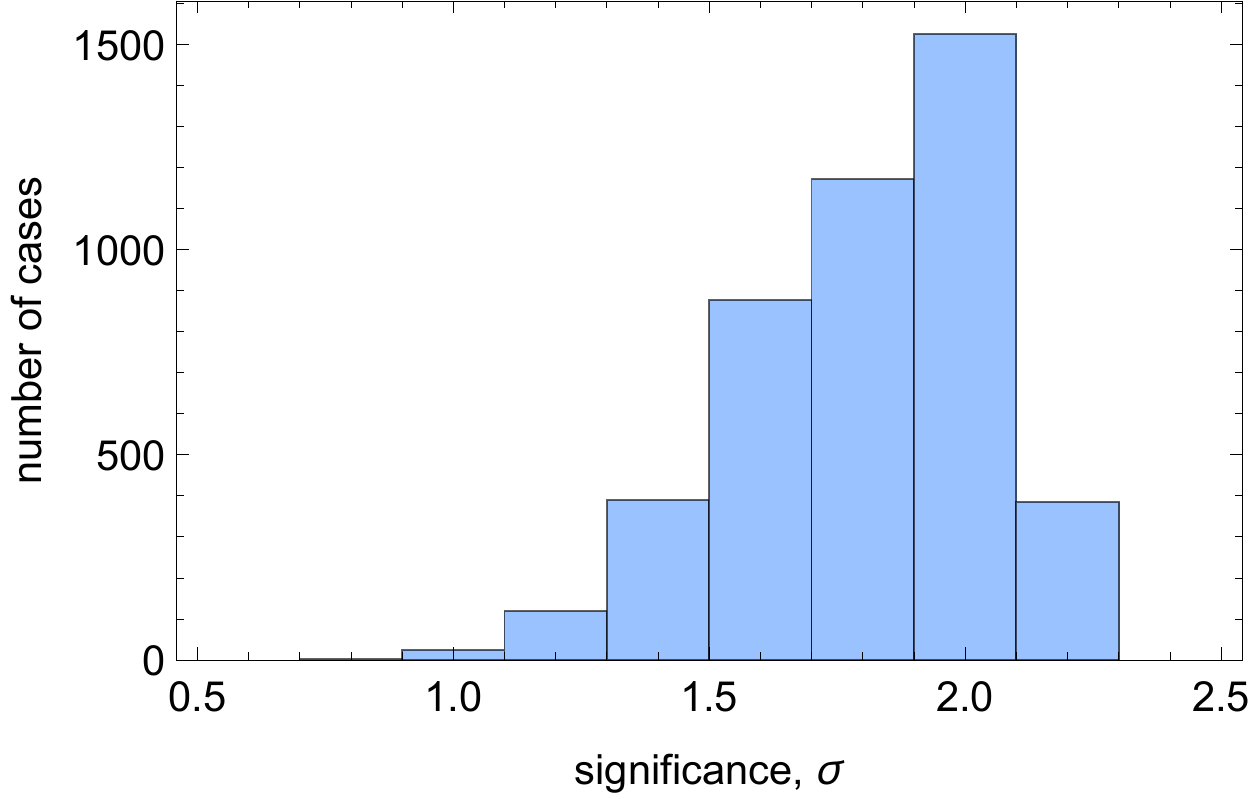}}
	\caption{\label{fig:drop-z}
			Distribution of significances for the samples with 2 objects
			artificially removed, imitating potential wrong redshifts.}
\end{figure}

Our study of distance-dependent spectral features is sensitive to correct
determination of distances to the sources. However, redshifts of distant
blazars are not always determined unambiguously. In particular (see
Sec.~\ref{sec:disc:comparison}), redshifts of \red 4 \black of 20 objects
in the sample we used for our previous study~\cite{gamma} in 2014 were
found to be unreliable by 2018, and the objects left our present
sample. Though for the present sample we performed a careful selection of
objects based on the quality of their redshifts, see criteria (R1),
(R2\textit{x}) in Sec.~\ref{sec:data:criteria}, one might imagine that
some of the \red 31 \black redshifts would still appear wrong in future
analyses. To understand how this might affect our results, we
artificially removed 1, 2 or 3 objects in all possible combinations from
the sample and repeated our analysis with the reduced samples. \black With 1
object removed, the lowest significance was \red 1.5\black$\sigma$ and the
highest one was \red 2.0 \black$\sigma$; for 2 objects removed the
interval was $(\red 1.1 \black \div \red 2.1\black)\sigma$; for 3
objects removed it was $(\red 0.8 \black \div \red 2.2\black)\sigma$.
Figure~\ref{fig:drop-z}
presents the distribution of significances among samples with 2 objects
removed. We see that while it is possible to select sources from the sample
whose removal would have a significant impact on the result, in
most cases this is not expected.

\subsection{Comparison to previous studies}
\label{sec:disc:comparison}
In this section, we give a detailed comparison of our methods and results
with those of key previous studies of the gamma-ray opacity of the
Universe with ensembles of blazar spectra.

\subsubsection{Rubtsov and Troitsky (2014)~\cite{gamma}}
\label{sec:disc:comparison:gamma}
The present work follows the same approach as our previous work~\cite{gamma}.
Advantages of the present work, besides the use of new observational data and
a more detailed report on systematics, are:

(i)~pre-selection of Fermi-LAT sources with the 3FHL catalog and the use
of Pass~8 data processing;

(ii)~use of new conservative EBL models \cite{Korochkin2018,
Franceschini2017};

(iii)~imposing strict criteria on the redshift quality;

(iv)~use of a more conservative method which is not based on the
bin-by-bin deabsorption;

(v)~use of a more general statistical test of the presence of spectral
breaks.

For the point (v), we stress here that the null hypothesis to be tested is
the absence of any spectral features at $E_{0}(z)$. This hypothesis is
disfavoured at \red 1.9\black $\sigma$ in the present work. In
Ref.~\cite{gamma}, we observed the linear growth of $\Delta \Gamma$ with
$\log z$, which does not have a known physical model behind it.

Next, the changes \black in criteria \black resulted in the removal of
\red 5 \black sources from the sample of Ref.~\cite{gamma}: redshifts of
\red 3 \black objects (RGB~J1448$+$361\red, 1ES~1101$-$232 \black and
1ES~0347$-$121) did not satisfy our new quality criteria; the reported
value of the redshift of B3~1307$+$433 changed in such a way that the
object no longer satisfies the (S2) criterion; for PKS~J0730$-$1141, new
Pass~8 Fermi-LAT reconstruction does not result in a significant detection
at $E>E_{0}$, contrary to the Pass~7REP (V15) used in Ref.~\cite{gamma}.
Removal of the most distant objects from the sample, together with a more
conservative analysis method, resulted in a considerable reduction of the
resulting significance (was 12.4$\sigma$ for the linear-growth test, now
\red 1.9\black$\sigma$ for the exclusion of the null hypothesis).

\begin{figure}
	\centerline{\includegraphics[width=0.67\columnwidth]{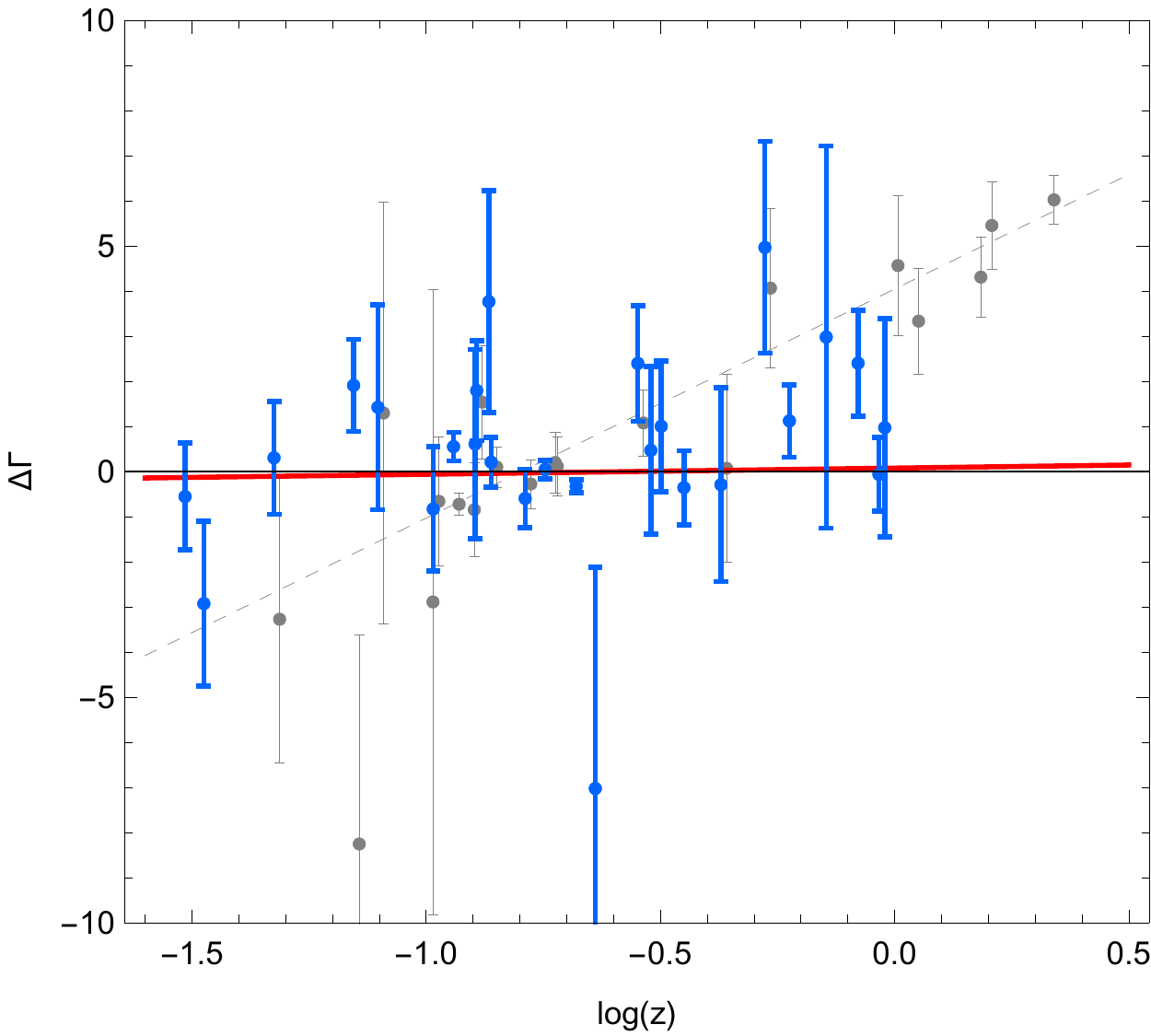}}
	\caption{\label{fig:comparison}
	\black Comparison of the present result with that of Ref.~\cite{gamma}:
	spectral breaks at $E_{0}(z)$ in deabsorbed (model of Gilmore et al. (2012)
	fixed) blazar spectra versus redshift. Individual data points for objects
	in the sample studied in the present work are shown in blue (thick error
	bars); the full red line represents the best fit of $\Delta \Gamma$ as a
	linear function in $\log z$. Data points of Ref.~\cite{gamma} are shown in
	gray (thin error bars), slightly shifted to the right for clarity; the
	corresponing best fit of Ref.~\cite{gamma} is presented as a thin dashed
	line.}
\end{figure}

It is interesting to note that the present study
does not confirm the linear growth of spectral breaks with $\log z$ observed in
Ref.~\cite{gamma} and never explained theoretically within any approach.
Figure~\ref{fig:comparison}
demonstrates that the significant distance dependence observed earlier was
dominated by the effect of distant sources whose redshifts do not pass the
quality cuts applied in the present study. Better measurements of redshifts
of distant blazars observed in VHE gamma rays are clearly welcome.

\subsubsection{Horns and Meyer (2012)~\cite{HornsMeyer}}
\label{sec:disc:comparison:Horns-Meyer}
Ref.~\cite{HornsMeyer} was the first study addressing quantitatively
spectral hardenings in the ensemble of distant gamma-ray sources.

(1)~Both Ref.~\cite{HornsMeyer} and our study address upward spectral
breaks \red at distance-dependent energies\black.

(2)~Only objects observed by IACTs ($z<0.536$) were included in the sample
of Ref.~\cite{HornsMeyer}. \red There,\black a stricter requirement of
detection at optical depths $\tau>2$ \red was used\black, resulting in the
sample of 7 objects only, while we require $\tau>1$ and include objects
for which the break was not significantly detected.

(3)~In Ref.~\cite{HornsMeyer}, bin-by-bin deabsorption was used but the
correction for the mean energy in the bin for deabsorption
was not implemented.

\subsubsection{Fermi-LAT collaboration (2012)~\cite{Fermi-opacity}}
\label{sec:disc:comparison:Fermi}
The interesting work~\cite{Fermi-opacity} presented an analysis of the
ensemble of BLLs detected by Fermi LAT at high redshifts and, for the
first time, presented a detection of the EBL absorption in stacked spectra
of these distant sources. The ensemble of 150 BLLs with redshifts
$\lesssim 1.6$ was split into three distance bins, and intrinsic spectra
were extrapolated from Fermi-LAT observations at low energies. The EBL
absorption correction was subsequently determined from spectra
stacked within the three redshift bins.

As we know from our present study, only a few blazars (5 with our criteria)
have been confidently observed by Fermi LAT above $E_{0}$. Constraints on
the amount of absorption obtained in Ref.~\cite{Fermi-opacity} were not very
tight. While available absorption models are consistent with the upper part of
the allowed band in $\tau(E)$ for $z=1$, cf.\ Fig.~1 of Ref.~\cite{Fermi-opacity},
considerably lower absorption is allowed for all energies. This is in agreement
with our results.

\subsubsection{Biteau and Williams (2015)~\cite{Biteau}}
\label{sec:disc:comparison:Biteau}
Ref.~\cite{Biteau} considered 106 spectra of 38 sources observed by IACTs.
Every spectrum was deabsorbed bin-by-bin, without applying the correction
for the mean energy of the bin, and then fitted by a model spectrum for which
a concave shape was assumed. This study did not reveal any anomaly in the
absorption on EBL.

\begin{figure}
	\centerline{\includegraphics[width=0.67\columnwidth]{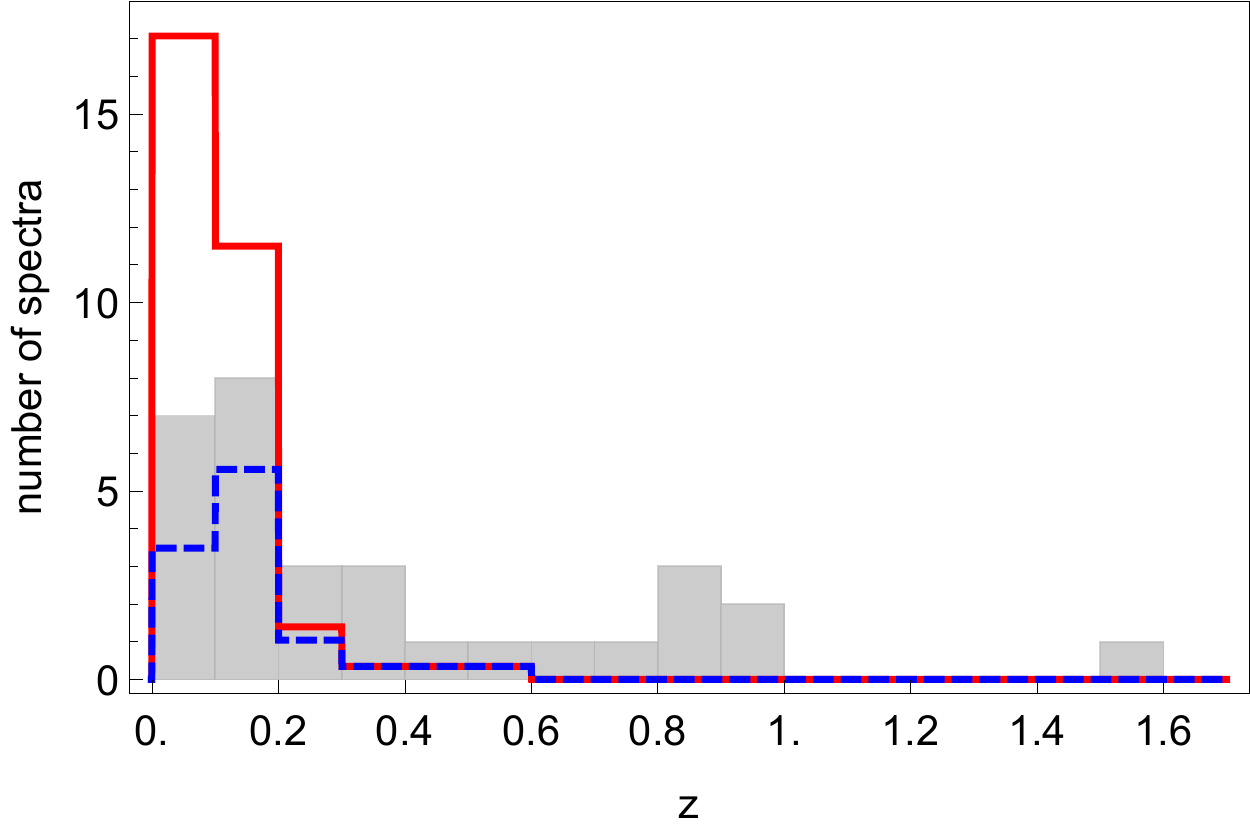}}
	\caption{\label{fig:bit-hist}
			Comparison of the redshift coverage of the sample used by Biteau and
			Williams~\cite{Biteau} and of our sample. Red full line: distribution of
			redshifts for individual spectra used in the analysis of
			Ref.~\cite{Biteau} (uncertain, lower-limit and erroneous redshifts are not
			included). Blue dashed line: independent (one spectrum per
			source) spectra from Ref.~\cite{Biteau}. Shaded histogram: redshifts in
			our sample. Red line and shaded histogram are normalized to the total
			number of spectra used, blue line gives a fraction of the red-line
			distribution.}
\end{figure}

\begin{figure}
	\centerline{\includegraphics[width=0.67\columnwidth]{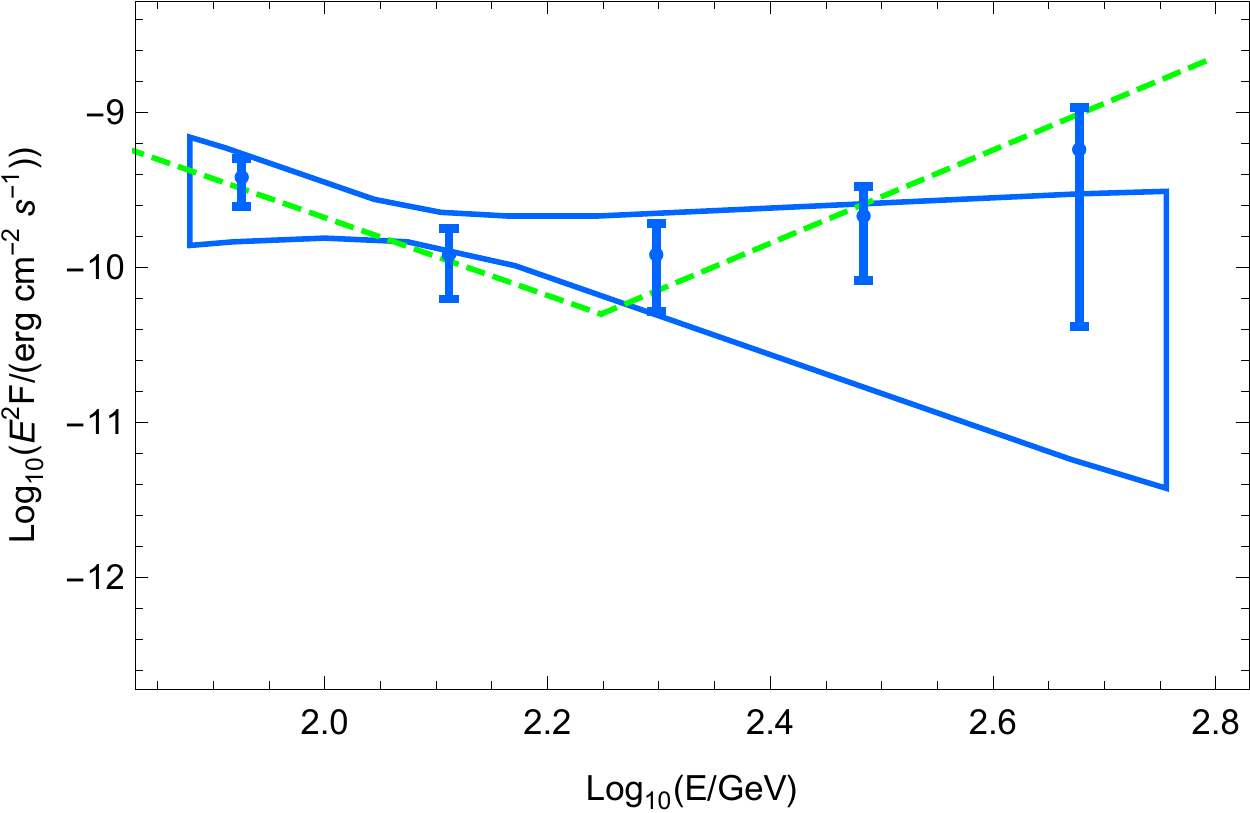}}
	\caption{\label{fig:bitplot}
			Comparison of the treatment of a single binned spectrum by Biteau and
			Williams~\cite{Biteau} and in the present work, using 3C~279 (spectrum
			from Ref.~\cite{0807.2822}) as an example. Data points: deabsorbed spectrum
			from Ref.~\cite{Biteau}, flux data points attributed to median log observed
			energy of the bins. The blue butterfly gives the allowed range of
			\textit{concave} spectra of Ref.~\cite{Biteau} while the dashed green line
			represents our best-fit spectrum which assumes the break at $E_{0}(z)$ for
			$z=0.536$, the redshift of 3C~279.}
\end{figure}

We argue that nonobservation of the absorption anomaly in Ref.~\cite{Biteau}
does not contradict to the results of the present work. The difference in
conclusions is related both to the sample of source spectra used and to the
assumptions on the intrinsic spectra. We point out the following differences:
\begin{enumerate}
\item
In the sample of Ref.~\cite{Biteau}, a different set of sources is used,
dominated by nearby ones, while the observation of anomalous features at
various $E_{0}(z)$ requires a good redshift coverage. The dominance of nearby
sources in Ref.~\cite{Biteau} was enhanced by the use of multiple spectra per
source, which gave higher statistical weights to better studied nearby objects,
see Figure~\ref{fig:bit-hist}. Different spectra of the same objects were treated
as independent data in the statistical analysis of Ref.~\cite{Biteau}, which may
introduce statistical \red biases\black.
\item
The individual-source analysis of Ref.~\cite{Biteau} assumed explicitly a
concave shape for the deabsorbed spectrum, thus not allowing for the
hardenings we study here by construction, see Figure~\ref{fig:bitplot}. In
fact, in most cases (91 of 106 spectra, in particular for all sources with
$z>0.12$), the best-fit model corresponded to a power law, that is to the
margin of the allowed concave spectra: convex shapes, which give better
fits for many of the spectra in our study, were not allowed.
\end{enumerate}
We conclude that the methods chosen by Biteau and Williams in Ref.~\cite{Biteau}
are insensitive to the anomalous effects found in Refs.~\cite{HornsMeyer, gamma}
and studied here, therefore Ref.~\cite{Biteau} and our work are not in a contradiction.

In addition to the main study, which did not allow for convex spectra by
definition, Biteau and Williams addressed the result of Horns and
Meyer~\cite{HornsMeyer} discussed above. While they confirmed that the
application of the method of Horns and Meyer to their sample results in a
significant dependence of the break position on the \red redshift \black
(with even stronger significance than in the original
Ref.~\cite{HornsMeyer}), Biteau and Williams claimed that the reason for
the effect found in Ref.~\cite{HornsMeyer} was in incorrect treatment of
statistical uncertainties of the flux measurements: the method of Horns
and Meyer, based on a nonparametric Kolmogorov--Smirnov
test\footnote{\black Ref.~\cite{Biteau} mentions the Anderson--Darling
method which was not used to obtain the main result of
Ref.~\cite{HornsMeyer}.}, does not take the uncertainties into account and
is sensitive to central values of the measured flux only. To illustrate
this, Biteau and Williams searched for observed excesses in residuals over
best-fit concave spectra at energies corresponding to large optical
depths. The results of the analysis, performed for their full set of
spectra, indicated no significant deviations. This part of the
study~\cite{Biteau}, however, also suffers from the use of multiple
spectra of the same objects treated as statistically independent data.
\red In addition, these residuals were calculated with respect to the
best-fit spectra deabsorbed with the EBL model determined from the very
same spectra, which makes the problem of independent treatment of data
points even more complicated. \black

\subsubsection{Galanti et al.\ (2015)~\cite{Roncadelli-ensemble}}
\label{sec:disc:comparison:Roncadelli}
It is interesting to note that distance-dependent hardenings can still
be noticed with the analysis based on a single power-law fit. Galanti et al.\
(2015)~\cite{Roncadelli-ensemble} considered a sample of 39 blazars observed
by IACTs ($z\le 0.536$), for all tested opacities. Deabsorbed spectra were
described by single power laws, and the distance dependence of the spectral
index was studied. Though this approach does not constructively include
spectral breaks, deabsorbed spectra with upward breaks look harder when
fitted by a power-law, and this hardness was found to be redshift-dependent.
Therefore, results of Ref.~\cite{Roncadelli-ensemble} are in agreement with
Ref.~\cite{gamma}.

\subsubsection{Studies of high-energy versus very-high-energy spectral indices}
\label{sec:disc:comparison:Dominguez}
Several previous studies used as their observables the difference between
spectral indices in the high-energy (roughly, GeV; definition varies) and
very-high-energy (roughly, hundreds of GeV) bands. These works include
Stecker et al.\ (2006) \cite{ApJ-648-774}, Stecker and Scully (2010)
\cite{ApJL-709-L124}, Sanchez et al.\ (2013) \cite{AnA-542-A59}, Essey and
Kusenko (2012) \cite{ApJL-751-L11}. The most elaborated study of this
kind, Dominguez and Ajello (2015) \cite{ApJL-813-L34}, was based entirely
on the Fermi-LAT data and was therefore insensitive to relative systematic
errors of the energy determination by different experiments. It was also
the only one working with spectra corrected for the absorption on EBL. In
all these studies, the energy of spectral breaks was a priory fixed to a
certain value, uniform for all sources. Therefore, the results of the
papers mentioned, some of which pointed to anomalies, some did not,
cannot be directly compared to ours, because we assumed breaks at energies
$E_{0}(z)$, different for different sources.

\section{Conclusions}
\label{sec:concl}
In the present work, we readdress the problem of the anomalous transparency
of the Universe for high-energy gamma rays, the most significant indications
to which~\cite{HornsMeyer,gamma} were based on unphysical spectral features
in deabsorbed spectra for an ensemble of blazars which showed up precisely
at the energies for which the correction for the absorption on EBL became
important. By making use of a more robust analysis procedure which avoids
bin-by-bin deabsorption and tests only the null hypothesis, of new gamma-ray
and redshift data and of most recent EBL models, we disfavour the absence of
anomalous distance-dependent spectral features with the \red modest \black
statistical significance of $\red 1.9 \black\sigma$. At the same time we
do not confirm the linear dependence of the break strength with $\log z$.
We discuss potential biases and systematic errors and, by performing
various tests, conclude that they cannot
\red seriously affect our results\black.

Interpretations of the anomalous transparency of the Universe have been widely
discussed in the literature, see e.g.\ Ref.~\cite{STrev} for a brief review
and a list of references. The simplest possible culprit for unphysical spectral
features might be an overestimated EBL intensity. However, reduction of the
intensity below the levels predicted by the models we use might be dangerous
because the model intensities are very close to the lower limits from galaxy
counting. With the help of the Markov-chain optimization behind the model~
\cite{Korochkin2018}, we plan to attempt to change the EBL model in such a
way that the effect\red, hints to which \black we observe in the present
work\red, is absent \black and to understand how much in conflict this
would be with underlying astrophysical data. This will be a subject of a
forthcoming work.

A number of proposals attempted to explain the apparent anomalous transparency
of the Universe by adding secondary emission to that arriving directly from
the source. If these secondary photons are born relatively close to the
observer, they do not have time to produce $e^{+}e^{-}$ pairs and arrive
unattenuated. With respect to the origin of this secondary emission, it is
useful to distinguish the cases when it is  caused by electromagnetic cascades
\cite{Timur} or from interactions of hadronic cosmic-ray particles
\cite{Kusenko, Kusenko1}. The main characteristic feature of both approaches
is that they require extremely low magnetic fields all along the propagation
path of the cascade, otherwise secondary photons would not point to the
sources because of deflections of their progenitor charged particles in the
cascade.

Other proposals require modification of the photon interactions.
We note that the pair-production cross section has been well determined
experimentally, and the only possibility to change it in the kinematic
regimes relevant for our study is to allow for small deviations from the
Lorentz invariance. However, this change would affect also the development
of photon-induced atmospheric showers \cite{noLIV, noLIV1} making the
work of IACTs impossible, so that the scenario, in which results of
Sec.~\ref{sec:anal} are explained by the Lorentz-invariance violation,
is in fact excluded by the fact that some high-energy photons have been
detected by these instruments (see also Ref.~\cite{Hassan2}). The remaining
explanation invokes a hypothetical axion-like particle (ALP). In external
magnetic fields, ALP mixes with photons \cite{raffelt} and, since ALP does
not attenuate on EBL, this makes it possible to detect photons from more
distant sources. In one of the scenarios, this mixing takes place in
intergalactic magnetic fields along the path from the source to the
observer~\cite{Csaba, DARMA}, while in the other, a part of photons is
converted to ALPs near the source and reconverted back to gamma rays in
our neighbourhood (in the magnetic fields of galaxies, clusters and
filaments)~\cite{Serpico, FRT}. Observational consequences of the two
scenarios are compared to each other and to data in Ref.~\cite{ST-2scen}.
The account of interactions of energetic photons with the axion-like
particle allows one to explain consistently all \red indications to the
anomalous transparency of the Universe \black without contradicting to any
other experimental or observational data. Our present results indicate
that further studies are required to prove or exclude the need for
new-physics effects in the propagation of energetic gamma rays through the
Universe.

\appendix
\section{Observed and best-fit intrinsic spectra of blazars in the main
sample.}
\label{sec:appendix}
\begin{figure}
	\centerline{%
		\includegraphics[width=0.5\columnwidth]{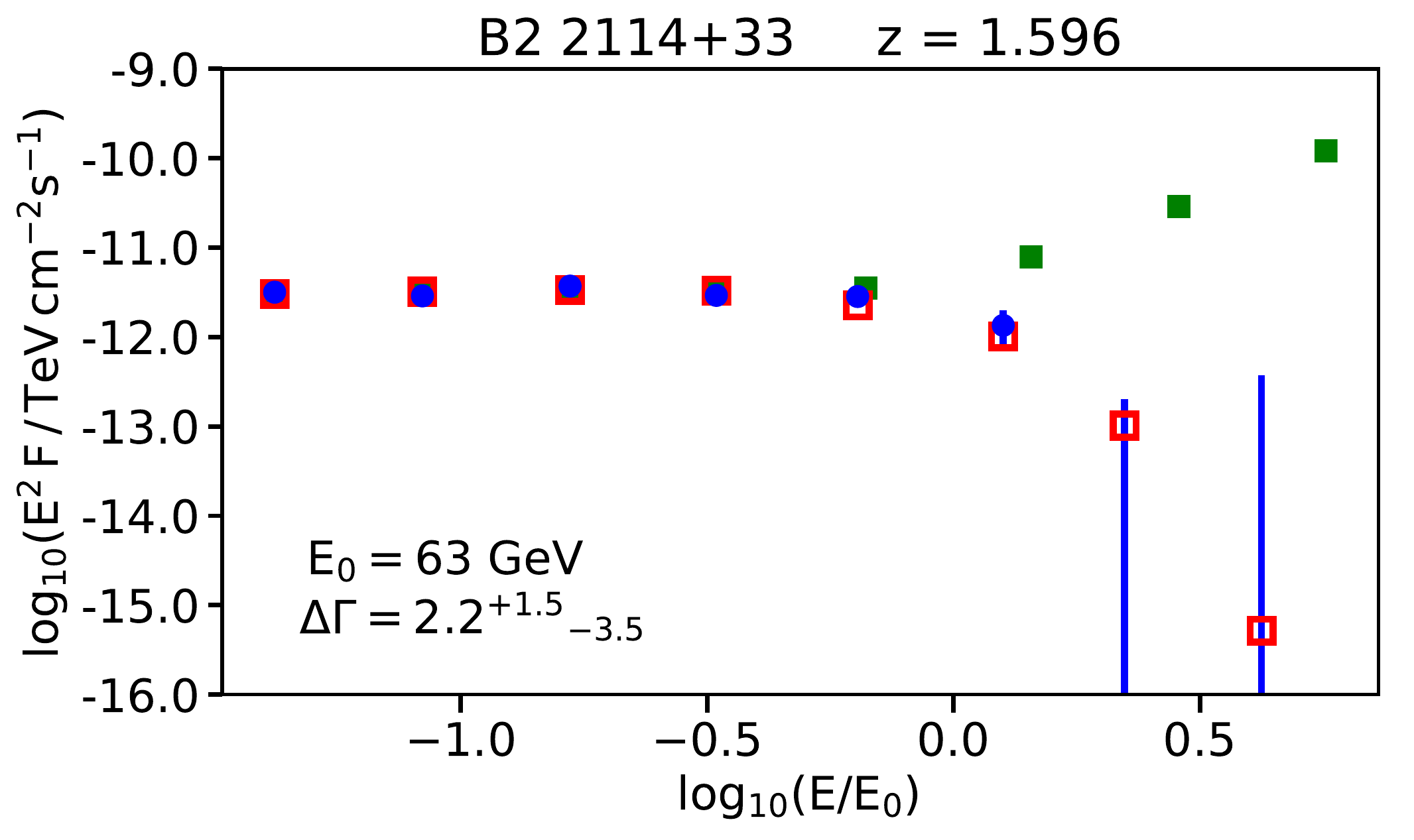}
		~~~~
		\includegraphics[width=0.5\columnwidth]{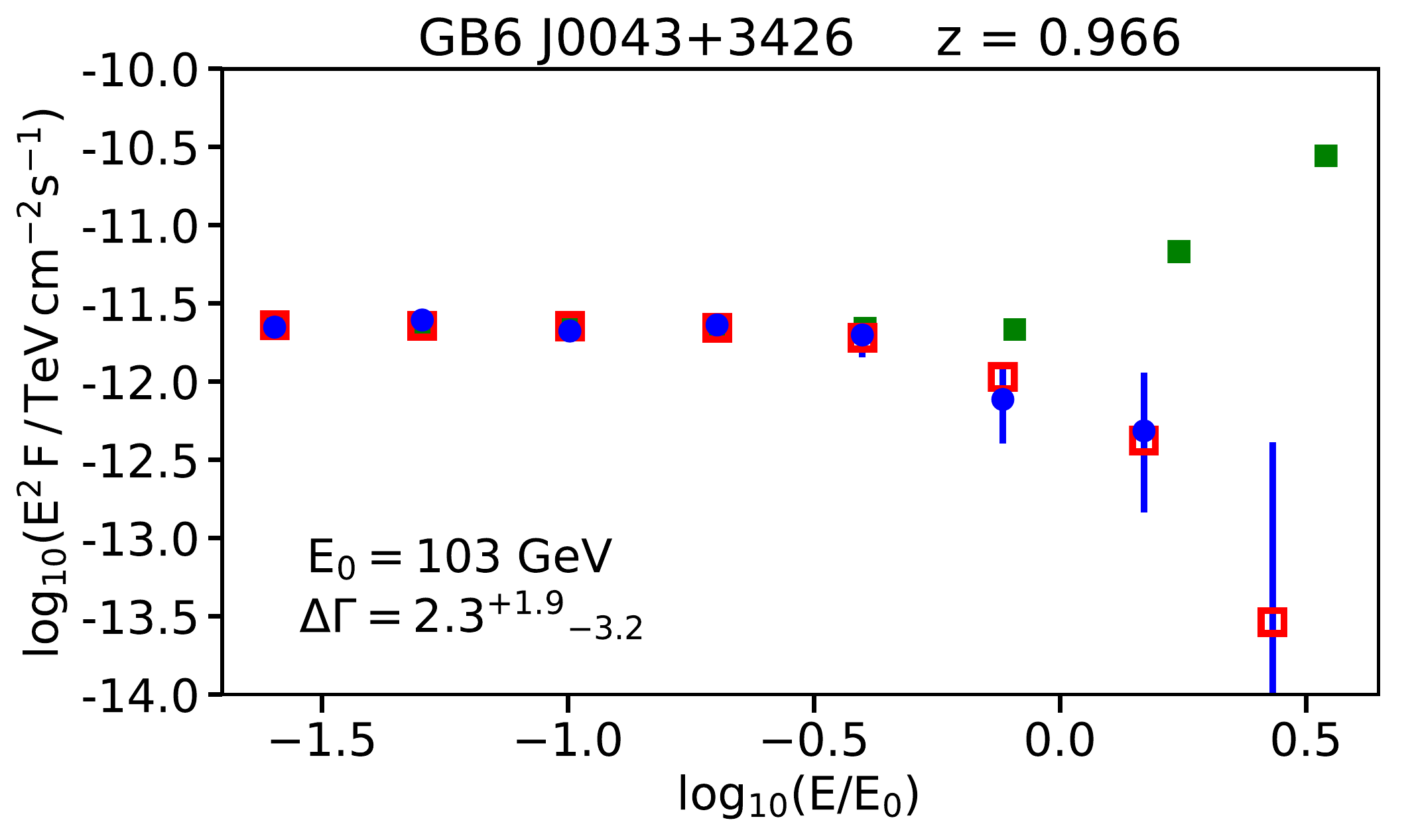}
	}
	\bigskip
	\centerline{%
		\includegraphics[width=0.5\columnwidth]{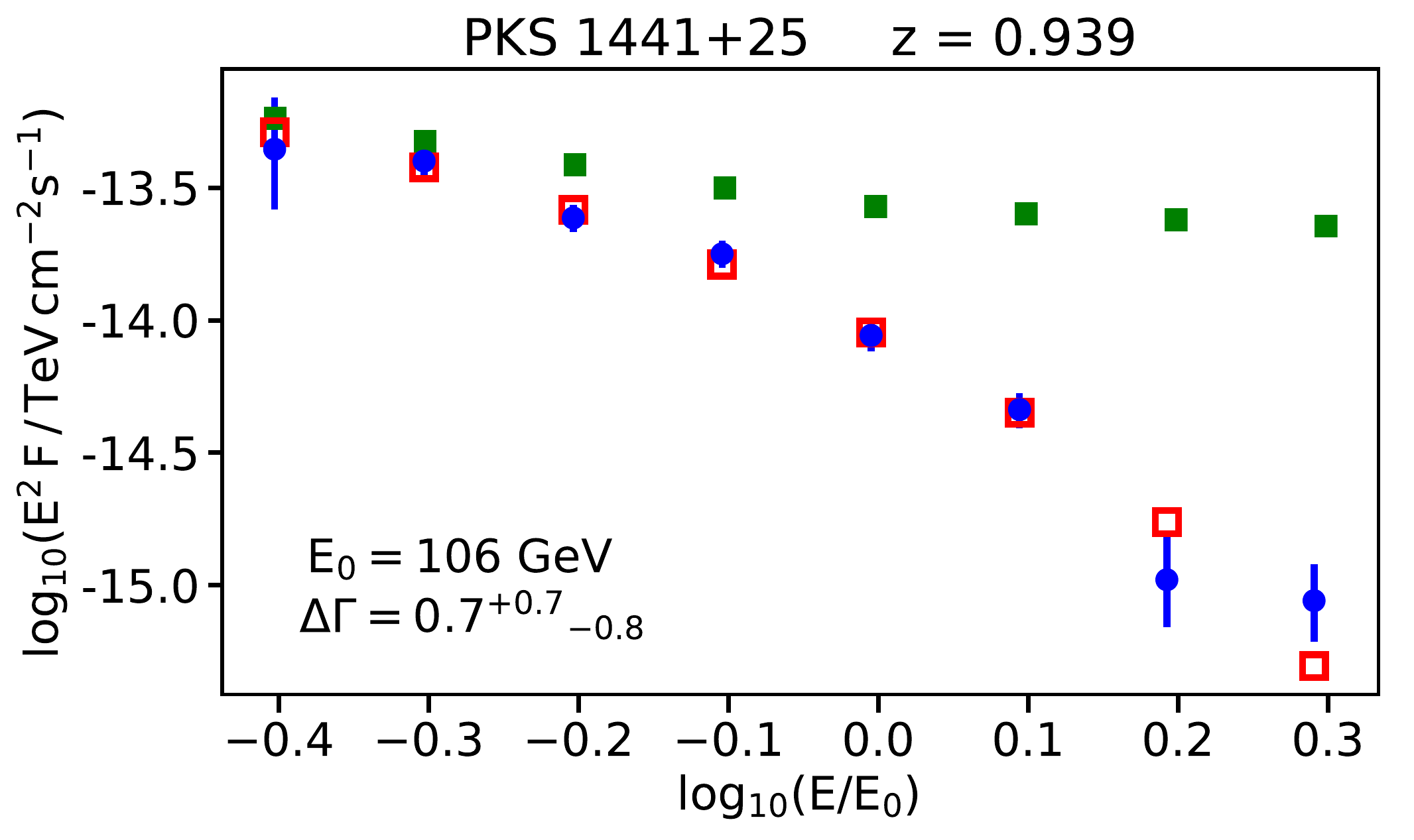}
		~~~~
		\includegraphics[width=0.5\columnwidth]{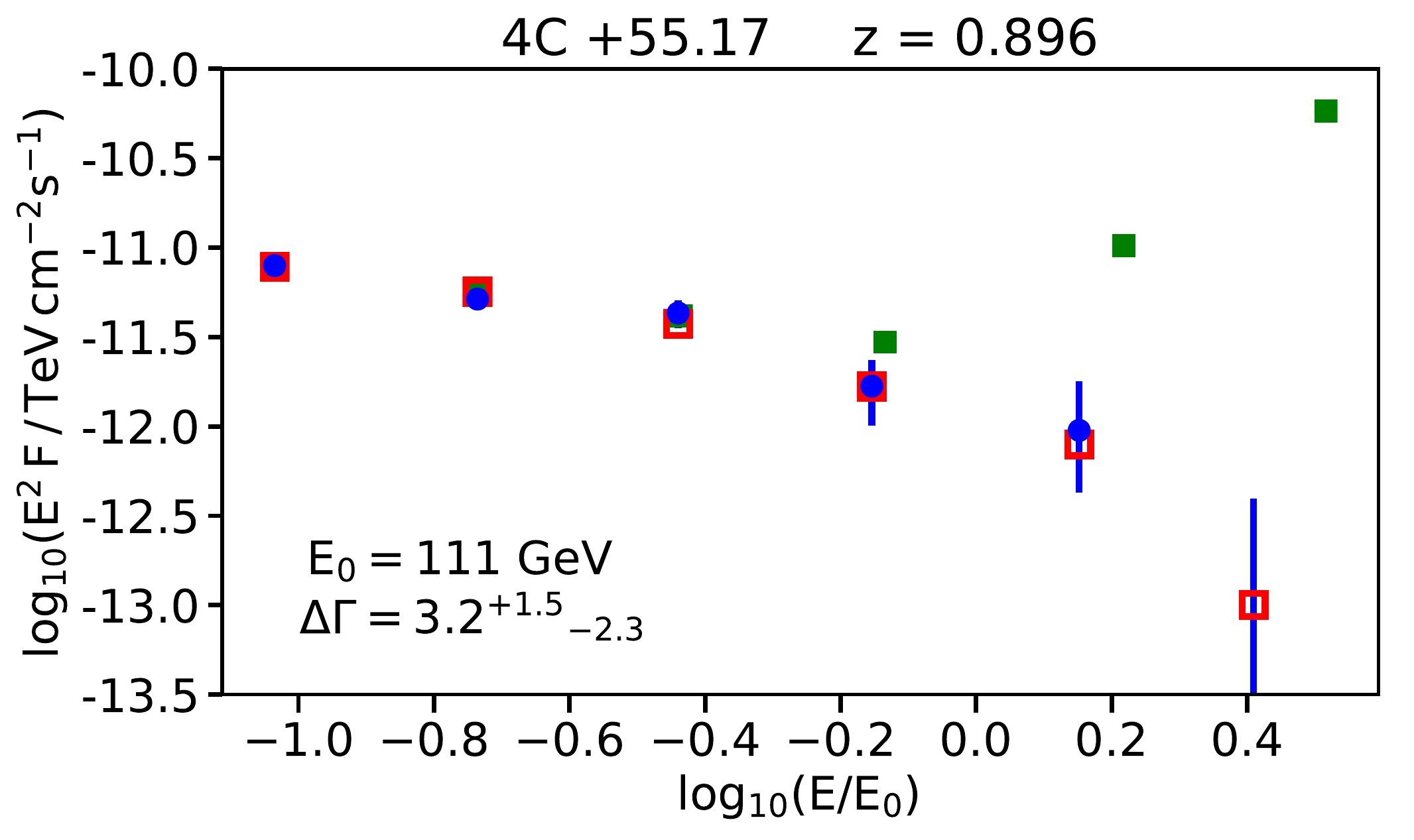}
	}
	\bigskip
	\centerline{%
		\includegraphics[width=0.5\columnwidth]{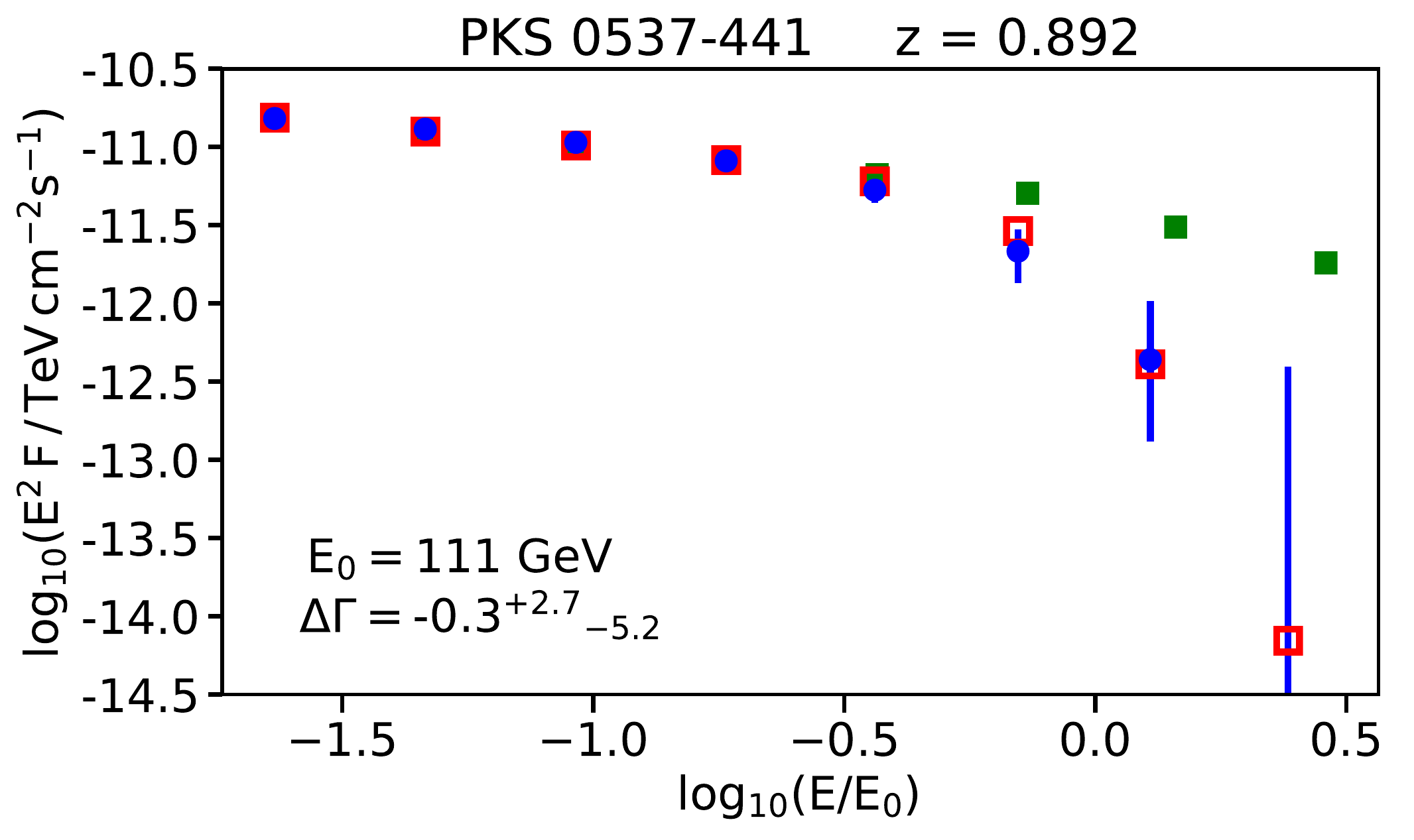}
		~~~~
		\includegraphics[width=0.5\columnwidth]{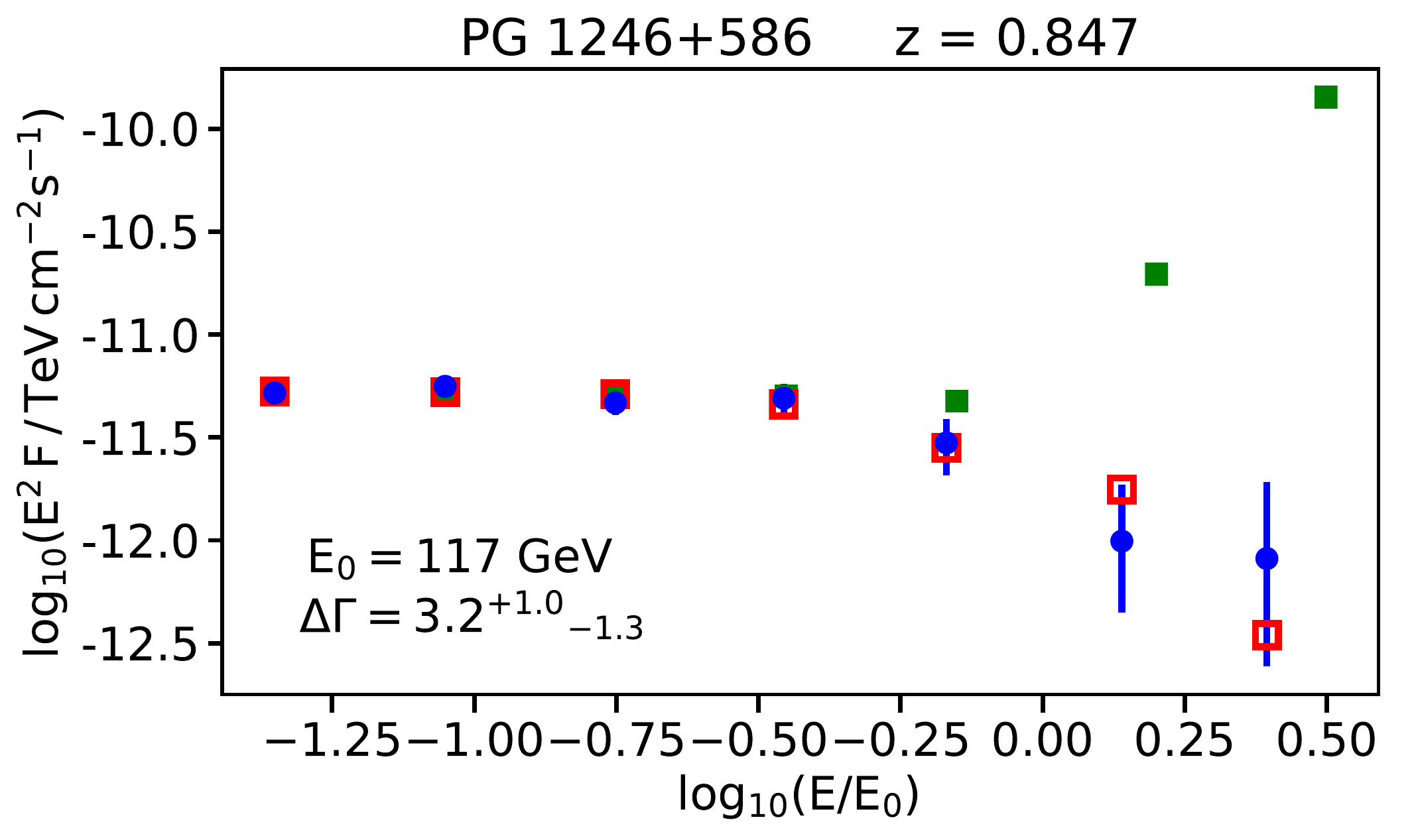}
	}
	\bigskip
	\centerline{%
		\includegraphics[width=0.5\columnwidth]{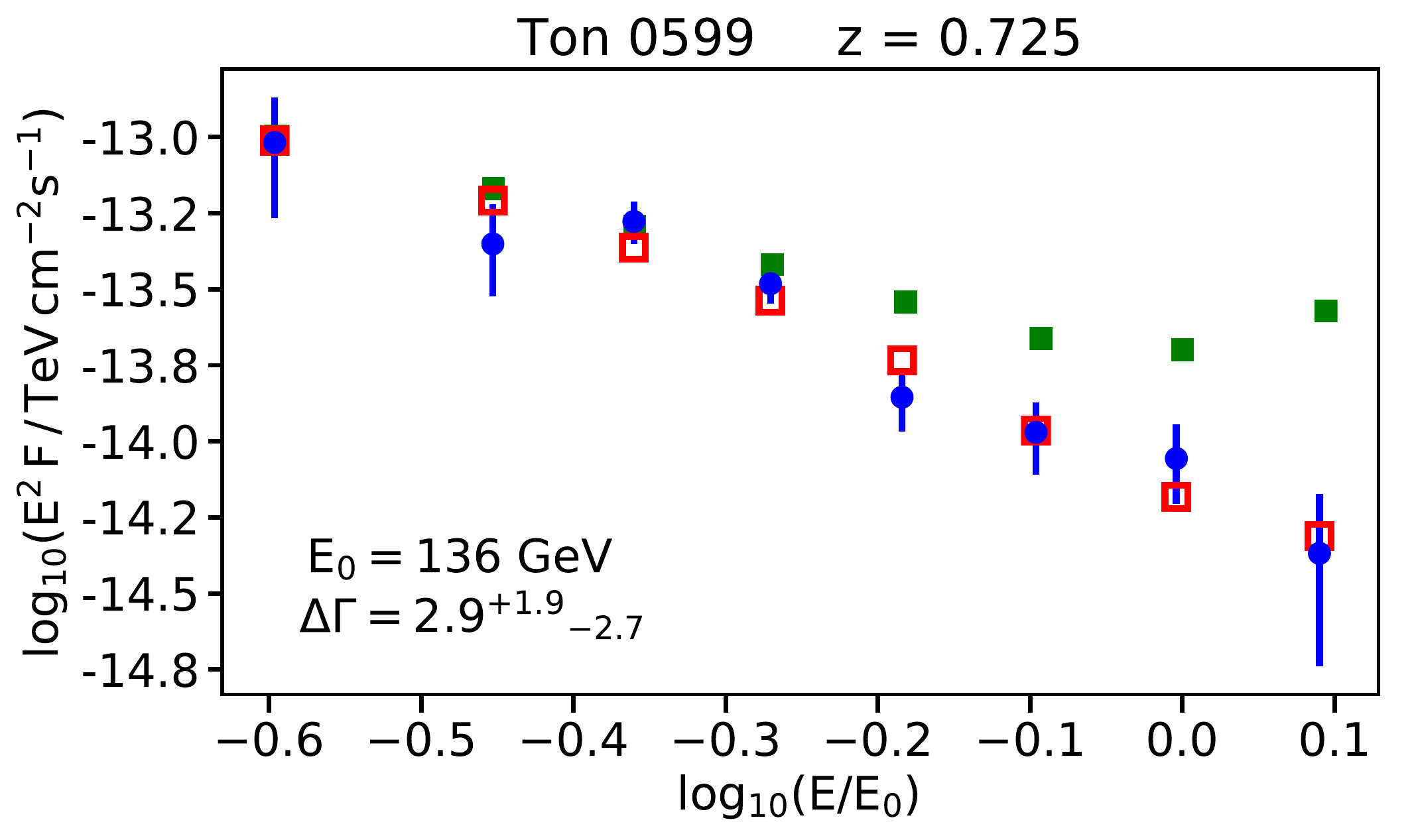}
		~~~~
		\includegraphics[width=0.5\columnwidth]{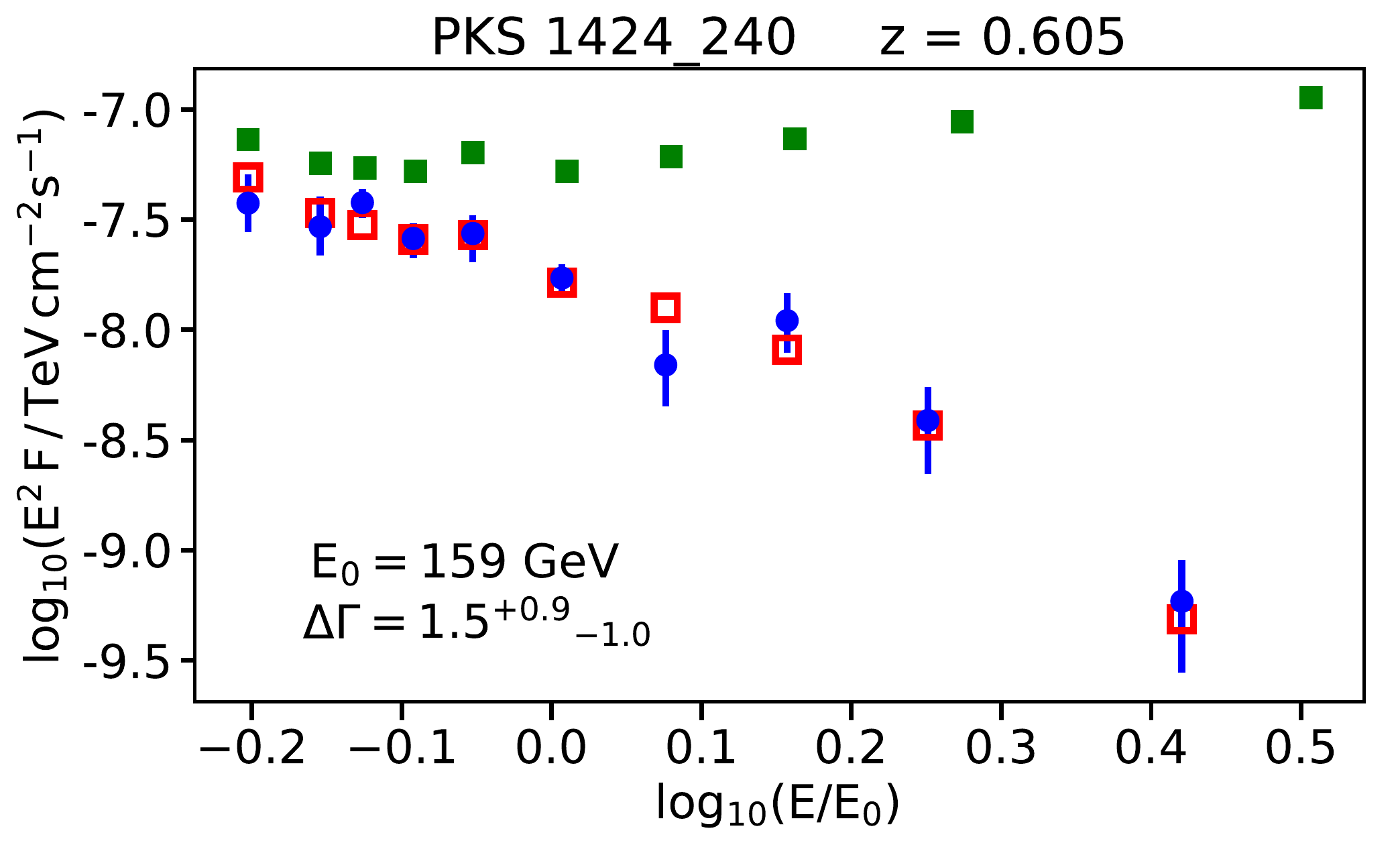}
	}
	\caption{\label{fig:spec1}
			Spectra of \green 31 \black blazars from the sample studied in this work
			(bin-by-bin fluxes). Blue circles with error bars: observed spectra; green
			boxes: best fit of the intrinsic spectrum with the break at the energy
			$E_{0}$ for which $\tau(E_{0},z)=1$. Red open boxes indicate bin-by-bin
			fluxes corresponding to the best-fit spectra after absorption. Names of
			the sources, values of their redshifts $z$, energies $E_{0}$ and break
			strengths $\Delta \Gamma$ are shown on the plots. See the main text for
			more details. }
\end{figure}

\begin{figure}
	\centerline{%
		\includegraphics[width=0.5\columnwidth]{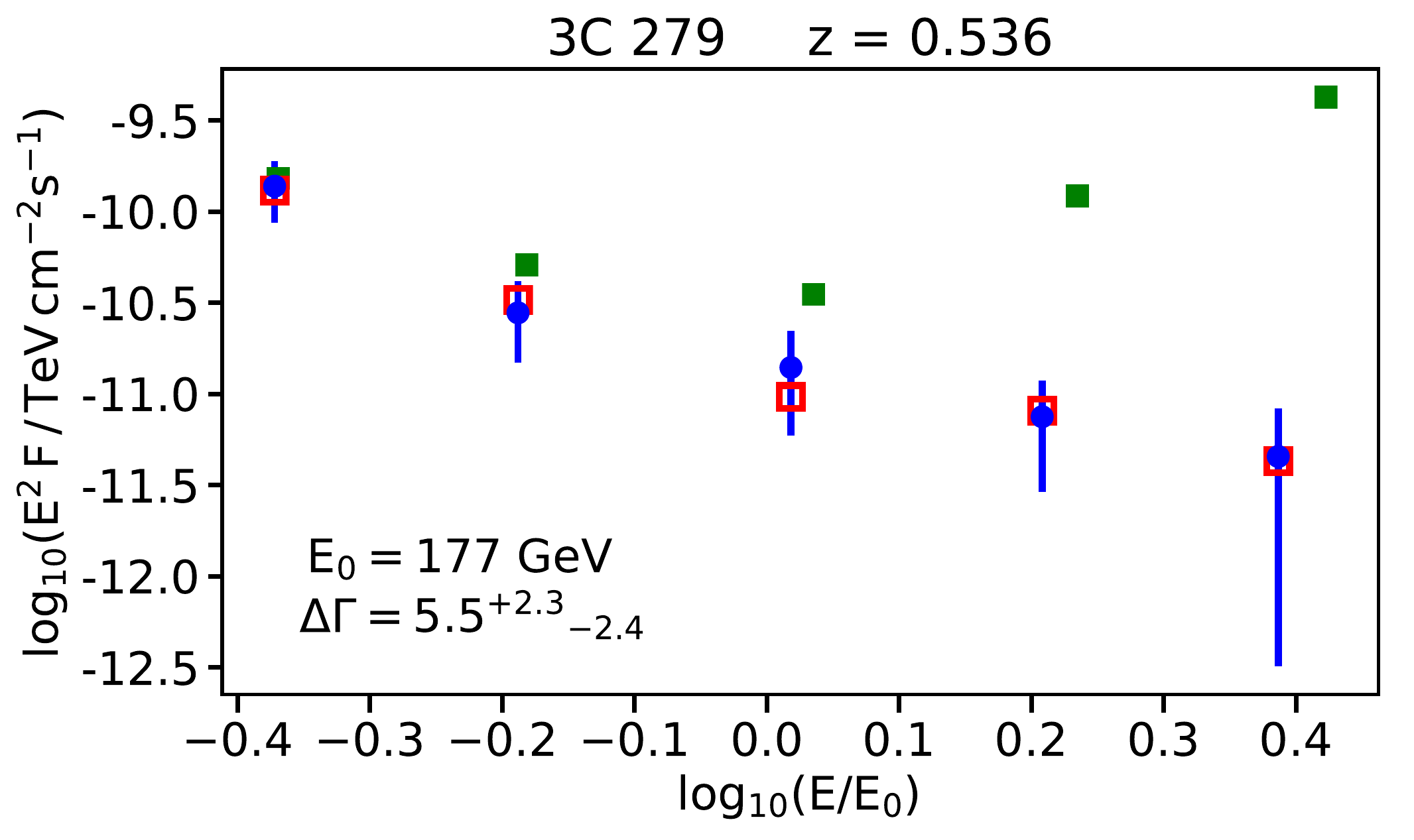}
		~~~~
		\includegraphics[width=0.5\columnwidth]{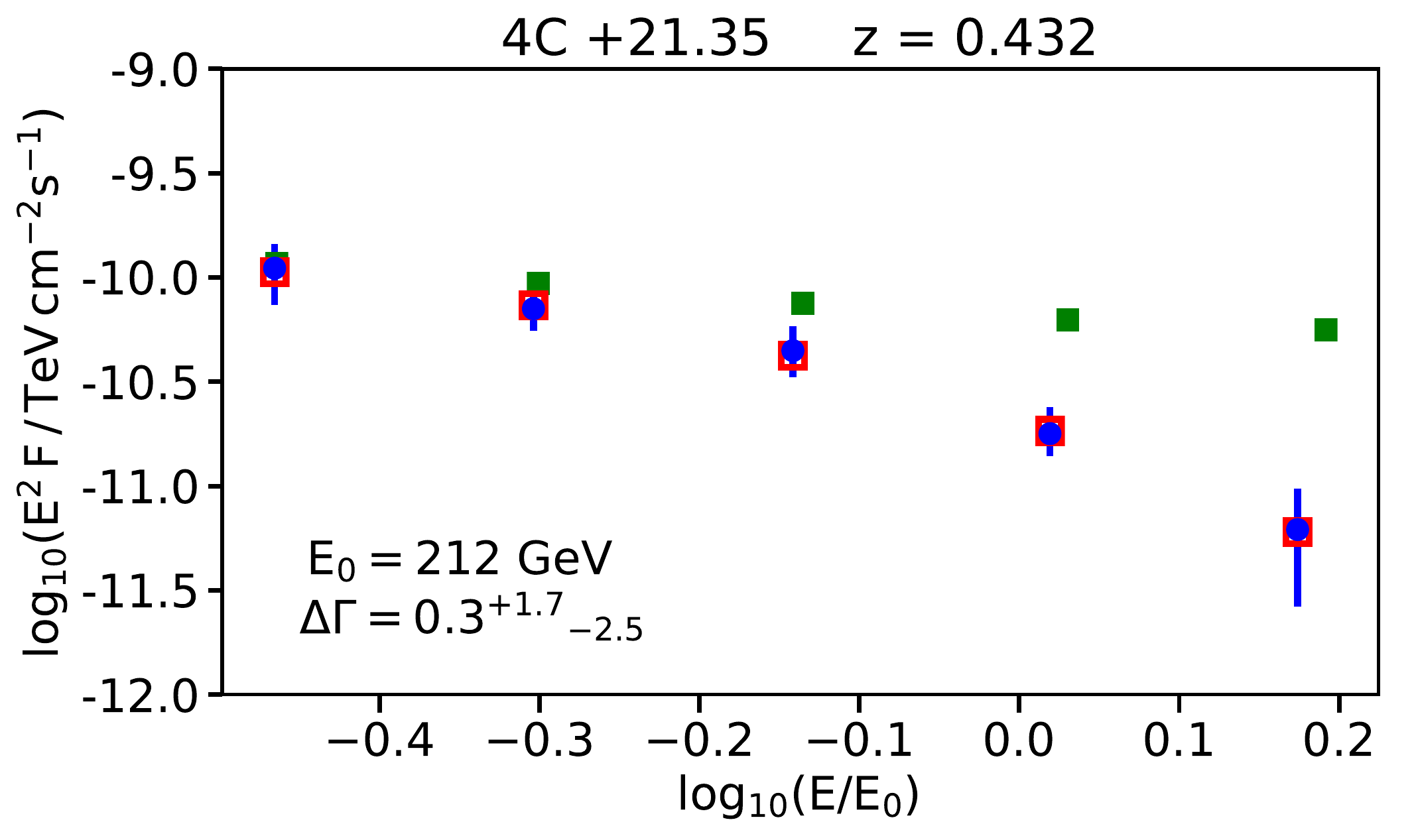}
	}
	\bigskip
	\centerline{%
		\includegraphics[width=0.5\columnwidth]{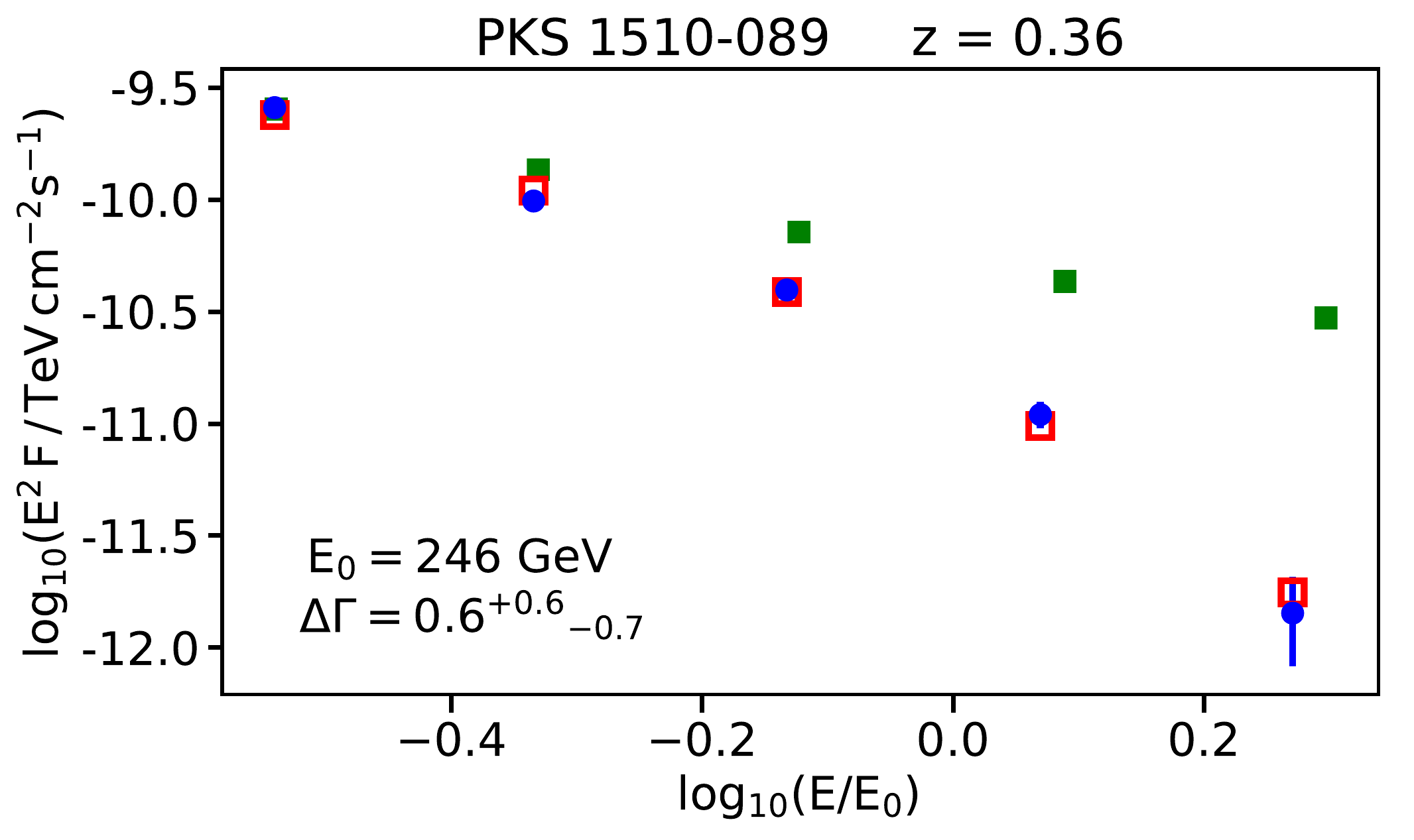}
		~~~~
		\includegraphics[width=0.5\columnwidth]{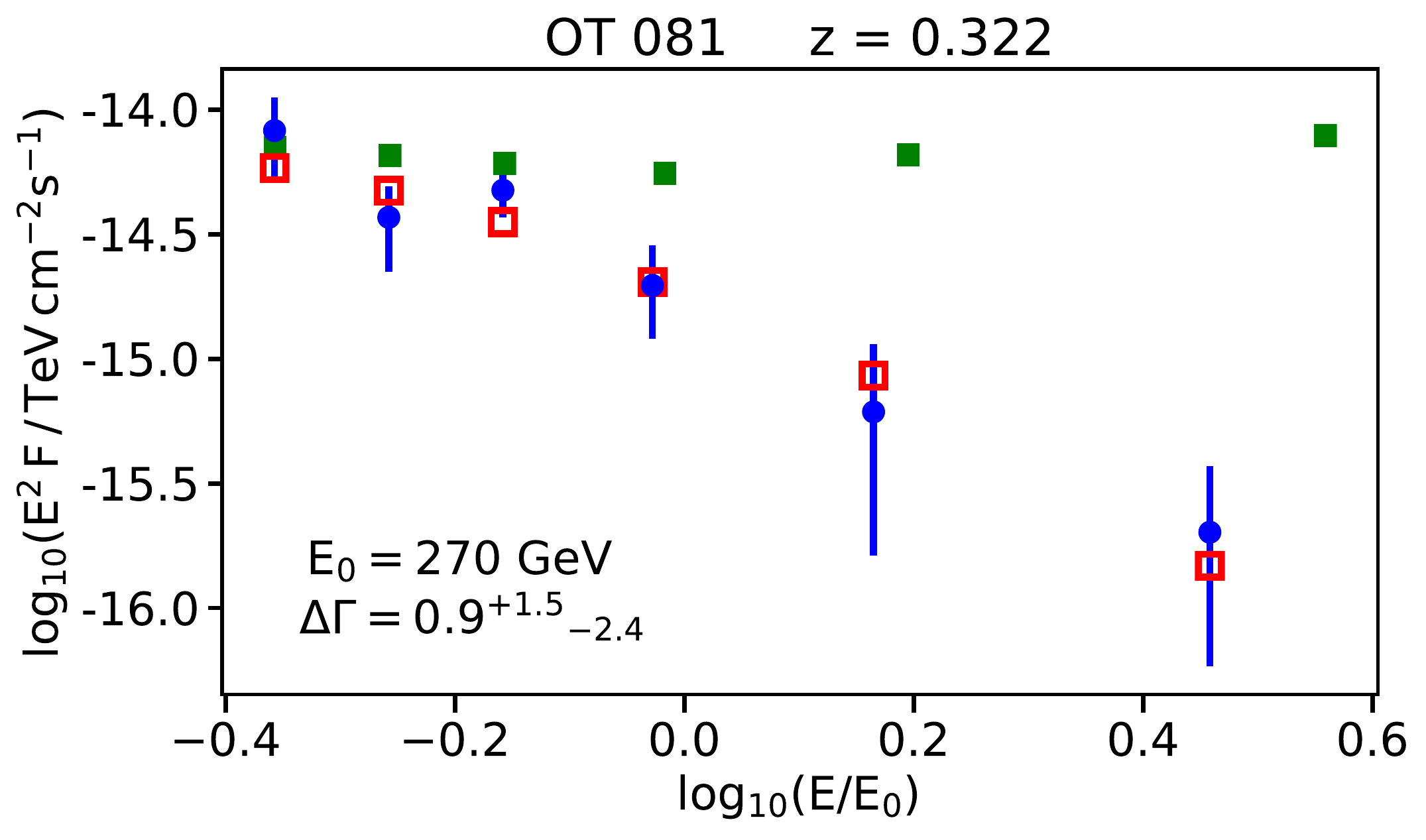}
	}
	\bigskip
	\centerline{%
		\includegraphics[width=0.5\columnwidth]{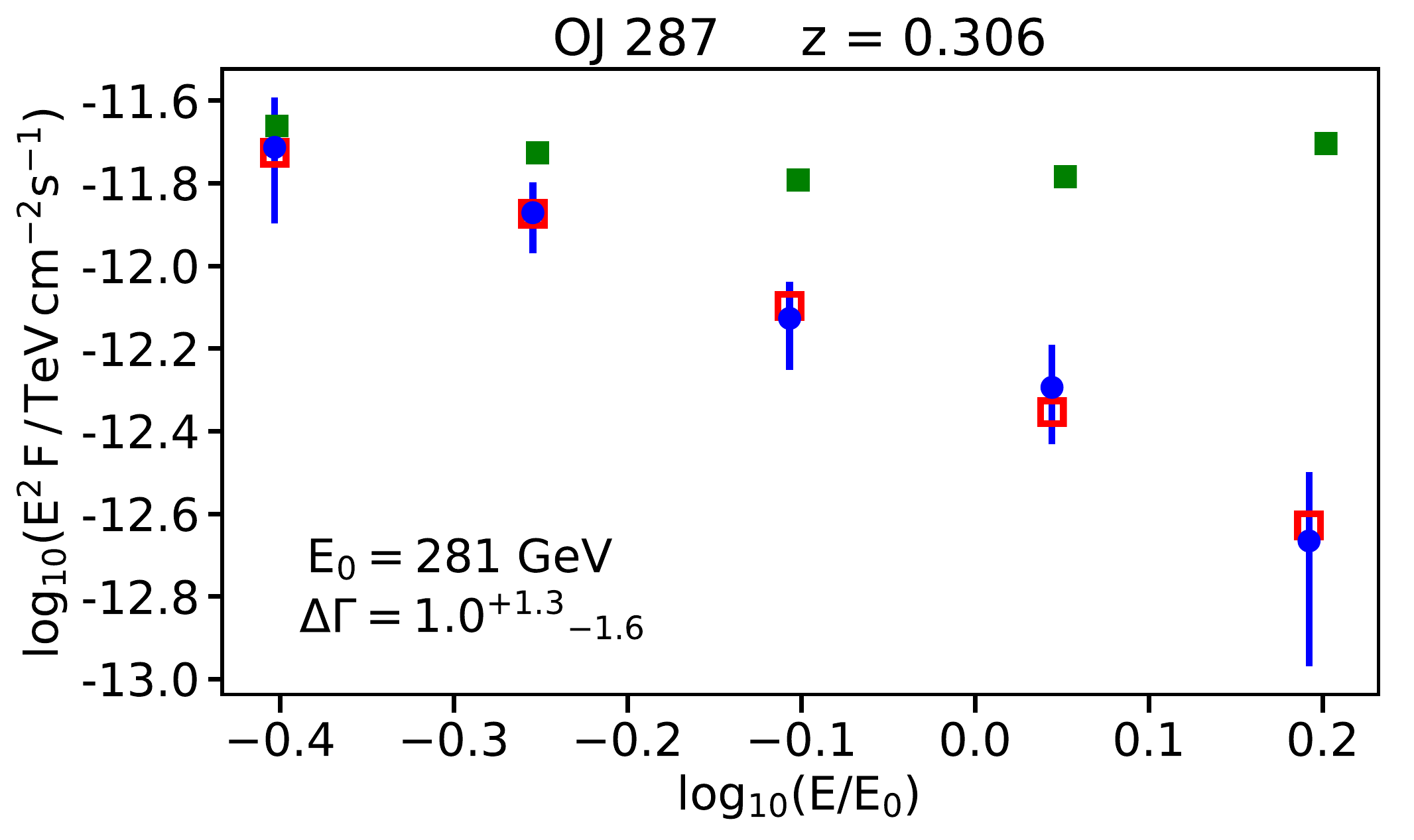}
		~~~~
		\includegraphics[width=0.5\columnwidth]{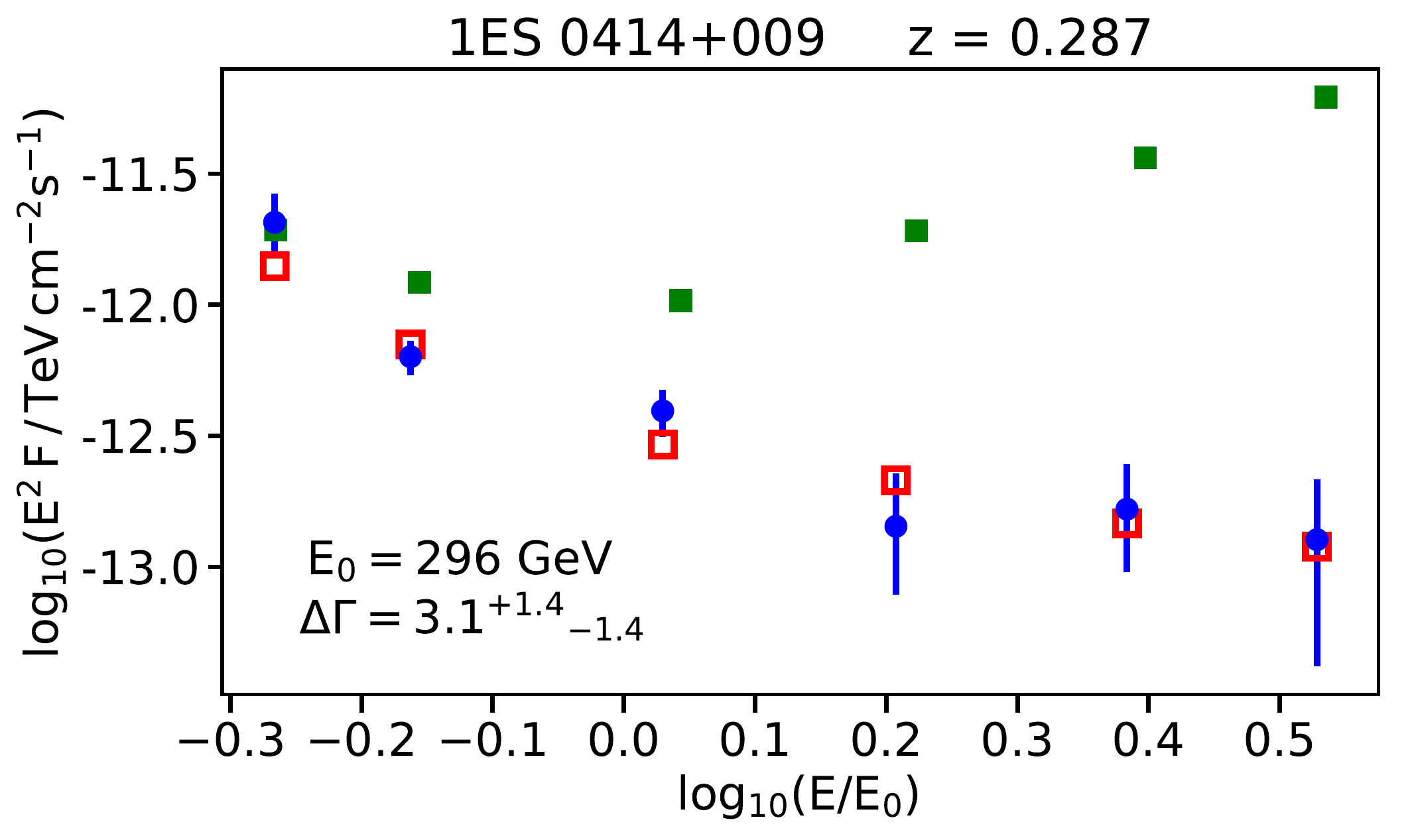}
	}
	\bigskip
	\centerline{%
		\includegraphics[width=0.5\columnwidth]{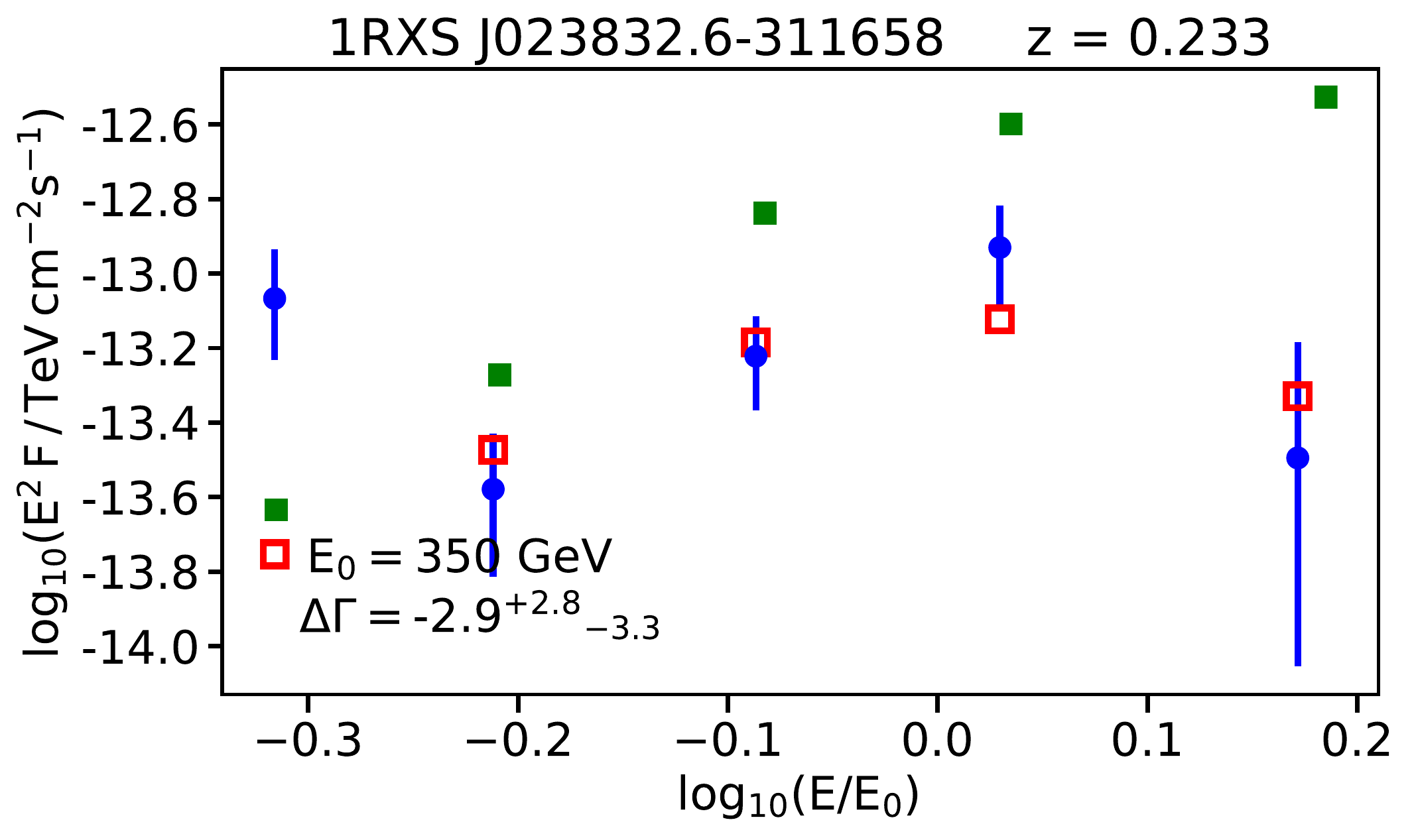}
		~~~~
		\includegraphics[width=0.5\columnwidth]{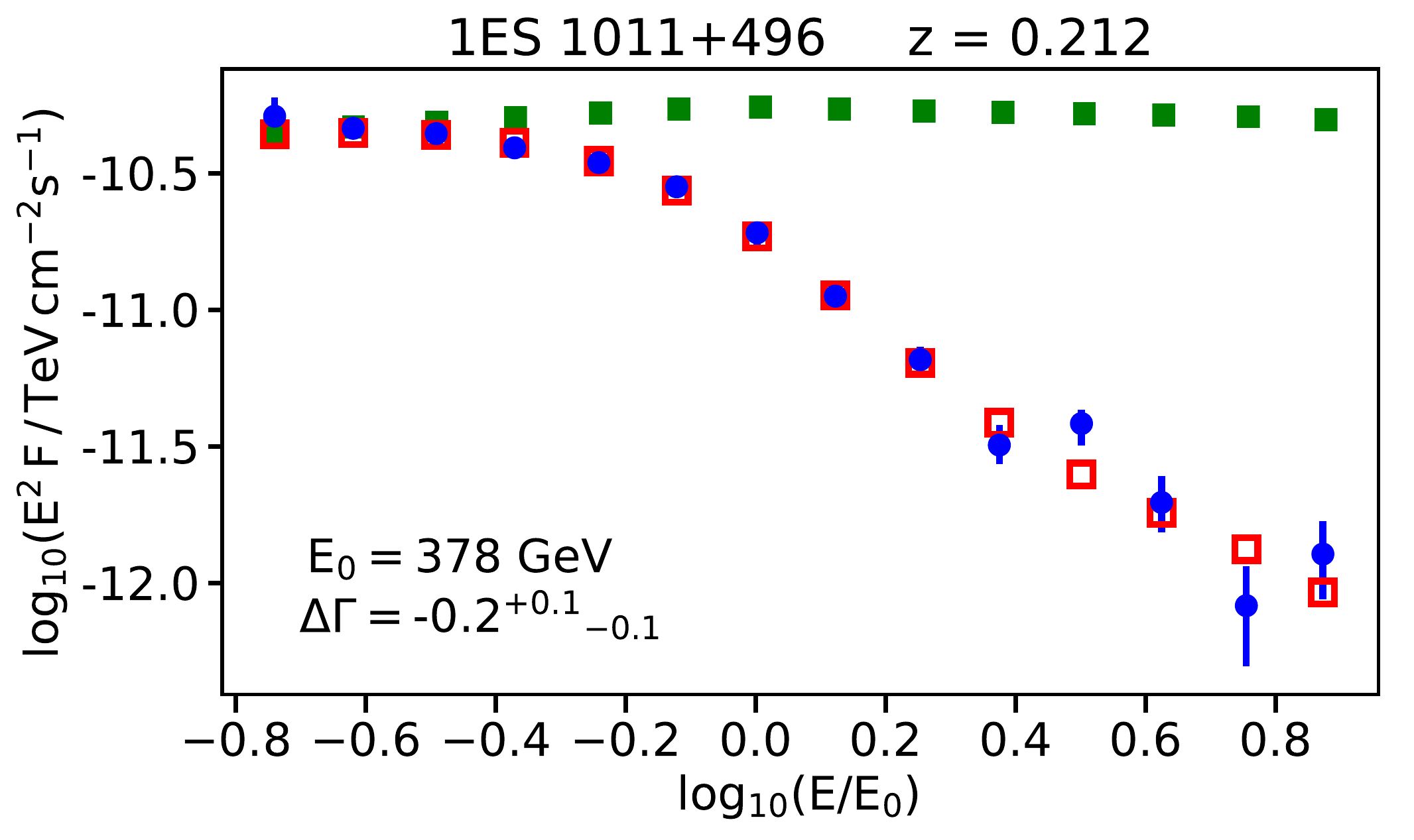}
	}
	\caption{\label{fig:spec2}
			Continuation of Fig.~\ref{fig:spec1}.}
\end{figure}

\begin{figure}
	\centerline{%
		\includegraphics[width=0.5\columnwidth]{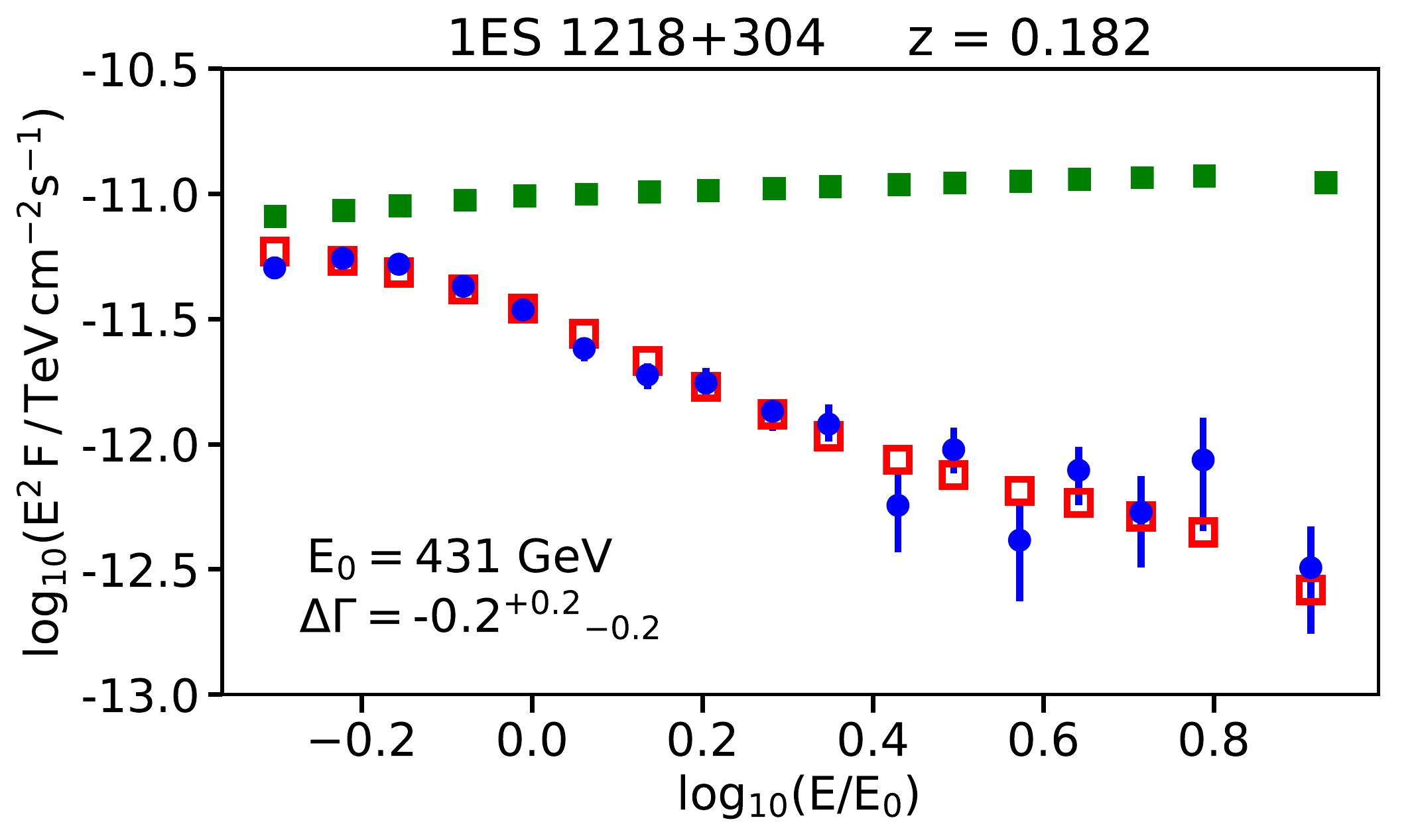}
		~~~~
		\includegraphics[width=0.5\columnwidth]{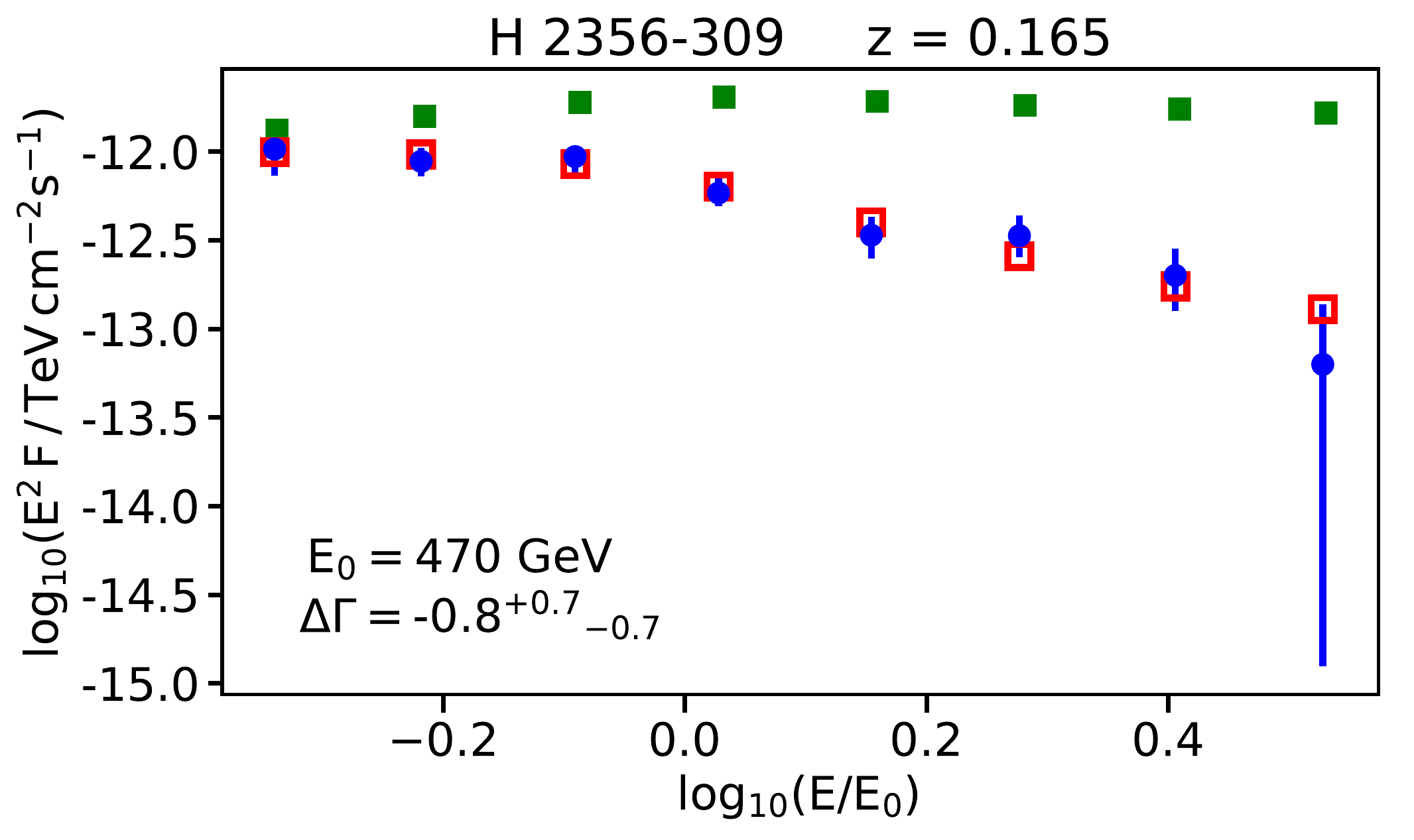}
	}
	\bigskip
	\centerline{%
		\includegraphics[width=0.5\columnwidth]{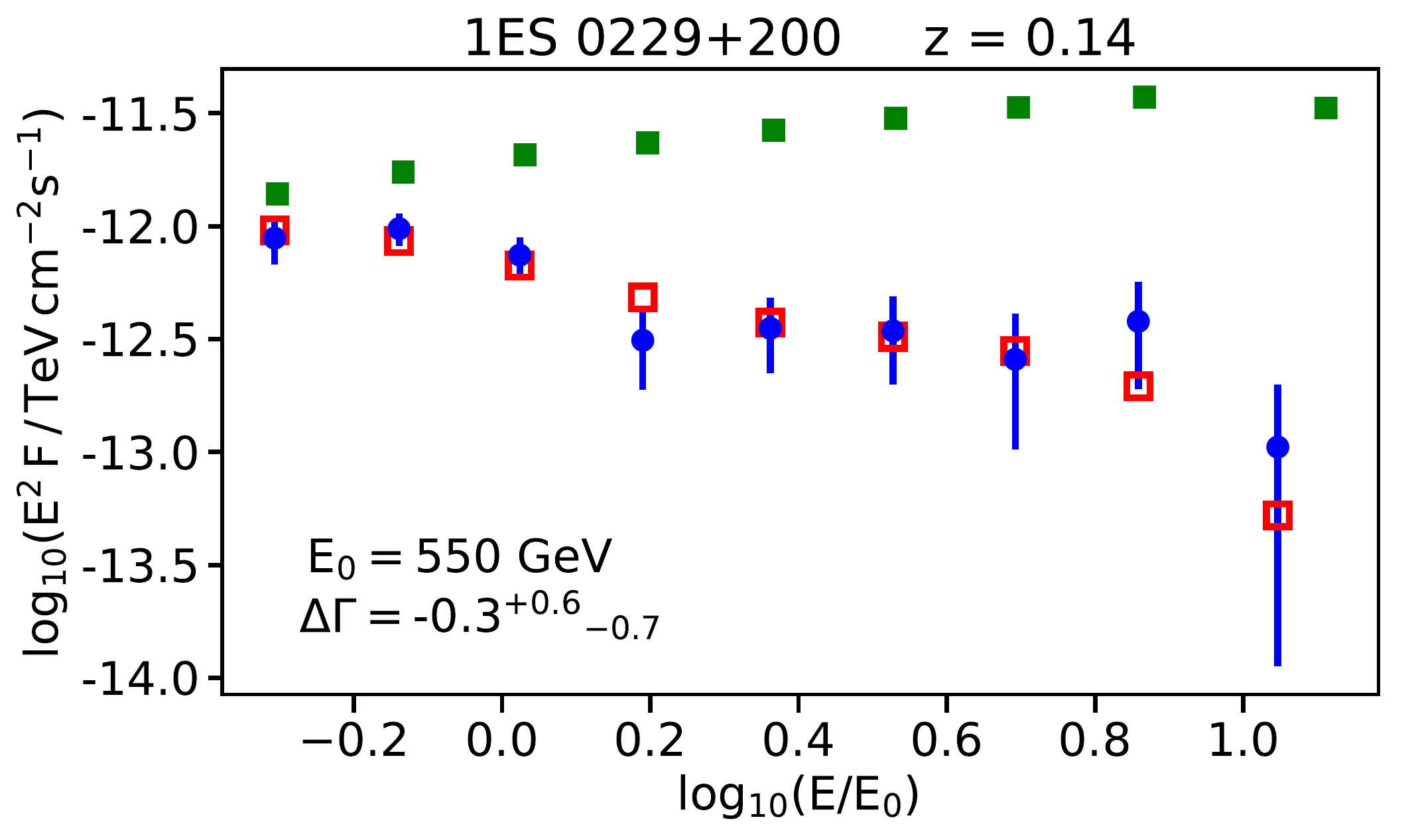}
		~~~~
		\includegraphics[width=0.5\columnwidth]{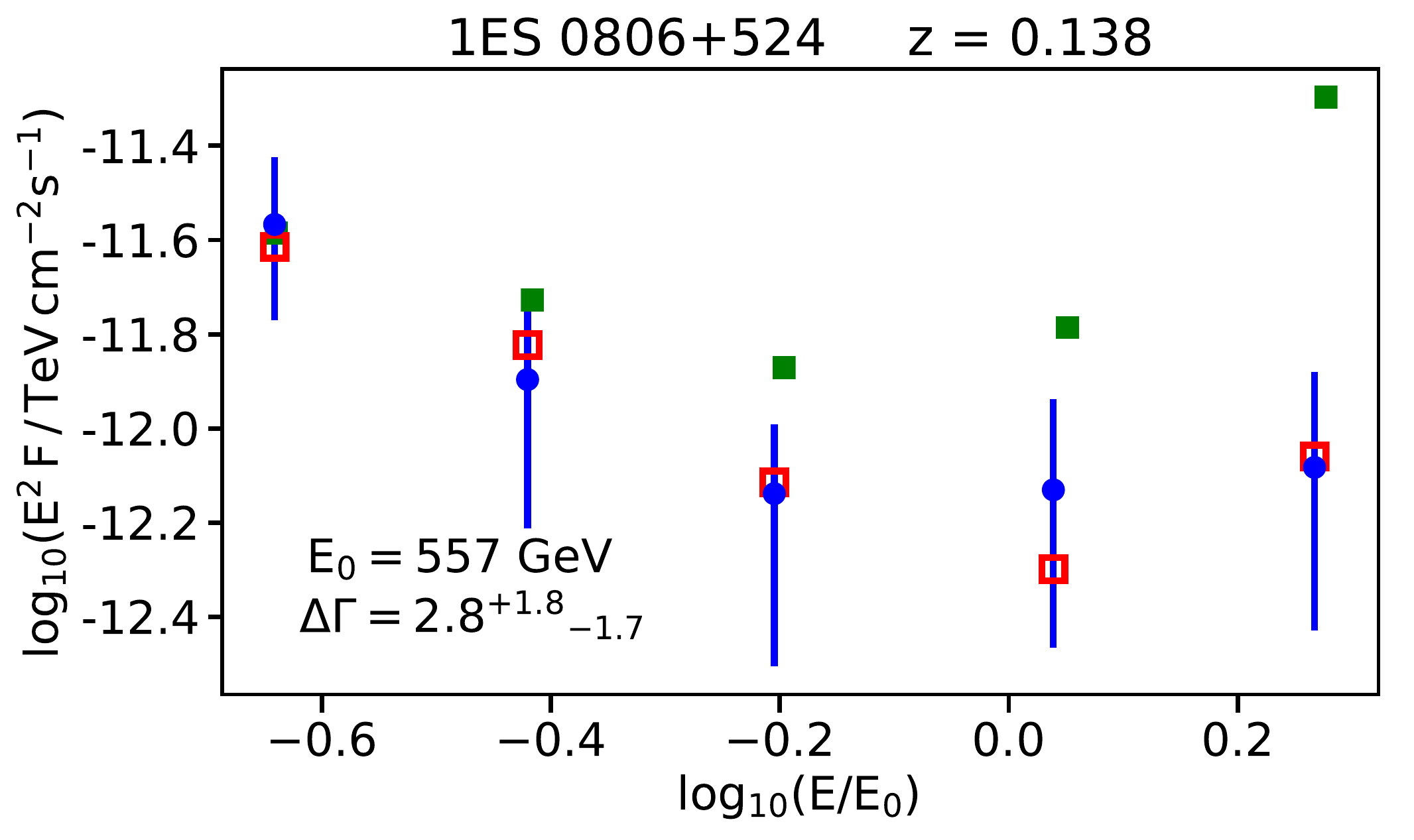}
	}
	\bigskip
	\centerline{%
		\includegraphics[width=0.5\columnwidth]{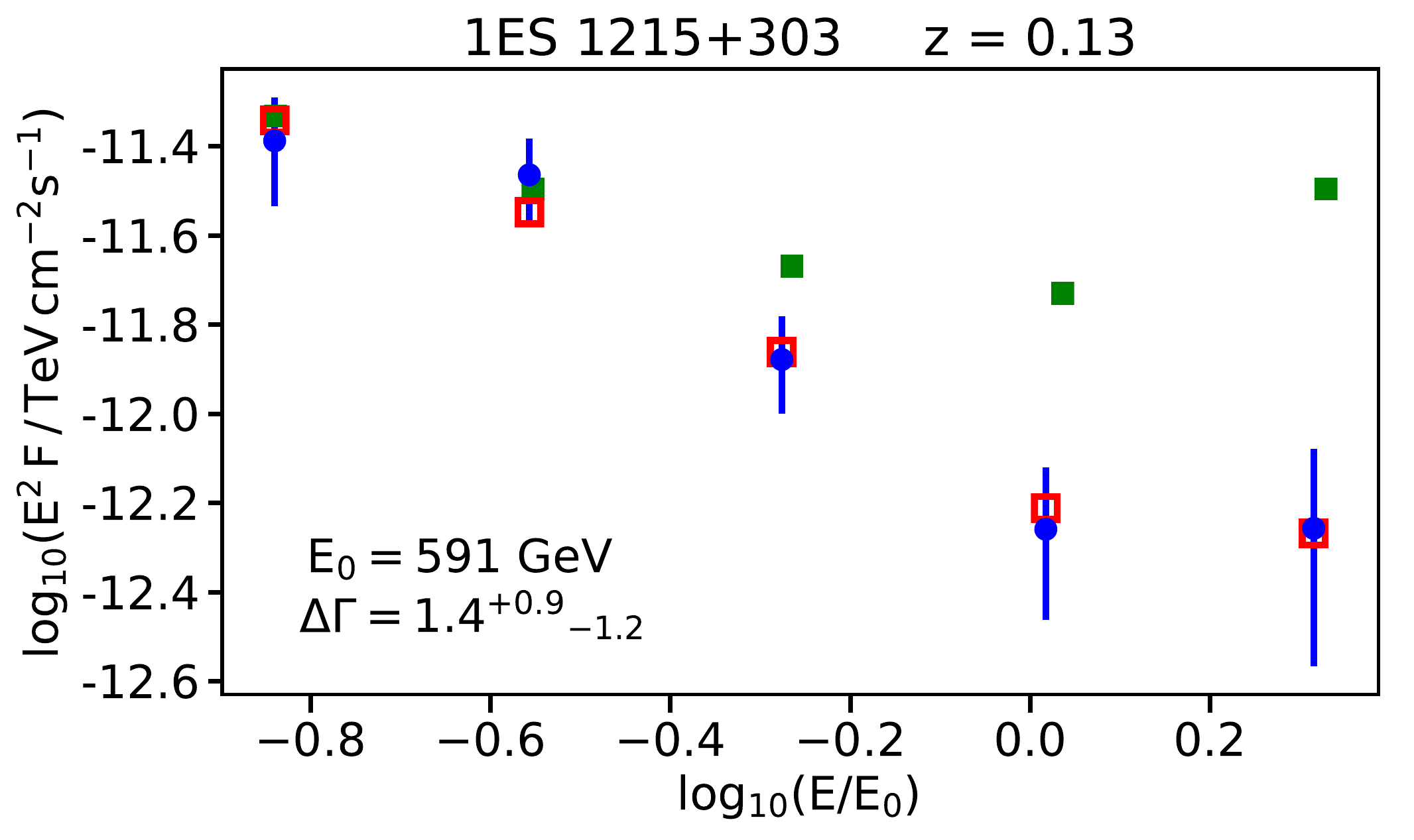}
		~~~~
		\includegraphics[width=0.5\columnwidth]{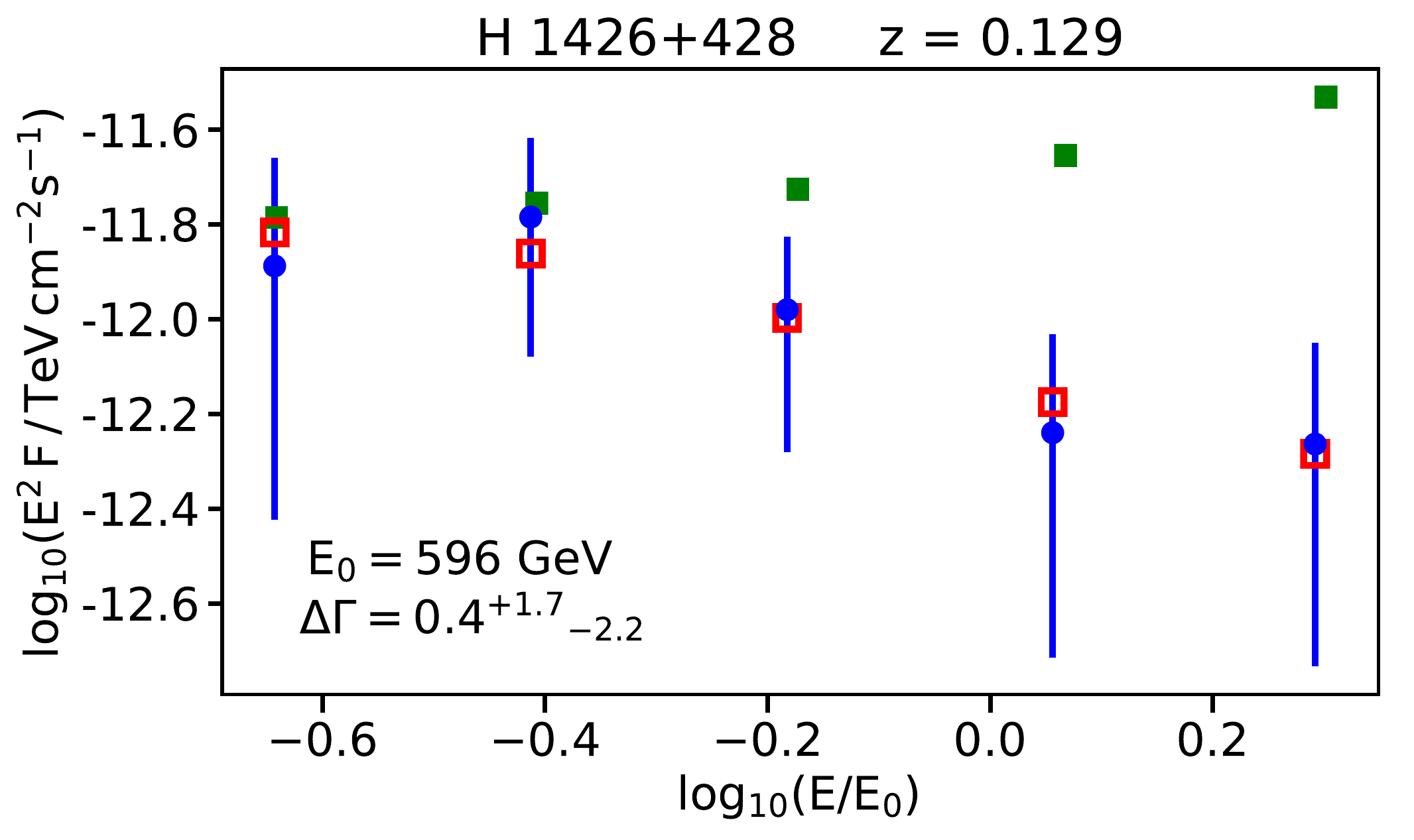}
	}
	\bigskip
	\centerline{%
		\includegraphics[width=0.5\columnwidth]{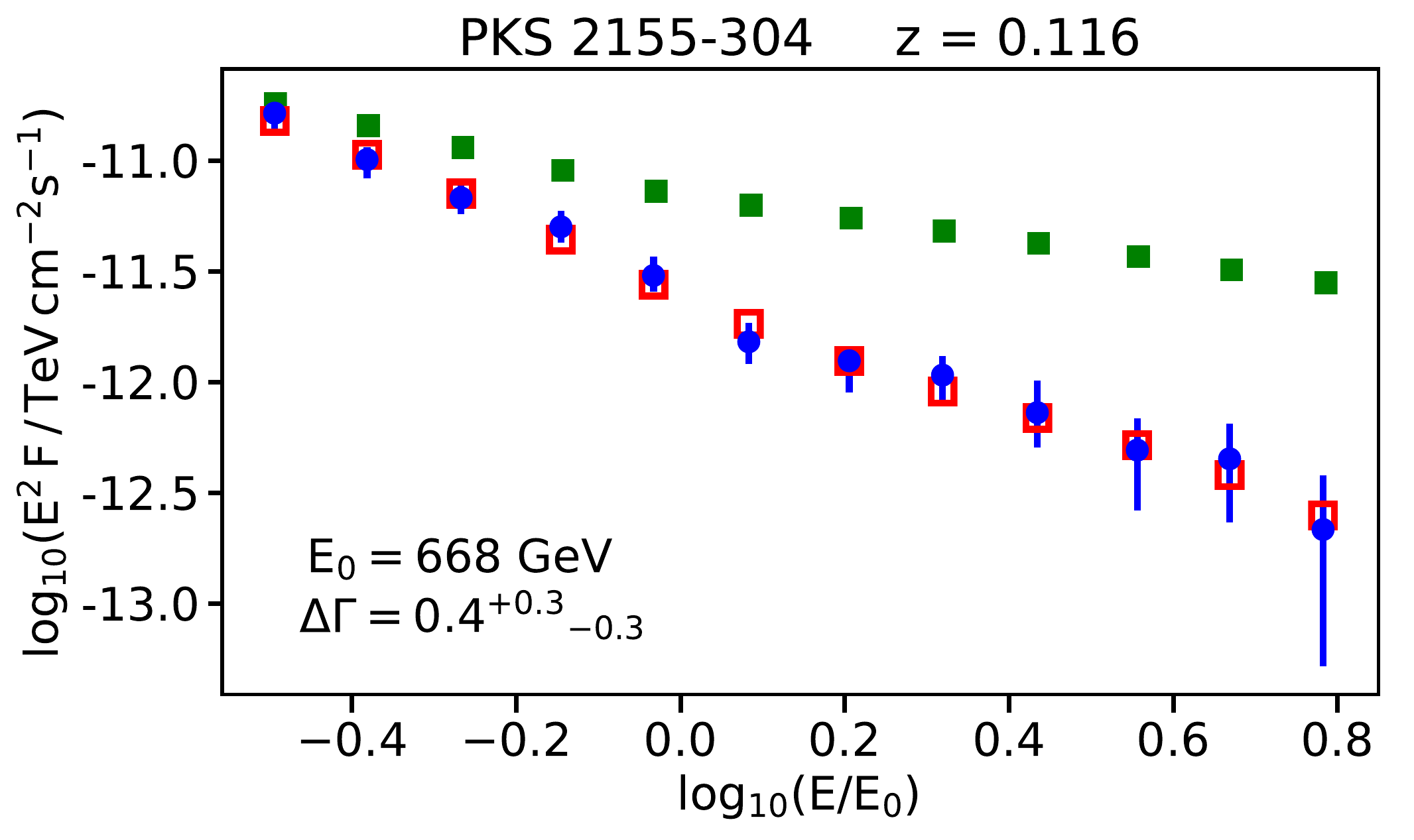}
		~~~~
		\includegraphics[width=0.5\columnwidth]{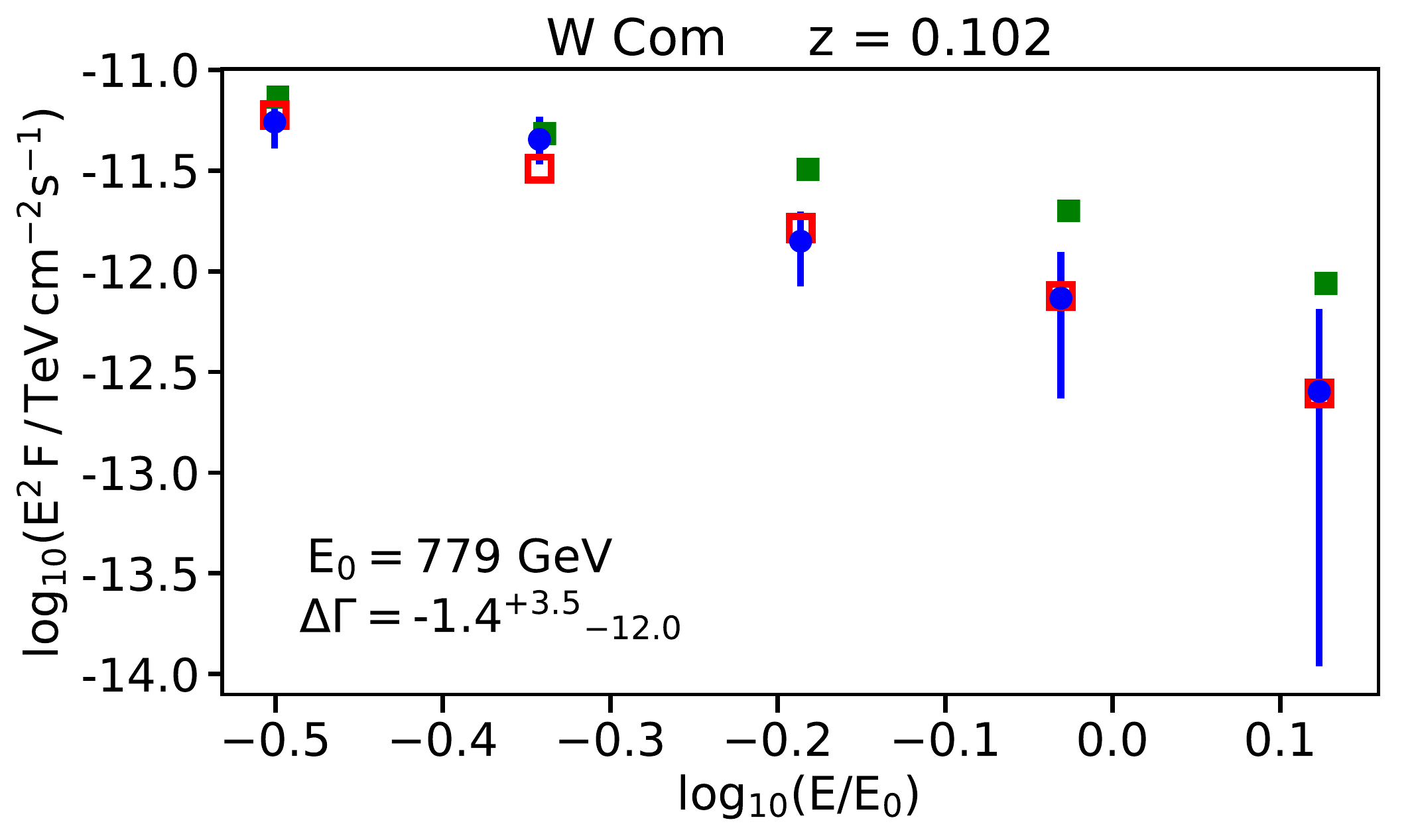}
	}
	\caption{\label{fig:spec3}
			Continuation of Fig.~\ref{fig:spec1}.}
\end{figure}

\begin{figure}
	\centerline{%
		\includegraphics[width=0.5\columnwidth]{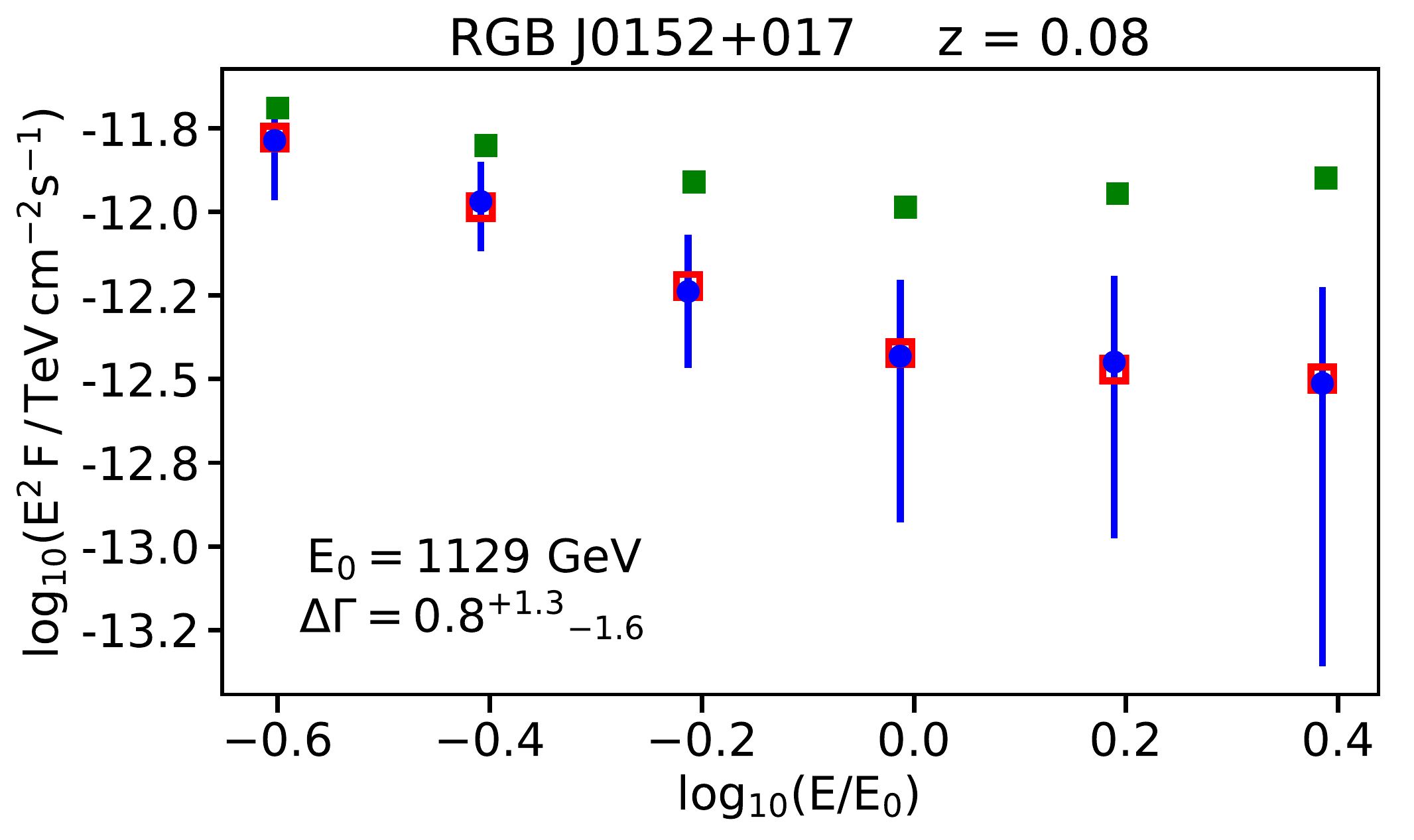}
		~~~~
		\includegraphics[width=0.5\columnwidth]{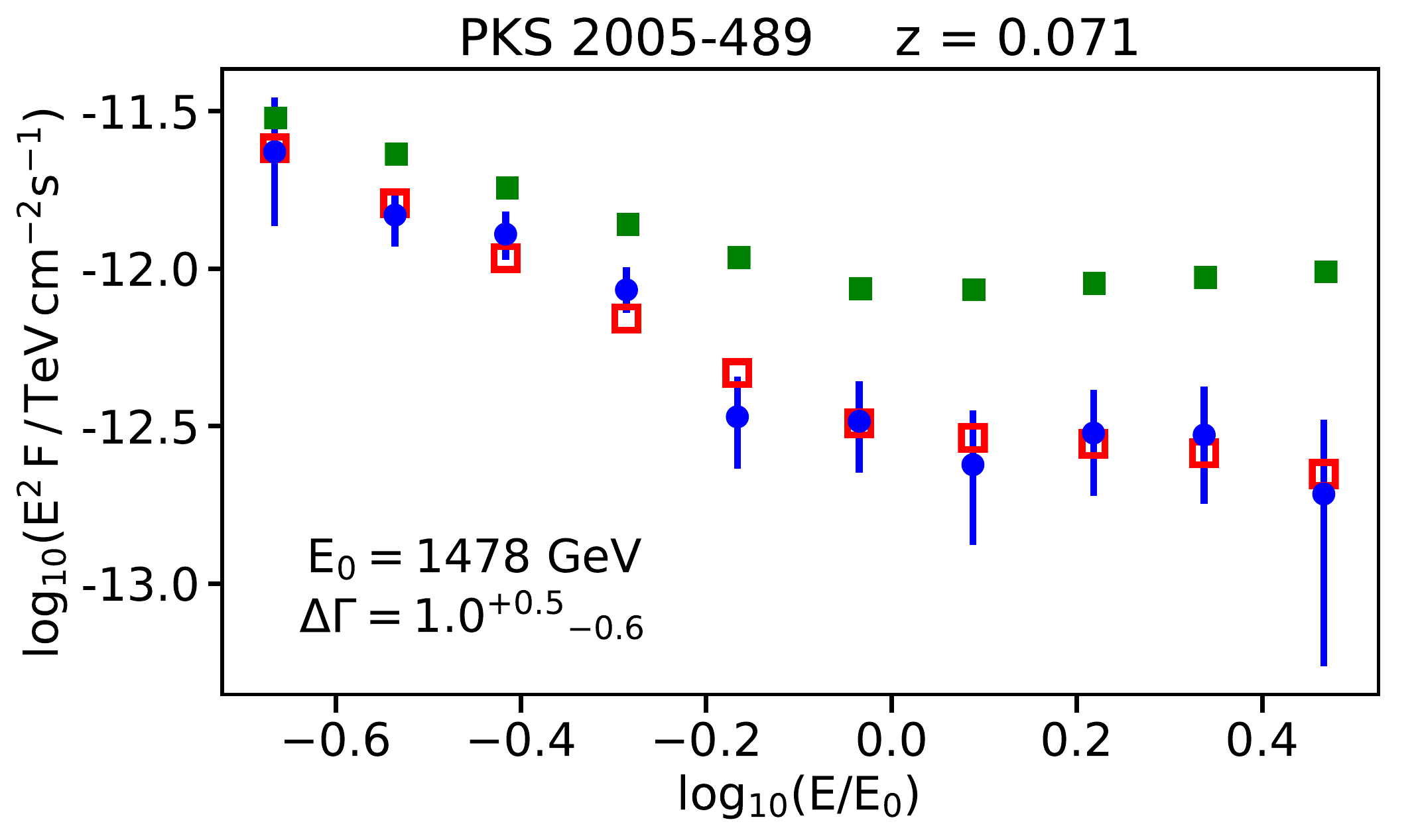}
	}
	\bigskip
	\centerline{%
		\includegraphics[width=0.5\columnwidth]{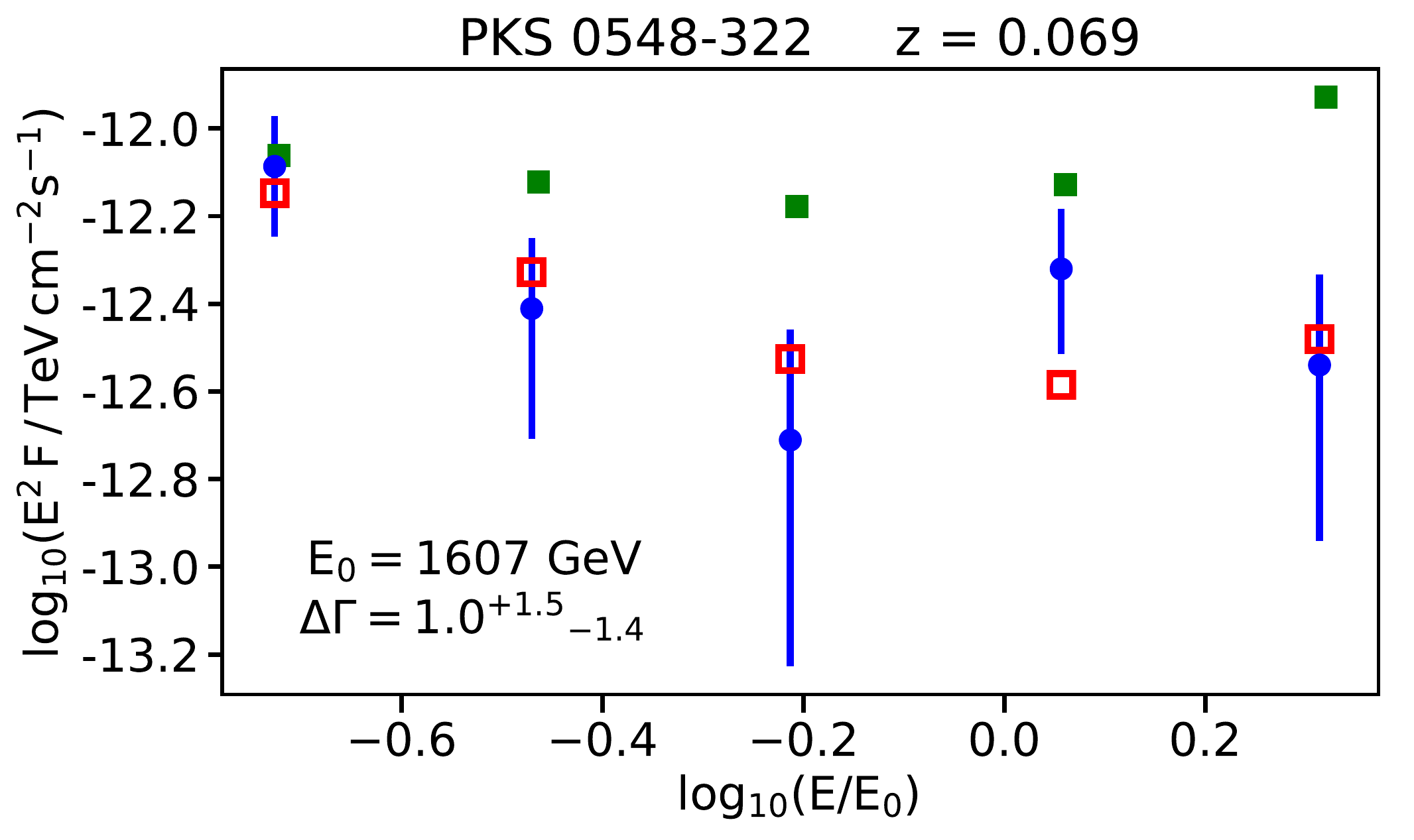}
		~~~~
		\includegraphics[width=0.5\columnwidth]{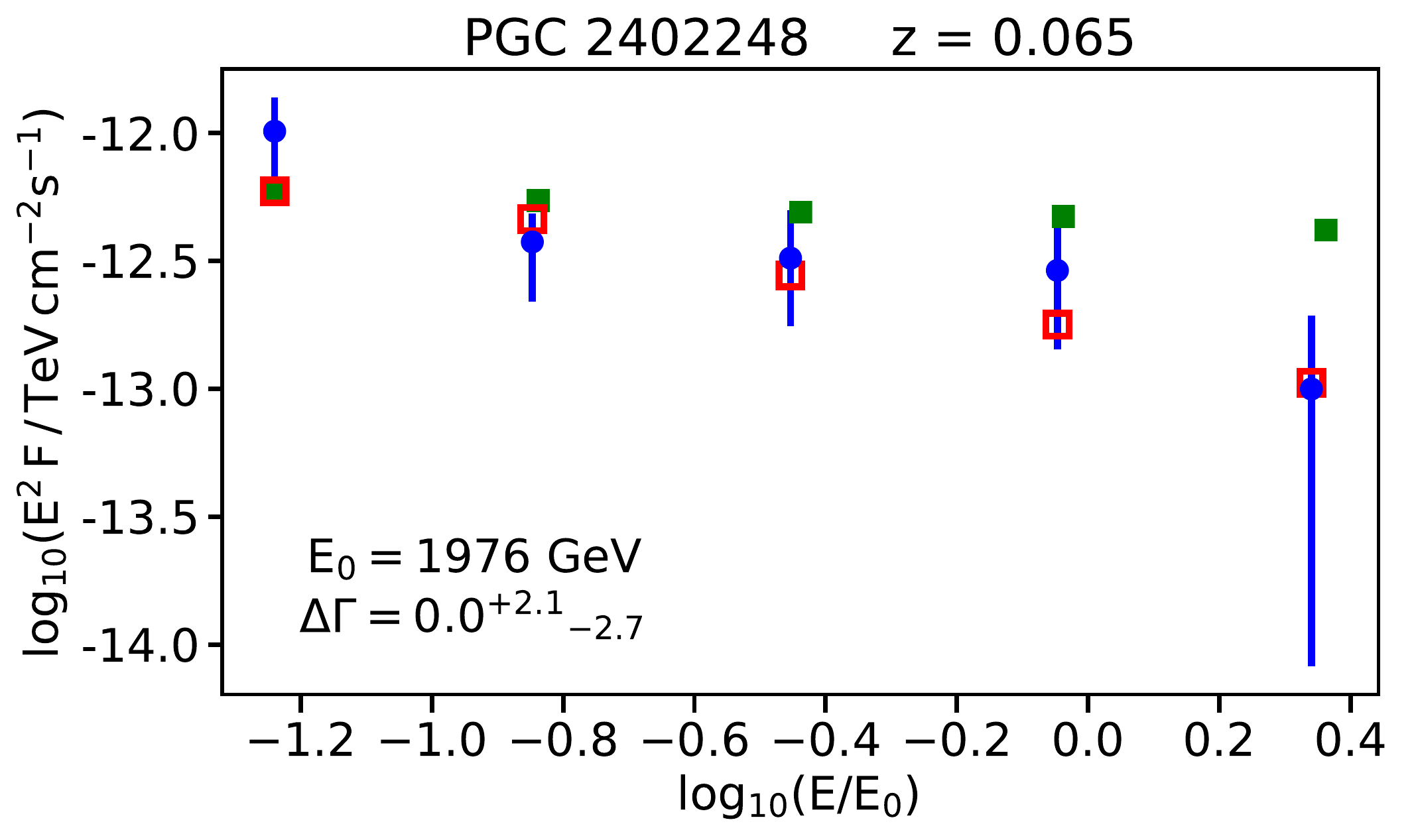}
	}
	\bigskip
	\centerline{%
		\includegraphics[width=0.5\columnwidth]{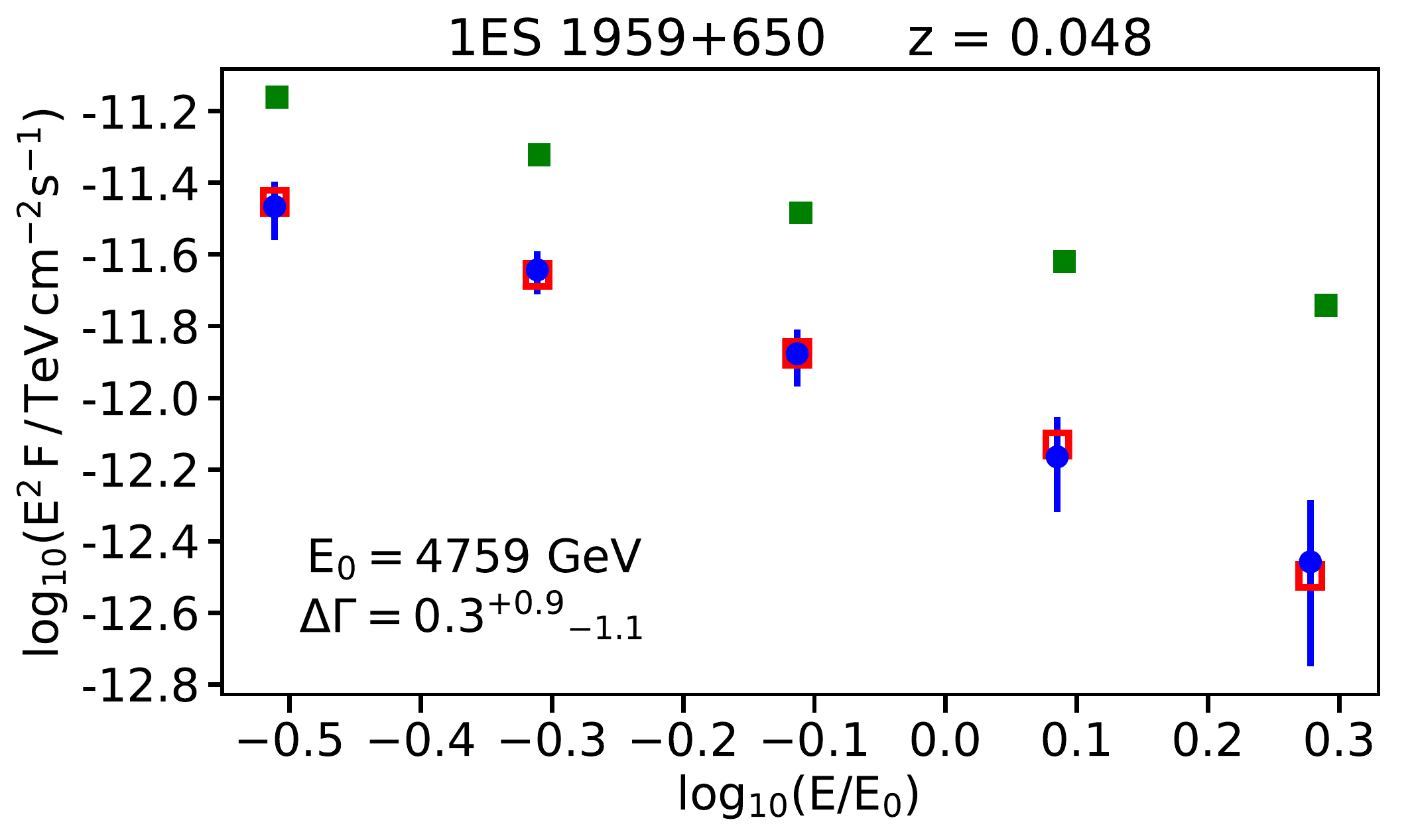}
		~~~~
		\includegraphics[width=0.5\columnwidth]{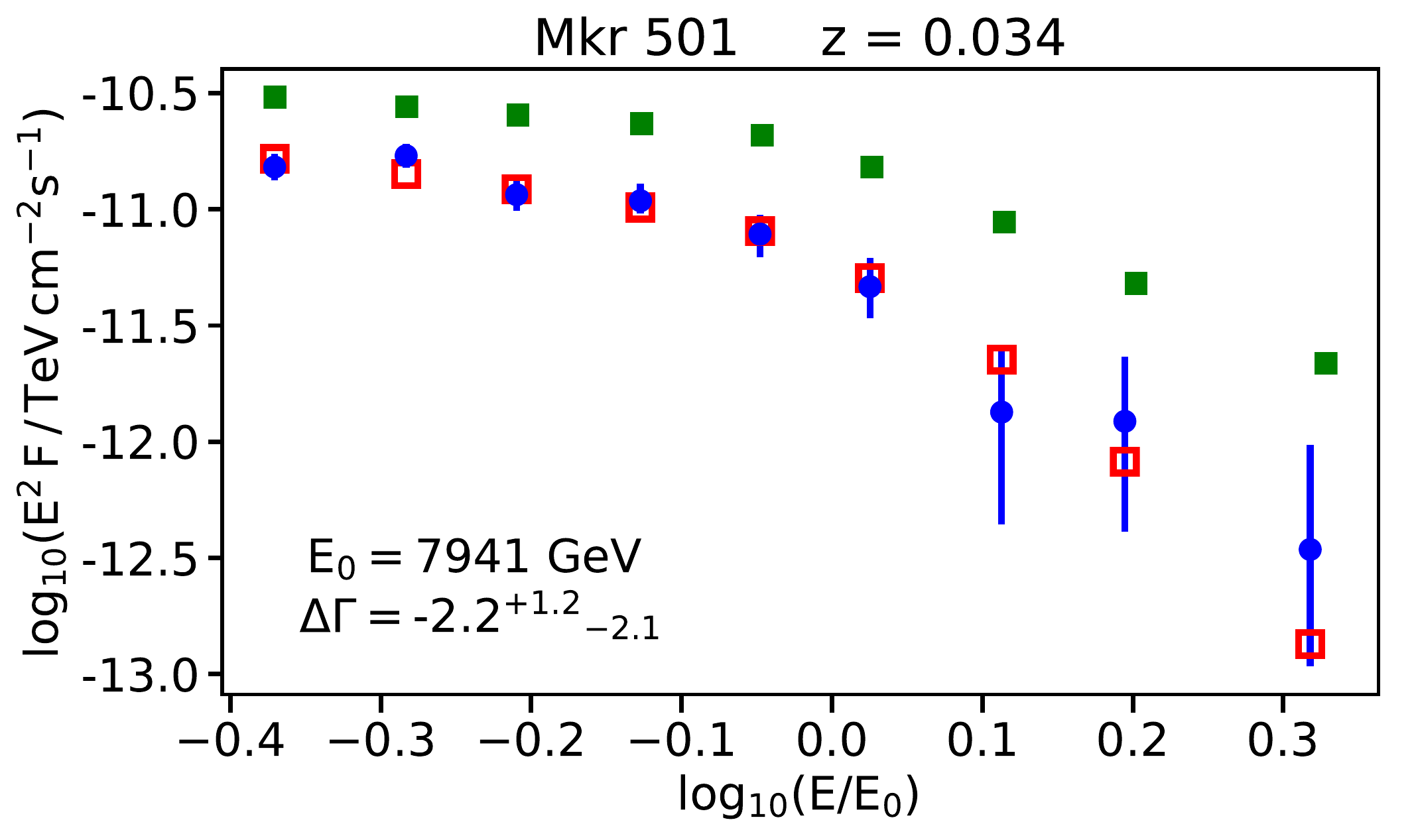}
	}
	\bigskip
	\centerline{%
		\includegraphics[width=0.5\columnwidth]{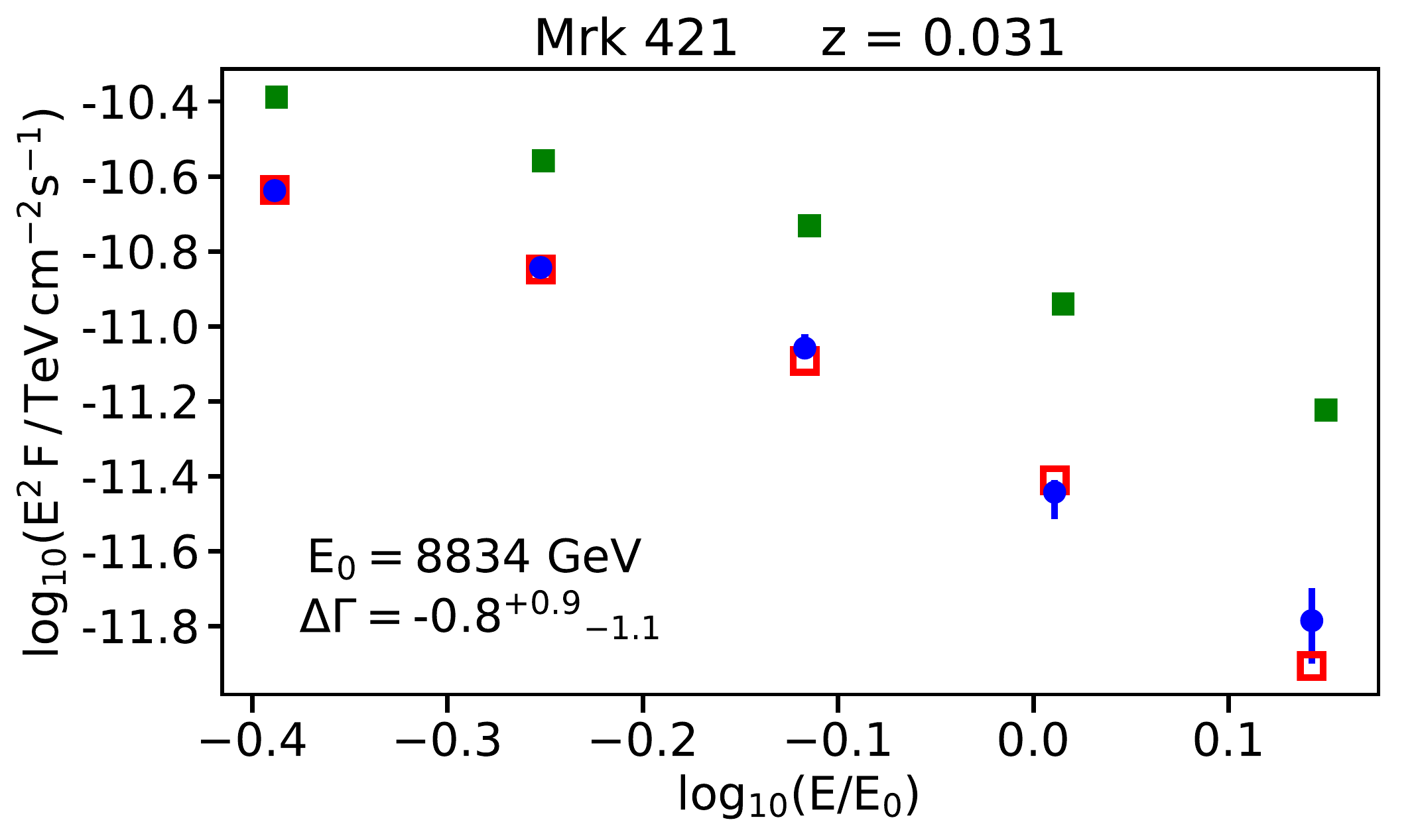}
	}
	\caption{\label{fig:spec4}
		Continuation of Fig.~\ref{fig:spec1}.}
\end{figure}

Figures~\ref{fig:spec1}--\ref{fig:spec4}
present the observed and deabsorbed spectra of \red 31 \black blazars from
our main sample, see Table~\ref{tab:list}, together with values of $E_{0}$
and break strengths $\Delta \Gamma$ on which the main result of the paper
is based. \red The observed spectra and numerical values of spectral
breaks are available online as a supplementary material for this paper.
\black

\acknowledgments
This research has made use of the NASA/IPAC Extragalactic Database
(NED), which is operated by the Jet Propulsion Laboratory, California
Institute of Technology, under contract with the National Aeronautics and
Space Administration. This research has made use of the TeVCat online
source catalog~\cite{TeVCat} and of the data and tools provided by the
Fermi Science Support Center.

We are indebted to Alberto Franceschini for interesting correspondence and
for sharing detailed numerical tables of the absorption
model~\cite{Franceschini2017} with us; to Manuel Meyer for numerous
interesting discussions and helpful comments;
to Timur Dzhatdoev, Oleg Kalashev, Mikhail Kuznetsov, Maxim
Libanov, Valery Rubakov and Igor Tkachev for interesting discussions.
We thank the anonymous reviewer for careful reading of the manuscript
\red and the attention to the study which went beyond the usual referee's
duties.
\black The work of AK and ST on constraining the intensity of
EBL and its applications to astrophysical manifestations of axion-like
particles was supported by the Russian Science Foundation, grant
18-12-00258.

\paragraph{Note added.} Several months after this study was completed and
submitted, an update of Ref.~\cite{Fermi-opacity} was published by the
Fermi LAT Collaboration \cite{Fermi-opacity-new}. We will compare our
results to those of Ref.~\cite{Fermi-opacity-new} in a forthcoming
publication based on Fermi-LAT data only.




\end{document}